\documentclass[twocolumn]{aastex631}

\usepackage[T1]{fontenc}
\usepackage{amssymb}
\usepackage{amsmath}
\usepackage{xspace}
\usepackage{graphbox}
\usepackage{graphicx,xparse,booktabs}

\newcommand{\xmax}{\ensuremath{X_{\mathrm{max}}}\xspace}%

\ExplSyntaxOn
\NewDocumentEnvironment{places}{mm}
 {% #1 is the desired width, #2 is the number of photos per line
  \setlength{\tabcolsep}{0pt} % no space between rows
  \dim_set:Nn \l_places_width_dim
   {
    (#1-\ht\strutbox-\dp\strutbox-0.2pt)/(#2)  % distance of plot rows
   }
  \begin{tabular}{r @{\hspace{2pt}} *{#2}{c}}  % distance of text measured from image
 }
 {
  \end{tabular}
 }

\NewDocumentCommand{\place}{mm}
 {% #1 is the name of the place, #2 is the comma separated list of images
  \seq_set_from_clist:Nn \l_places_images_in_seq { #2 }
  \seq_set_map:NNn \l_places_images_out_seq \l_places_images_in_seq { \places_set_image:n {##1} }
  \seq_put_left:Nn \l_places_images_out_seq
   {
   \scriptsize
    \begin{tabular}{c}\rotatebox[origin=c]{90}{\strut#1}\end{tabular}
   }
  \seq_use:Nn \l_places_images_out_seq { & } \\ \addlinespace
 }

\dim_new:N \l_places_width_dim
\seq_new:N \l_places_images_in_seq
\seq_new:N \l_places_images_out_seq

\cs_new_protected:Nn \places_set_image:n
 {
  \makebox[\l_places_width_dim]
   {
    \begin{tabular}{c}
    \includegraphics[
      width=\l_places_width_dim,
      height=\l_places_width_dim,
      keepaspectratio,
    ]{#1}
    \end{tabular}
   }
 }

\ExplSyntaxOff

\begin{document}
%\title{Tracing cosmic structures through ultra-high-energy cosmic ray data}
\title{Constraints on UHECR sources and extragalactic magnetic fields from directional anisotropies}

%Reflections of cosmic structures in UHECRs

\author[0000-0003-4005-0857]{Teresa Bister}
\email{teresa.bister@ru.nl}
\affiliation{Institute for Mathematics, Astrophysics and Particle Physics, Radboud University Nijmegen, Nijmegen, The Netherlands}
\affiliation{Nationaal Instituut voor Kernfysica en Hoge Energie Fysica (NIKHEF), SciencePark, Amsterdam, The Netherlands}

\author[0000-0003-2417-5975]{Glennys R. Farrar}
\email{gf25@nyu.edu}
\affiliation{Center for Cosmology and Particle Physics, New York University, New York, NY 10003, USA}

% \author{include Chen / Noemie?}

\begin{abstract}
A dipole anisotropy in ultra-high-energy cosmic ray (UHECR) arrival directions, of extragalactic origin, is now firmly established at energies $E>8\,\mathrm{EeV}$. Furthermore, the UHECR angular power spectrum shows no power at smaller angular scales than the dipole, apart from hints of possible individual hot or warm spots for energy thresholds $\gtrsim$ 40 EeV.  
Here, we exploit the magnitude of the dipole and the limits on smaller-scale anisotropies to place constraints on two quantities: the extragalactic magnetic field (EGMF) and the number density of UHECR sources or the volumetric event rate if UHECR sources are transient. We also vary the {\it bias} between the extragalactic matter and the UHECR source densities, reflecting whether UHECR sources are preferentially found in over- or under-dense regions, and find that little or no bias is favored. 

We follow~\citet{ding_imprint_2021} in using the Cosmic Flows 2 density distribution of the local universe as our baseline distribution of UHECR sources, but we improve and extend that work by employing an accurate and self-consistent treatment of interactions and energy losses during propagation.  Deflections in the Galactic magnetic field are treated using both the full JF12 magnetic field model, with random as well as coherent components, or just the coherent part, to bracket the impact of the GMF on the dipole anisotropy. This Large Scale Structure (LSS) model gives good agreement with both the direction and magnitude of the measured dipole anisotropy and forms the basis for simulations of discrete sources and the inclusion of EGMF effects.   
\end{abstract}


\section{Introduction} \label{sec:intro}
Ultra-high-energy cosmic rays (UHECRs), characterized by energies exceeding $10^{18}\,\mathrm{eV}$, continue to challenge our understanding of astrophysical processes in the Universe. Large numbers of UHECRs have been detected by the giant Pierre Auger Observatory~\citep{auger_2015} (Auger) in the southern hemisphere and the Telescope Array~\citep{KAWAI2008221} (TA) in the northern hemisphere, leading to robust determinations of the UHECR energy spectrum and arrival directions, and approximate information on the mass composition.

A major obstacle to finding the sources of UHECRs arises from their interactions with magnetic fields during their propagation from sources to Earth, which -- given the uncertainty on the UHECR charge, $Z$, and the intervening magnetic fields -- impedes straightforward source identification. Understanding UHECR trajectories demands a good knowledge of the Galactic magnetic field (GMF) within our Milky Way. For that, several models have been in the literature; see, e.g., the  review by~\citet{Jaffe:2019iuk}.  A comprehensive new suite of models encompassing the range of possible coherent GMF models consistent with the latest data has very recently been released (\citet{UF23}; UF23 below). %The deflections predicted with the classic JF12 model~\citep{jansson_galactic_2012} are found to lie centrally within the range found for all allowed models~\citep{UF23}, justifying our use of simulations of UHECR deflections using JF12 coherent and coherent+random fields for this study.
In this paper, we use the deflections predicted with the classic JF12 model including random component~\citep{jansson_new_2012}, and bracket the sensitivity to the random field by using the JF12 coherent field only. \citet{UF23} checked that deflections in the JF12 coherent field generally fall within the range found for the UF23 suite of updated models, for the rigidities tested, so the results of the present study should be qualitatively robust. %At the time of writing, the UF23 ensemble of coherent fields was not yet available. We will update this work in the near future, using these models. %When the UF23 field models including the random component become available in the near future, the approach introduced here will be updated.  Fits to the direction of the UHECR dipole may help select among the different UF23 models.

In addition to the GMF, the extragalactic magnetic field (EGMF) plays a role at some level in UHECR propagation. The strength, spatial distribution, and even origin of the EGMF is unclear both theoretically and observationally. Limits on parameters of the EGMF exist from theoretical expectations and observations of the CMB or $\gamma$-rays (see review~\citet{Durrer_2013}), but large parts of the parameter space are still unconstrained. One of the accomplishments of the present work is placing new constraints on key parameters of the EGMF through a combination of its field strength and coherence length.  %These constraints should improve further when the UF23 models including individualized random components become available.

The observation of large-scale anisotropies in the arrival directions of UHECRs allows for valuable insights into the distributions of sources, and at the same time into the interactions with cosmic magnetic fields. The Pierre Auger collaboration made the important discovery of a large-scale dipole anisotropy of the UHECR arrival directions above 8 EeV~\citep{Auger_dipole_2017_Science, Auger_dipole_2018}, with a magnitude of $7.3\%$ above 8 EeV and a significance of $6.9\sigma$~\citep{Golup_ADs_2023}. The Telescope Array data supports these findings, but without being able to significantly differentiate from isotropy due to limited statistics~\citep{Abbasi_2020}. The dipole strength increases with the energy threshold. Its direction is well away from the Galactic center, adding to evidence for an extragalactic origin of UHECRs at these energies. Another important Auger result is that all higher multipole moments are compatible with isotropy~\citep{Almeida_dipole_2021}. This is also corroborated by results of the combined working group between Auger and TA~\citep{Caccianiga_ICRC2023}.

In the absence of an obvious dominant source, a natural hypothesis is that UHECR sources follow the local matter density, and thereby the large-scale structure (LSS) of the Universe. This scenario was investigated in e.g.~\citet{Harari_2010, Harari_2015, Globus_2017, globus_cosmic_2019, Auger_dipole_2017_Science, ding_imprint_2021} with promising results indicating that the dipole amplitude and direction may be due to the LSS. Here, we make significant technical improvements on the work of~\citet{ding_imprint_2021}. Our results corroborate the conclusion that the LSS naturally can explain the observed dipole. We then exploit the dipole and higher multipole anisotropy observations to draw new and important conclusions about UHECR sources and extragalactic magnetic fields. Specifically, we place limits on the \emph{bias} between UHECR sources and the LSS (whether UHECR sources populate predominantly matter-dense or underdense regions), and we constrain extragalactic magnetic fields. This enables us to place improved constraints on the source density, for steady sources, or the volumetric rate if sources are transients.  These constraints allow the possible classes of UHECR sources to be further delimited. 

% Finally, we make predictions of the model regarding the distribution of composition and rigidity ($R=E/Z$, defined here as energy $E$ divided by charge $Z$) over the sphere (Sec.~\ref{sec:predictions}).

Some of these ideas were investigated in~\citet{allard_what_2022}, but without a fit of the injection parameters and with a different treatment of propagation effects. A distinction between~\citet{allard_what_2022} and our analysis is the different approaches to modeling the UHECR source distribution. Determination of the local galaxy or matter distribution is ultimately strongly reliant on the 2MRS flux-limited catalog, which excludes 8$^\circ$ from the Galactic plane (5$^\circ$ in the outer Galaxy). To avoid distance-dependent selection biases, users must create a volume-limited sub-catalog. To do that to $176\,\mathrm{Mpc}$, given the 2MRS flux limit, requires restricting to such bright sources that the number density is $1.1 \times 10^{-4}\, {\rm Mpc}^{-3}$ ~\citep{allard_what_2022}, just 1\% the density of ordinary galaxies~\citep{Conselice_2016}.  %limiting the range of possible source number densities that can be explored to $n \lesssim 7.6 \times 10^{-3}\, {\rm Mpc}^{-3}$. 
This biases the source distribution to very bright galaxies and limits the ability to make independent samples of sources except for low source densities, which they address by some procedure to generate synthetic galaxy catalogs at higher density. 
%This also requires that large peculiar velocities, leading to noise in the determination of galaxy distances, be corrected which is challenging. Instead of trying to do those things ourselves, we adopt the
Instead, we adopt the \textit{CosmicFlows2}~\citep{Hoffman_2018} 3D matter density, which was determined by self-consistently modeling galaxy peculiar velocities as well as distances, angular positions, and redshifts, including very careful analysis of all the available distance measures for each galaxy used in their modeling.  To our mind, this is the optimal conceptual approach to finding the matter density distribution in the angular region observed by 2MRS, and moreover thanks to self-consistent modeling of the potential in which observed galaxies move, and inclusion of more observations, the mass distribution in the region behind the Galactic plane is also approximately described. Note that the computed CosmicFlows matter density field is a smooth representation of the LSS and hence is not comprised of individual galaxies.
We find that approximating the UHECR source density by the CosmicFlows matter density without a bias gives a good fit to the observations, and that restricting to regions of high mass density leads to significantly worse agreement with the dipole. 

Our paper is structured as follows. First, the analysis pipeline is introduced in Sec.~\ref{sec:pipeline}, including the modeling of the source setup, propagation, magnetic fields, and the fit procedure. The results for the baseline case are presented in Sec.~\ref{sec:fit_results}, and the influence of possible model uncertainty is elaborated in Appendix~\ref{sec:model_uncertainties}. 
In Sec.~\ref{sec:bias}, we investigate the possibility of a bias between the UHECR source distribution and the matter density distribution.
Then, the constraining power of the model regarding the extragalactic magnetic fields as well as the source number density is investigated in Sec.~\ref{sec:constraints}, and astrophysical implications of the constraints are presented. Finally, we conclude with a summary in Sec.~\ref{sec:conclusions}.
%\textit{Parts of this work have already been presented in the form of a conference proceeding in~\citet{ICRC_2023}.}

\section{Analysis pipeline: modeling UHECRs from sources to Earth} \label{sec:pipeline}
In this section, the analysis pipeline is described, i.e. the injection, propagation, and the deflections in magnetic fields. The fit method and likelihood function are presented in Sec.~\ref{sec:fit_method}.

\subsection{Source distribution and injection} \label{sec:injection}
The baseline distribution of UHECR sources is taken to follow the (predominantly dark) matter density of the nearby universe, as determined in by \textit{CosmicFlows2}~\citep{Hoffman_2018} which is based on self-consistently combining peculiar galaxy velocities with redshifts and angular positions. 
We use the CosmicFlows densities between 0 and 360\,Mpc in shells of 1.3 Mpc with an angular binning using healpy~\citep{healpy} with \texttt{nside=32} (leading to roughly 12000 voxels per shell).  At larger distances, we assume a homogeneous distribution of sources up to 5000\,Mpc using a linear extrapolation of the density.
We take the injection of UHECRs to be common to all sources and model the injection rate (number of nuclei injected per unit of energy, volume, and time) to follow a Peters cycle:
\begin{equation}
\label{eq:emission}
    Q(E, A) \propto \ a_A \ E^{\gamma} \ f_\mathrm{cut}(E, Z_A R_\mathrm{cut}).
\end{equation}
It is characterized by a spectral index $\gamma$ and a cutoff at the maximum rigidity $R_\mathrm{cut}$, described by a variable cutoff function $f_\mathrm{cut}(x=\frac{E}{Z_A R_\mathrm{cut}})$. The cutoff function is either a broken exponential (b.e.) of the form $f_\mathrm{cut}^\mathrm{b.e.}(x) = \exp(1-x)$ (that sets in for $x>1$) to compare to Auger results~\citep{Auger_CF_2017}, or a continuous hyperbolic secans function with variable softness $f_\mathrm{cut}^\mathrm{sec}(x, k) = 1/\text{cosh}(x^k)$ with $k=1, 2, 3$ as in \citet{Mollerach_2019}.
We inject and track five representative elements with charge numbers $Z_A$ and contributions below the cutoff $a_A$: H, He, N, Si, and Fe. Instead of the fraction $a_A$ defined below the cutoff, we quote below the integral fractions $I_A=a_A \ Z_A^{2-\gamma} / (\sum_A a_A \ Z_A^{2-\gamma})$
as these represent the total emission better.

\subsection{Propagation and cosmic magnetic fields} \label{sec:propagation}
\begin{figure}[ht]
    \includegraphics[width=0.23\textwidth]{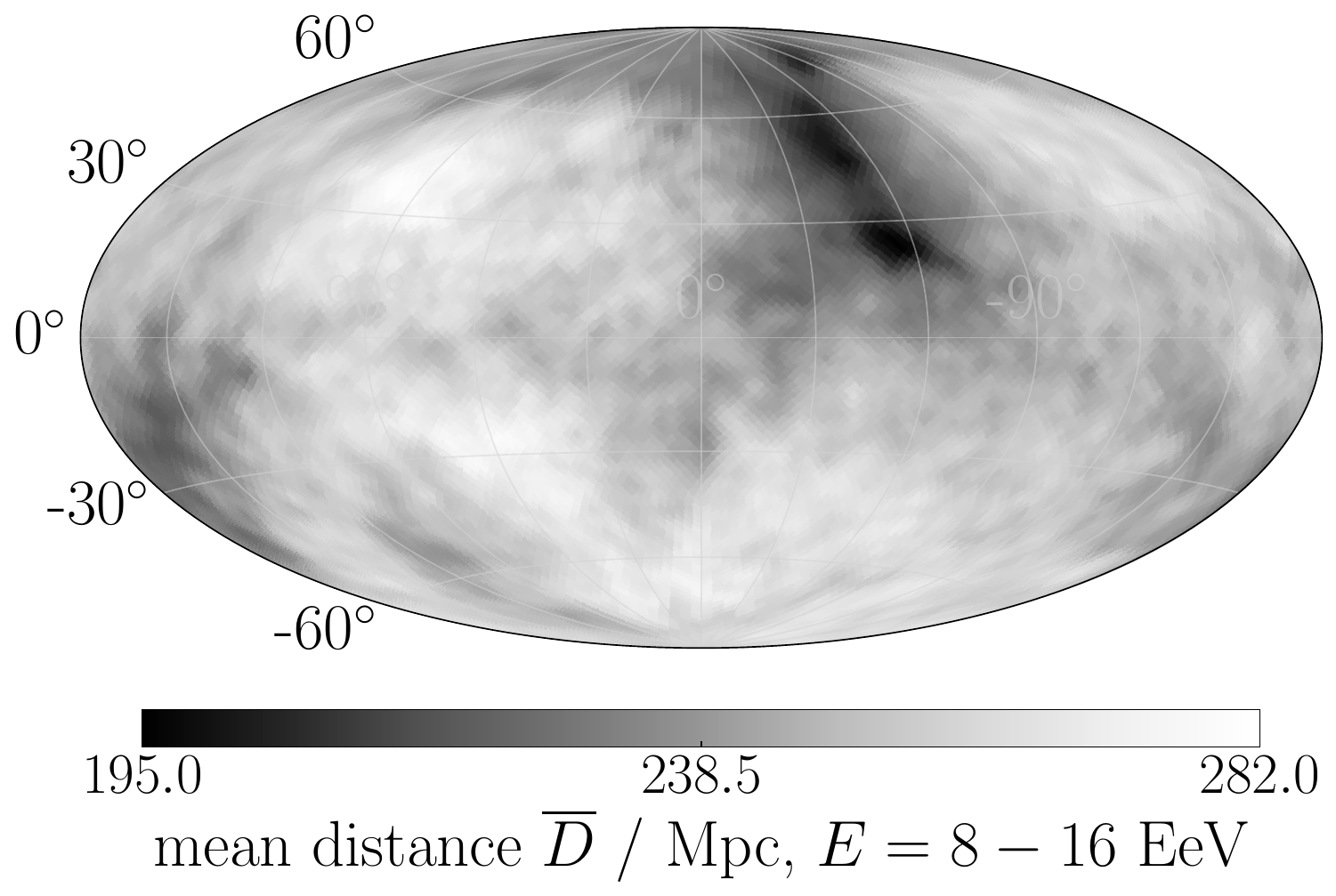}
    \includegraphics[width=0.23\textwidth]{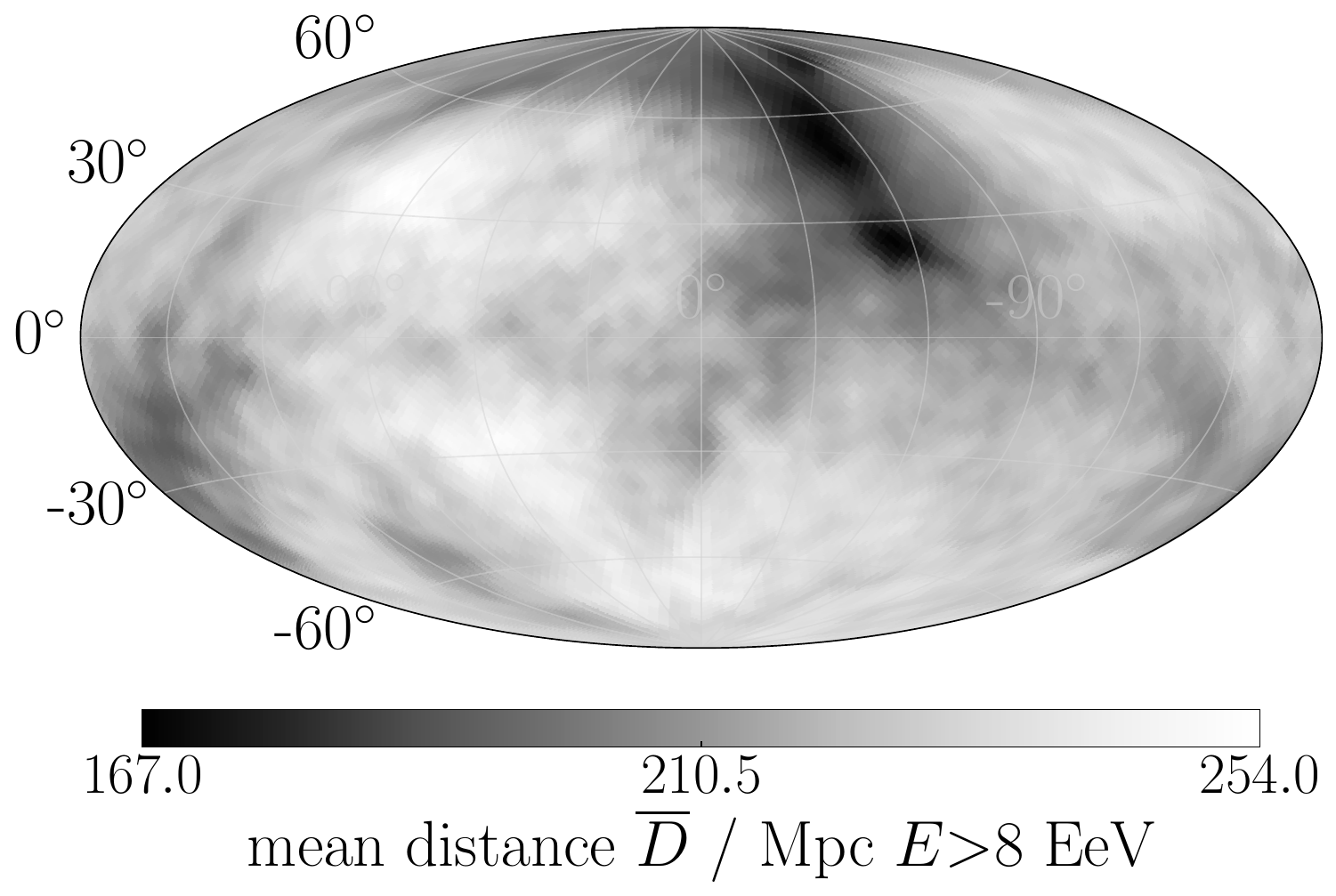}
    \includegraphics[width=0.23\textwidth]{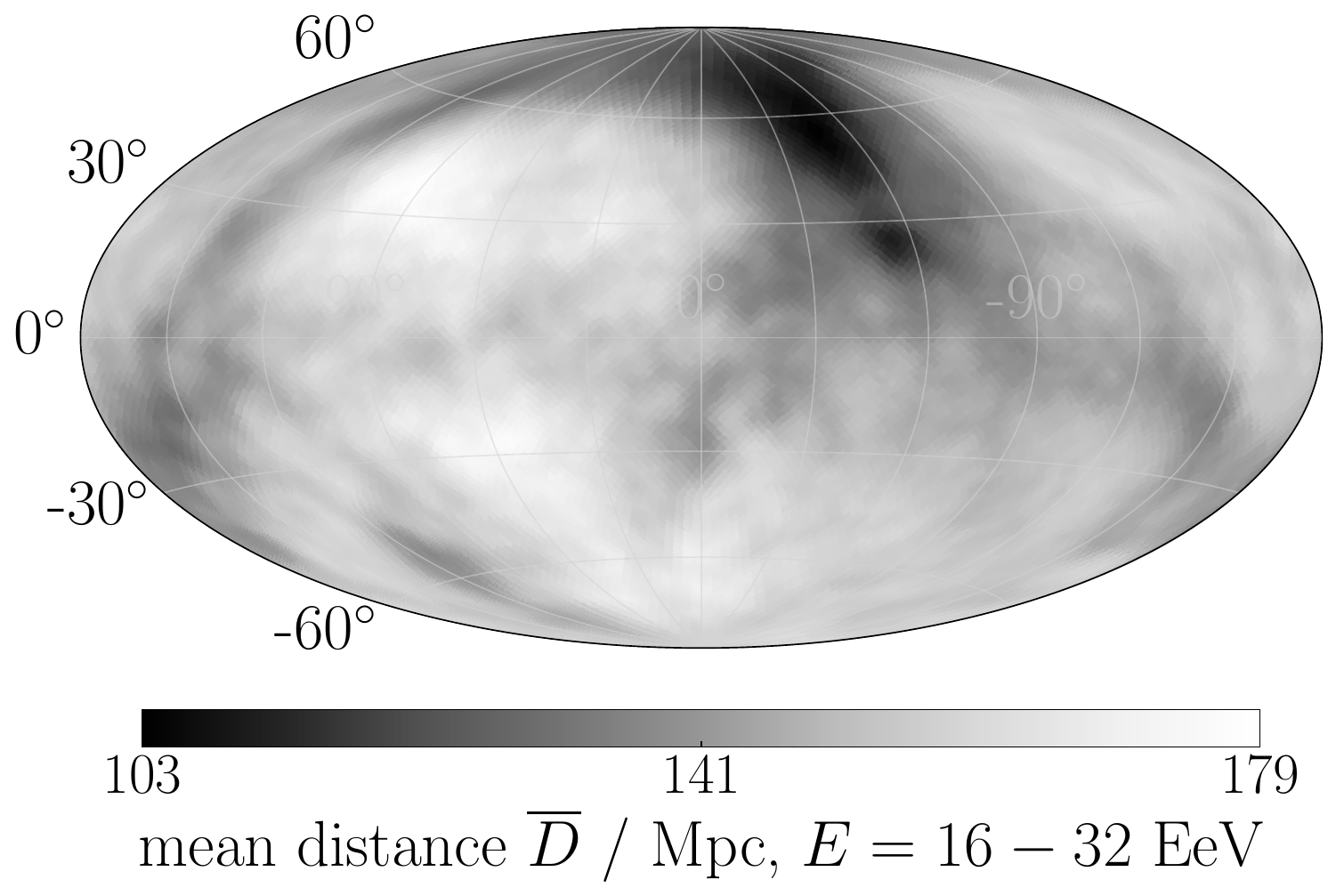}
    \includegraphics[width=0.23\textwidth]{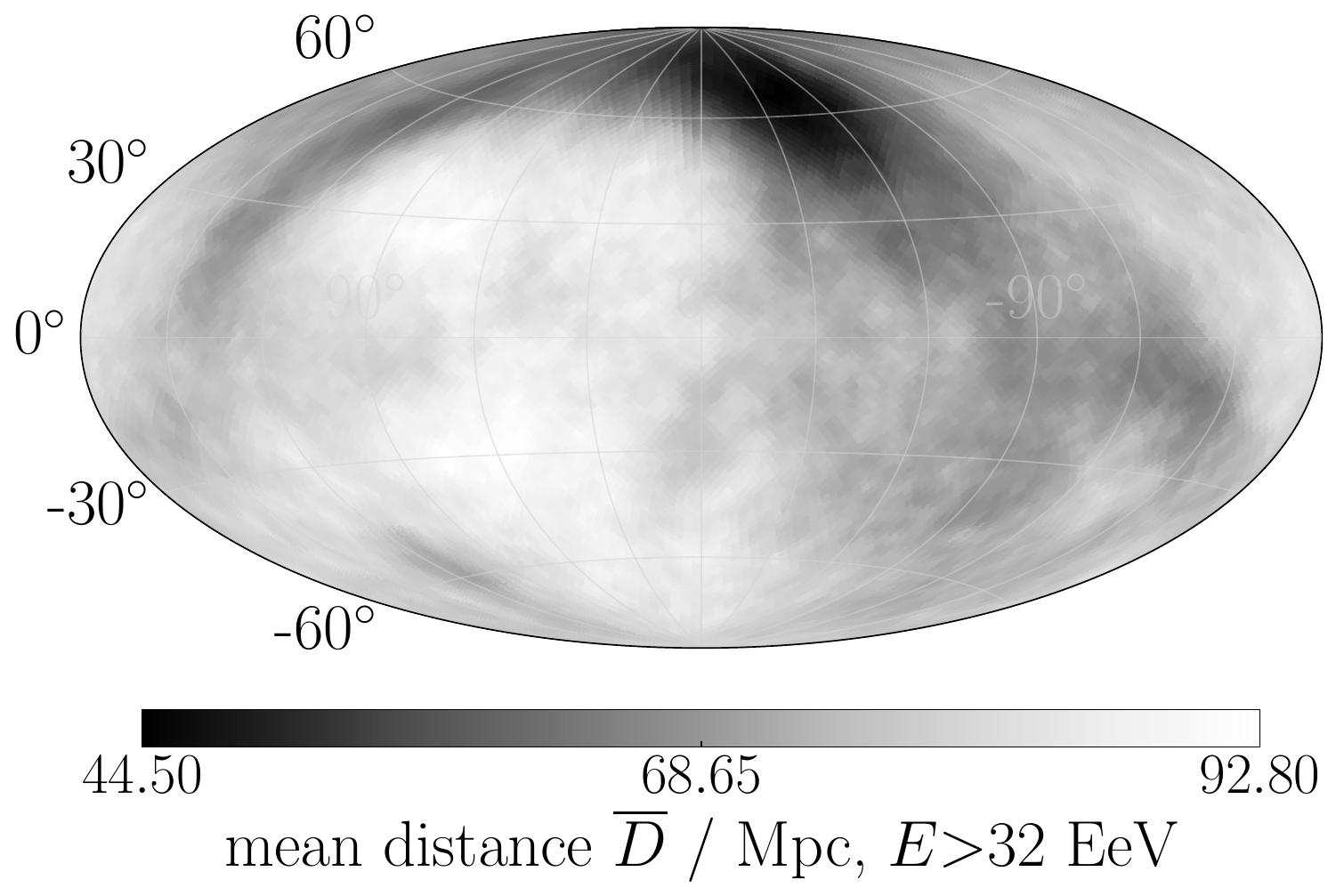}
    \caption{Map in Galactic coordinates showing the mean distance from which a UHECR originates in the baseline model, for various energy bins or energy thresholds. Both the anisotropic extragalactic matter distribution as well as the effects of propagation are visible in the maps. With increasing energy, the overall propagation horizon shrinks. At the highest energies, many CRs come from nearby, predominantly from the Virgo cluster (compare to Fig.~\ref{fig:illum}).} 
    \label{fig:propagation}
\end{figure}

% Following the results from~\citep{ding_imprint_2021}, we assume a negligible extragalactic magnetic field.
For the propagation, we use a database of 1-dimensional CRPropa3~\citep{alves_batista_crpropa_2016} simulations for which all available interactions are considered. For the extragalactic background light, the Gilmore model~\citep{Gilmore_2012} is used, and for the photodisintegration the TALYS model~\citep{TALYS}. The interactions with background photon fields lead to a shrinking propagation horizon for increasing energies, as can be seen in Fig.~\ref{fig:propagation}. 

\begin{figure}[ht]
    \includegraphics[width=0.23\textwidth]{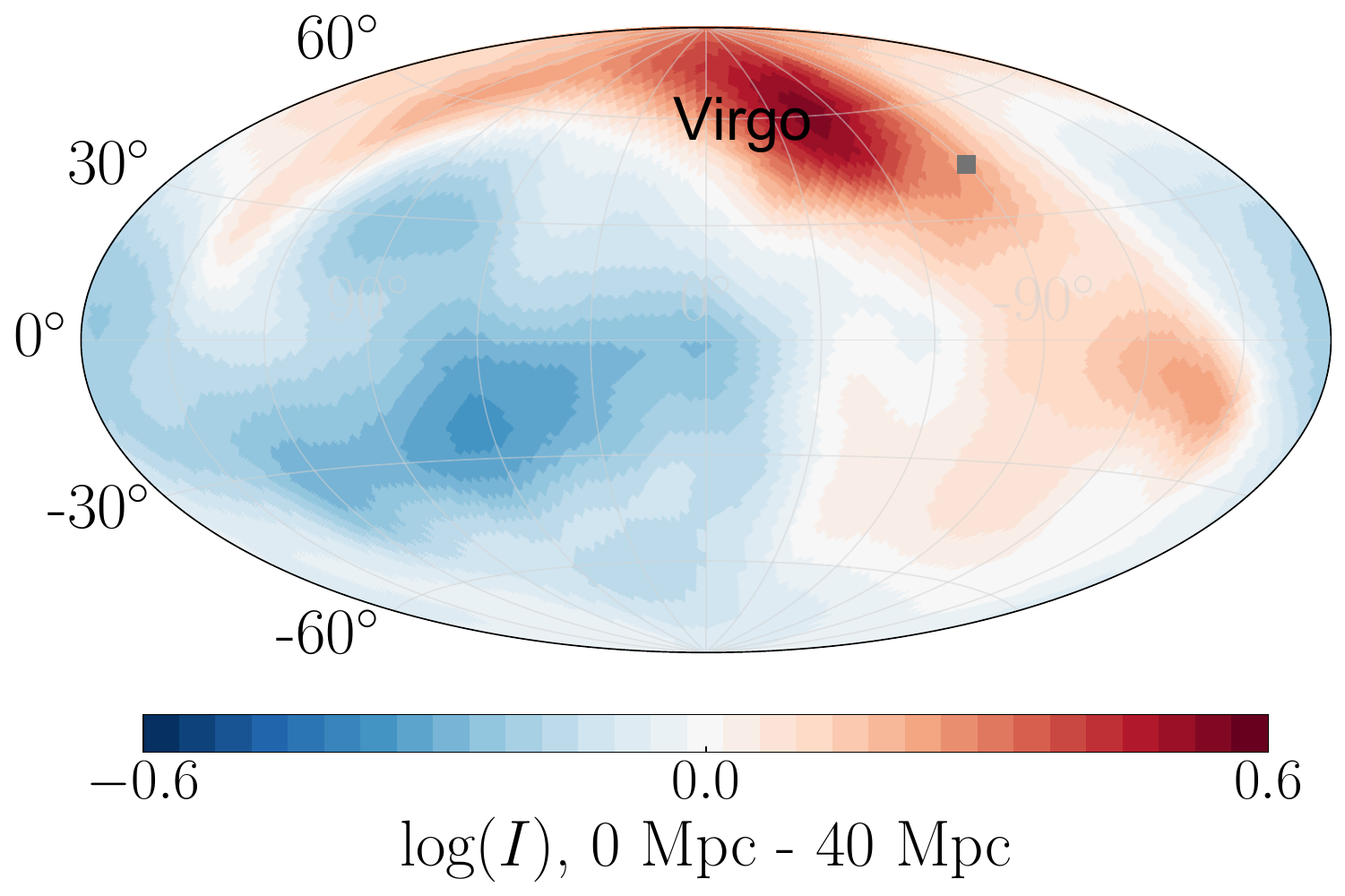}
    \includegraphics[width=0.23\textwidth]{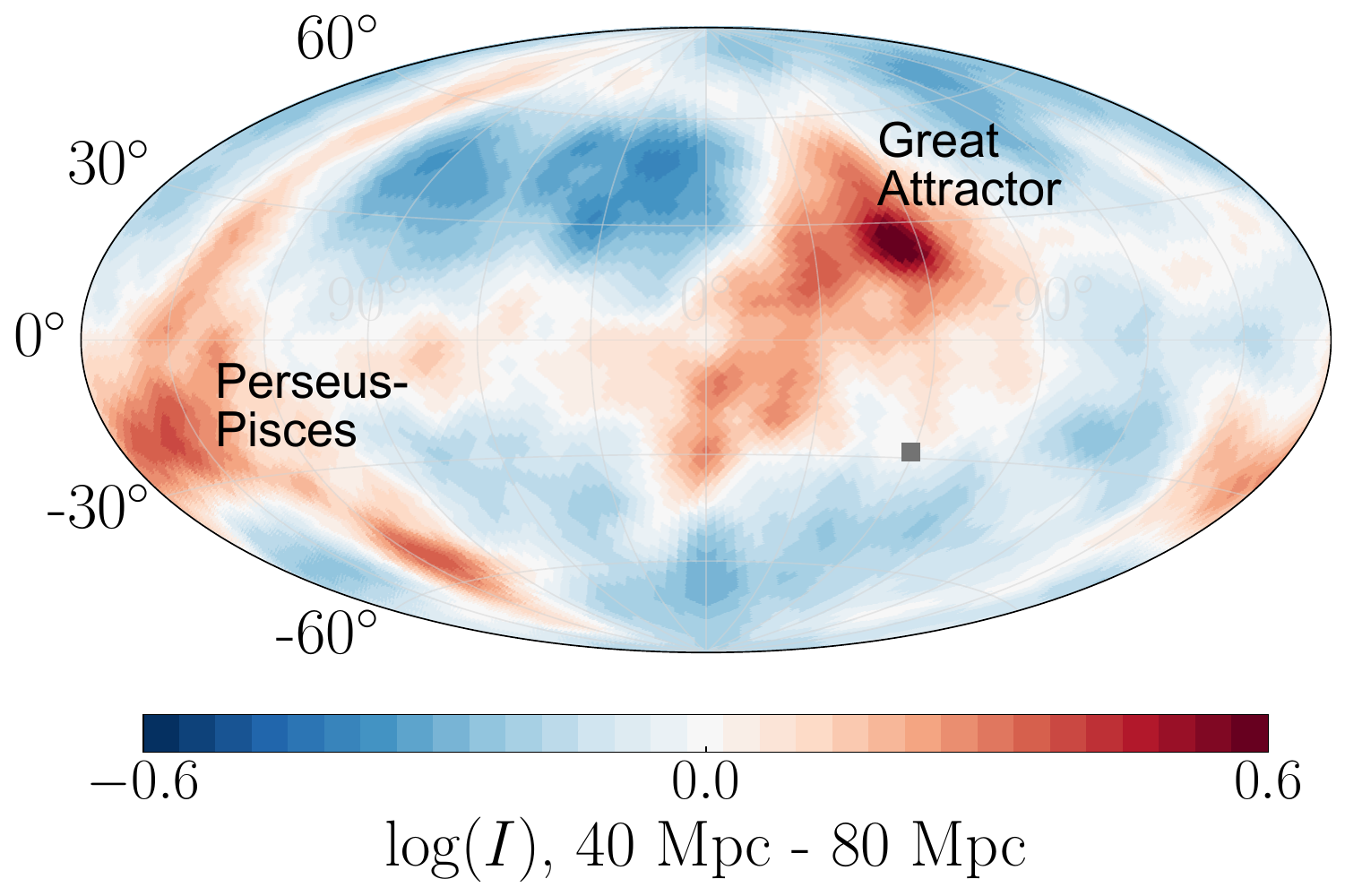}
    \includegraphics[width=0.23\textwidth]{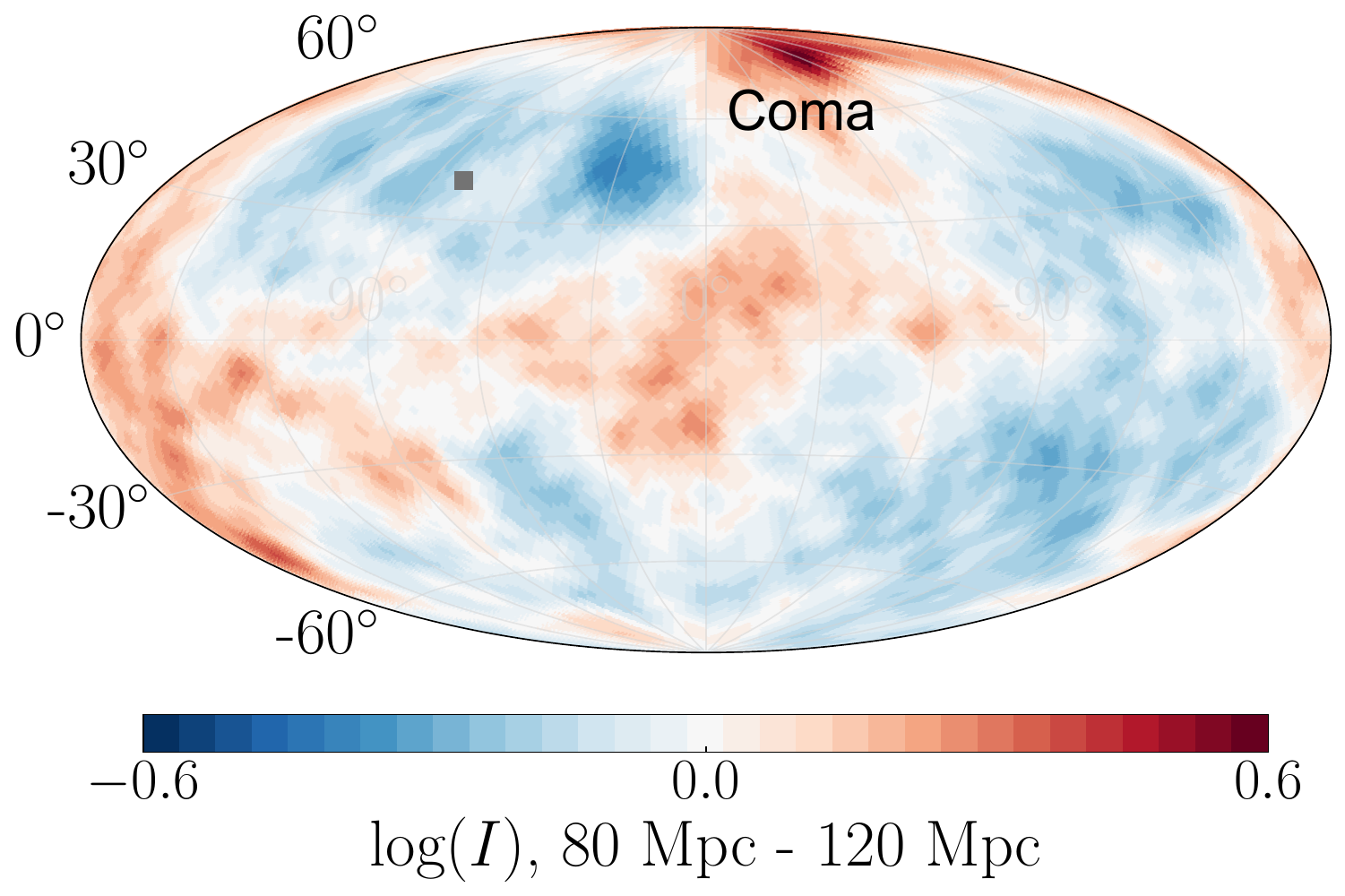}
    \includegraphics[width=0.23\textwidth]{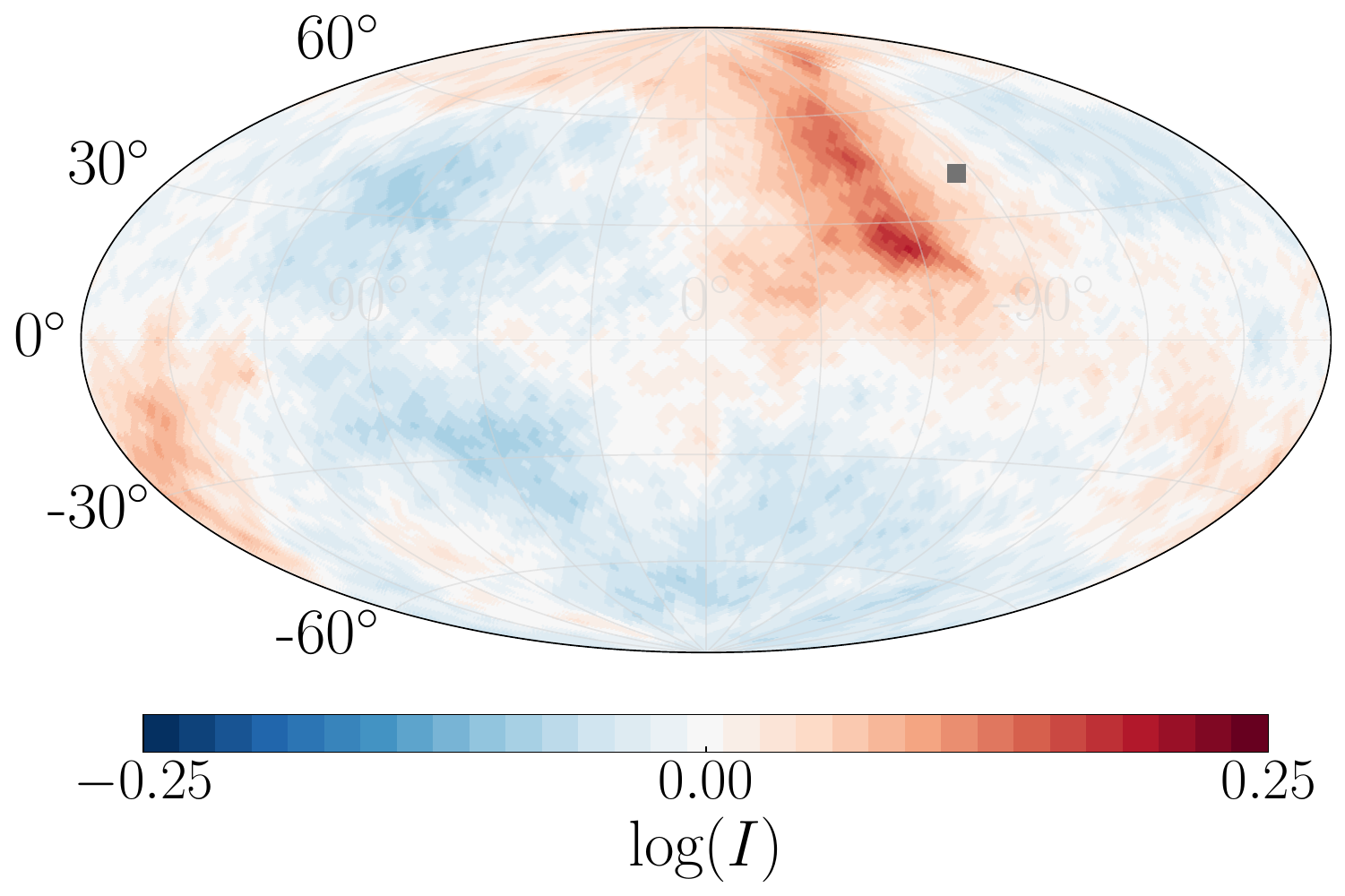}
    \caption{Logarithmic illumination maps (see text) for different distance shells, and a cumulative one (\textit{lower right}), for the best-fit model and $E>8$\,EeV. The direction of the full-sky dipole component of the map is indicated with a grey marker. For the closest distance bin, $0-40$\,Mpc, the influence of the Virgo cluster in the Galactic north is clearly visible. For the further away distances, mainly Perseus-Pisces and the Great Attractor become important at $40-80$\,Mpc, and Coma at $80-120$\,Mpc.} 
    \label{fig:illum}
\end{figure}

As in~\citet{ding_imprint_2021}, we calculate \textit{illumination maps} $I$. These encode the UHECR flux at the edge of our Galaxy induced by the local anisotropic source distribution and more distant homogeneous distribution, shaped by propagation effects. The illumination maps for our best-fit model, assuming zero EGMF, are displayed in Fig.~\ref{fig:illum}, for different distance bins as well as one cumulative bin. Evidently, the illumination map is peaked in the Galactic North, where the Virgo cluster and several other massive clusters reside.
% \begin{figure}[ht]
%     \includegraphics[width=0.49\textwidth]{img/best_fit/D_DEP2.pdf}
%     \caption{Flux contribution of different distance shells up to 500\,Mpc, for three energy bins, for the best-fit model. The distribution of the LSS following CosmicFlows2 is shown as a grey line in arbitrary units.} \label{fig:propagation}
% \end{figure}

The local arrival direction distribution can be calculated from the illumination maps by using a \textit{lens} of the Galactic magnetic field (GMF). As a baseline case, we use the same high-resolution backtracking simulations as~\citet{ding_imprint_2021} of the JF12 model~\citep{jansson_new_2012, jansson_galactic_2012} with a 
coherence length of 30\,pc for the random component.
As pointed out in~\citet{fComptRend14}, the turbulent component of the GMF is likely overestimated in the original JF12 random model~\citep{jansson_galactic_2012} since JF12 was based on the original WMAP synchrotron intensity map which has subsequently been revised downward due to better modeling of dust contributions~\citep{Planck_2016}. 
To bracket the effect of the random field, we therefore also consider the case with only the JF12 regular magnetic field without a random component, denoted as JF12 reg.

Following the results from~\citet{ding_imprint_2021} who find that very weak extragalactic magnetic field (EGMF) strengths lead to the best agreement with the data, we neglect the extragalactic magnetic field in the baseline case. This is backed up by evidence for small field strengths in voids~\citep{Planck_2016, Hackstein_2016}. In Sec.~\ref{sec:egmf} we constrain the maximum allowed strength of a turbulent EGMF. The effect of magnetic confinement by large magnetic field strengths in denser galaxy clusters is investigated in the context of bias in Sec.~\ref{sec:bias}.

\subsection{Fit of model parameters} \label{sec:fit_method}
After propagation and magnetic field deflections, the arriving (detected) CRs are binned into a matrix whose elements are $(E_\mathrm{det}, A_\mathrm{det}, D, \mathrm{pixel})$. 
Here, $\mathrm{pixel}$ labels the healpix sky-direction and $D$ the propagation distance, binned as described above. $E_\mathrm{det}$ labels bins in energy of width $10^{0.1}\,\mathrm{eV}$, between $10^{18.0}\,\mathrm{eV} = 1\, \mathrm{EeV}$ and $10^{20.4}\,\mathrm{eV} \approx 250 \, \mathrm{EeV}$. The masses $A_\mathrm{det}$ are binned as $A_\mathrm{det} \in (1,\ 2\text{--}4,\ 5\text{--}22,\ 23\text{--}38,\ $\textgreater39), to compare to the fitted Auger composition~\citep{Auger_CF_2023, Auger_CFAD_2023}. 

We use a Poissonian likelihood $\mathcal{L}_E$ as given in~\citet{Auger_CF_2017} to compare the modeled energy spectrum above $10^{18.9}\,\mathrm{EeV} \approx 8\,\mathrm{EeV}$ to the measured one. (This is the same energy threshold as for the dipole components, see below). We use the unfolded spectrum from~\citep{Auger_spectrum_2020}, so that no detector effects on the energy spectrum have to be considered in the model. Below $10^{18.9}\,\mathrm{EeV}$, we ensure that the model does not overshoot the data by a one-sided Poissonian likelihood.

We use the peak of the shower profile, $\xmax$, as the mass estimator. %It can be calculated from $E_\mathrm{det}$ and $A_\mathrm{det}$ by using the Gumbel parameterizations~\citep{de_domenico_reinterpreting_2013}. For that, 
EPOS-LHC~\citep{EPOS_2013} is used as the baseline hadronic interaction model and Sibyll2.3d~\citep{Sybill} as an alternative. The effects induced by the detector during measurement~\citep{the_pierre_auger_collaboration_a_aab_et_al_depth_2014-1} are included in the modeling of \xmax. As in~\citep{Auger_CFAD_2023}, we consider a possible shift of the $\xmax$ scale as a nuisance parameter in the model. That parameter $\nu_{\xmax}$ is given in units of standard scores that refer to multiples of the experimental systematic uncertainty, so that the likelihood function $\mathcal{L}_\mathrm{syst}$ is a Gaussian (see~\citet{Auger_CFAD_2023} for details). The most important experimental systematic uncertainties on the  \xmax scale arise from the detector calibration, the reconstruction of the shower maximum with a fit algorithm, and atmospheric variations influencing the fluorescence yield~\citep{the_pierre_auger_collaboration_a_aab_et_al_depth_2014-1}.
Finally, the modeled shower depth distributions binned in energy $>10^{18.9}\,\mathrm{EeV}$ are compared to the data from~\citet{a_yushkov_for_the_pierre_auger_collaboration_mass_2019} via a Multinomial likelihood $\mathcal{L}_{\xmax}$ as in~\citet{Auger_CF_2017}.

To compare to the measurements of the large-scale anisotropy, the directional exposure of the Observatory is taken into account.
Afterwards, we acquire the modeled arrival directions in energy bins $E_\mathrm{bins} = (8-16\,\mathrm{EeV}, 16-32 \,\mathrm{EeV}, >32\,\mathrm{EeV})$, for which we calculate the 3D dipole components $d_x$, $d_y$, and $d_z$ considering the effect of the limited exposure (\citep{Deligny_2013}; see also K-inverse method explained in~\citet{Ding_ICRC_2019}). The dipole components $d_{e, i\in(x,y,z)}$ for each of the three energy bins are then compared to the measured $d_{e, i\in(x,y,z)}^\mathrm{data}$ (including the uncertainties $\sigma_{e, i\in(x,y,z)}^\mathrm{data}$) from Auger~\citep{Almeida_dipole_2021} using a Gaussian likelihood function:
\begin{equation}
    \mathcal{L}_d = \prod_{e \in E_\mathrm{bins}} \prod_{i \in (x, y, z)}  \frac{1}{\sqrt{2 \pi} \sigma_{e, i}^\mathrm{data}} e^{-\frac{1}{2}\Big(\frac{d_{e, i} - d_{e, i}^\mathrm{data}}{\sigma_{e, i}^\mathrm{data}} \Big)^2}.
\end{equation}
That way, the energy evolution of both the direction and amplitude of the dipole are taken into account. 

Finally, all parts of the log-likelihood are added to get the total one:
\begin{equation} \label{eq:lik_tot}
    \log \mathcal{L}_\mathrm{tot} = \log \mathcal{L}_E + \log \mathcal{L}_{\xmax} + \log \mathcal{L}_d+ \log \mathcal{L}_\mathrm{syst}
\end{equation}

\section{Fit results} \label{sec:fit_results}
\subsection{Composition and spectrum}

\begin{figure}[ht]
\includegraphics[width=0.49\textwidth]{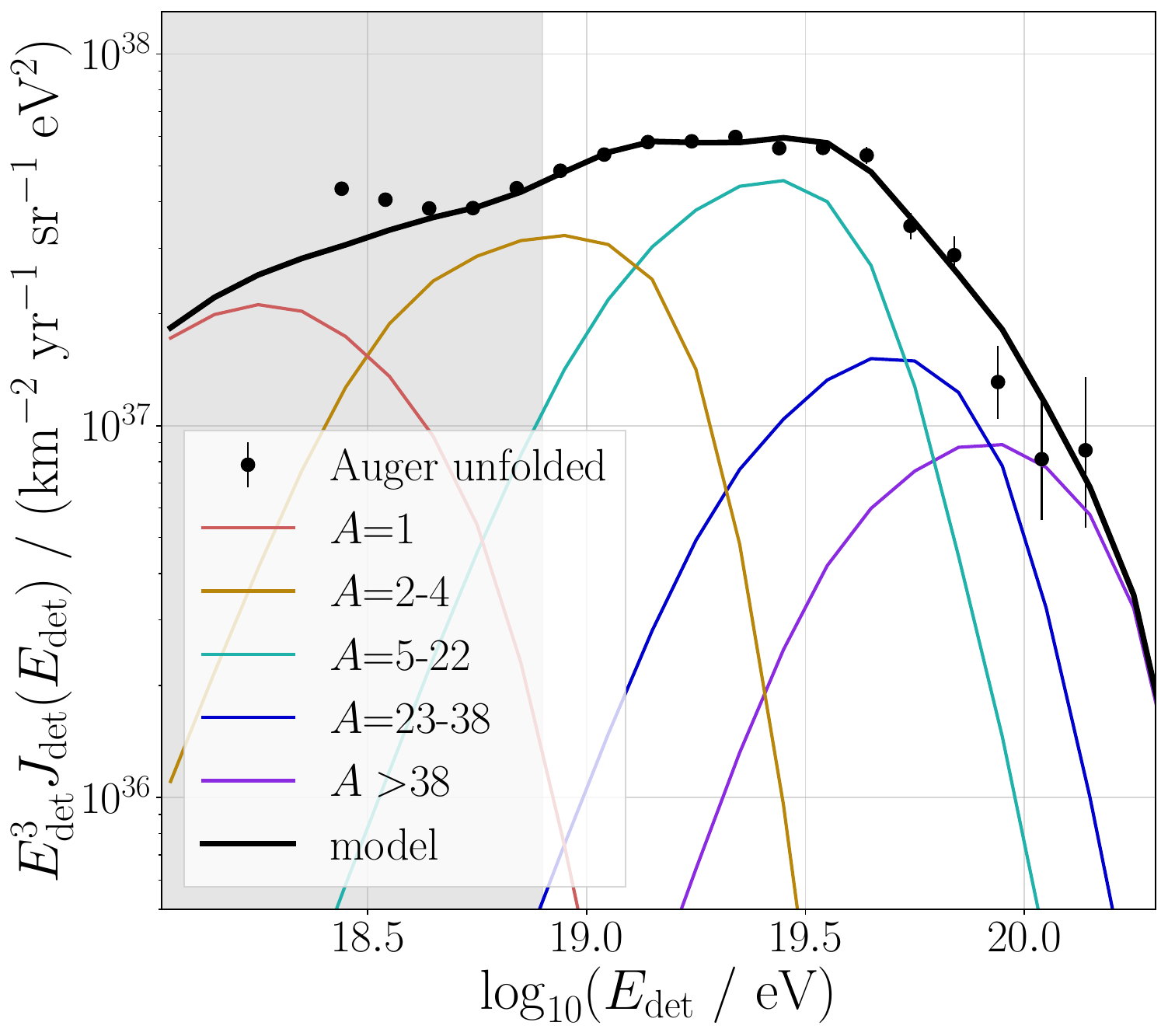}
\includegraphics[width=0.49\textwidth]{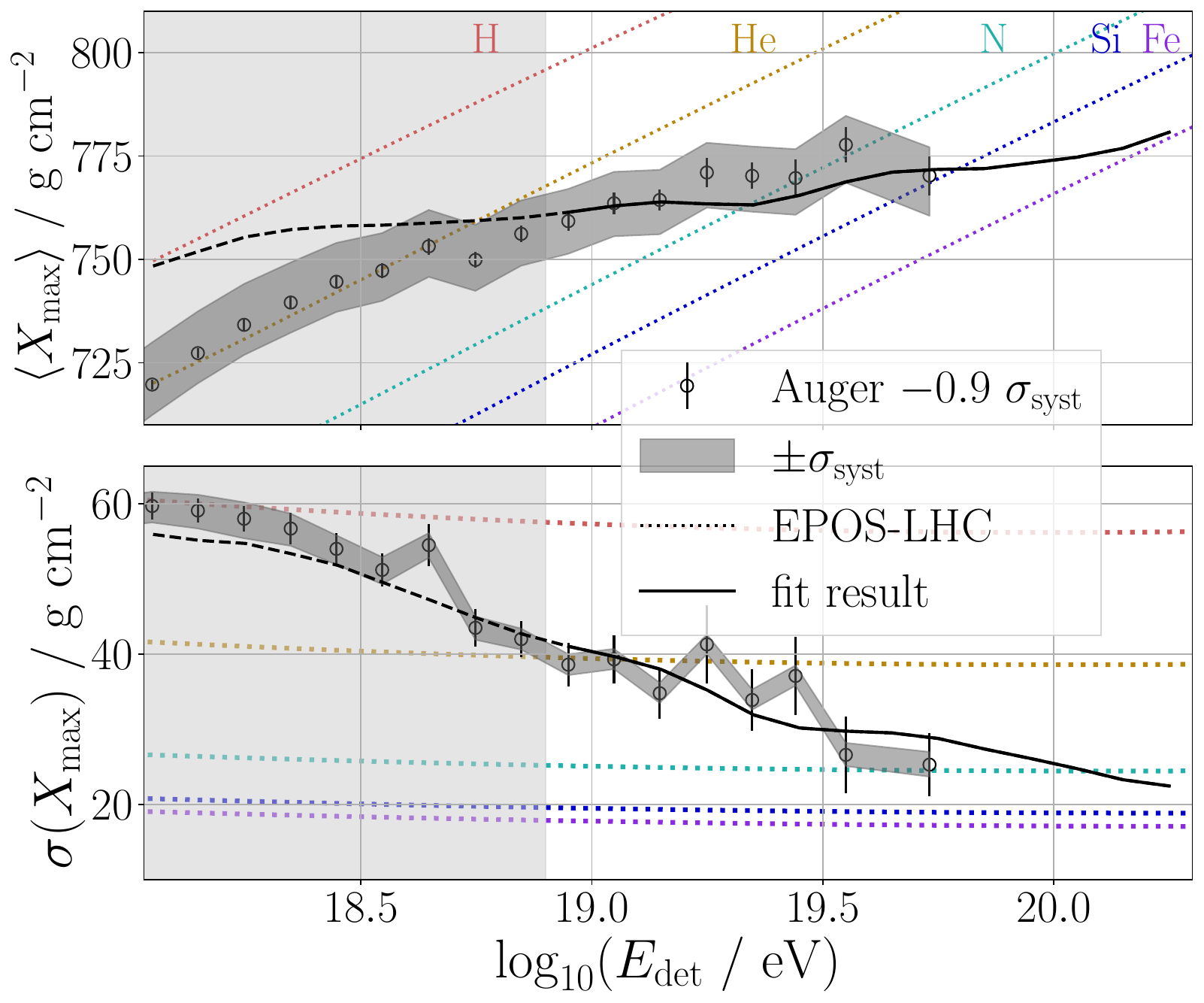}
\caption{Energy spectrum (upper) and first two moments of the \xmax distributions (lower) for the \textit{baseline model}. The grey area below 8\,EeV is not included in the fit. The measured \xmax distributions denoted by black markers have been moved by the best-fit \xmax scale shift (see text).}
\label{fig:best_fit_EXmax}
\end{figure}

\begin{figure}[ht]
    \includegraphics[width=0.228\textwidth]{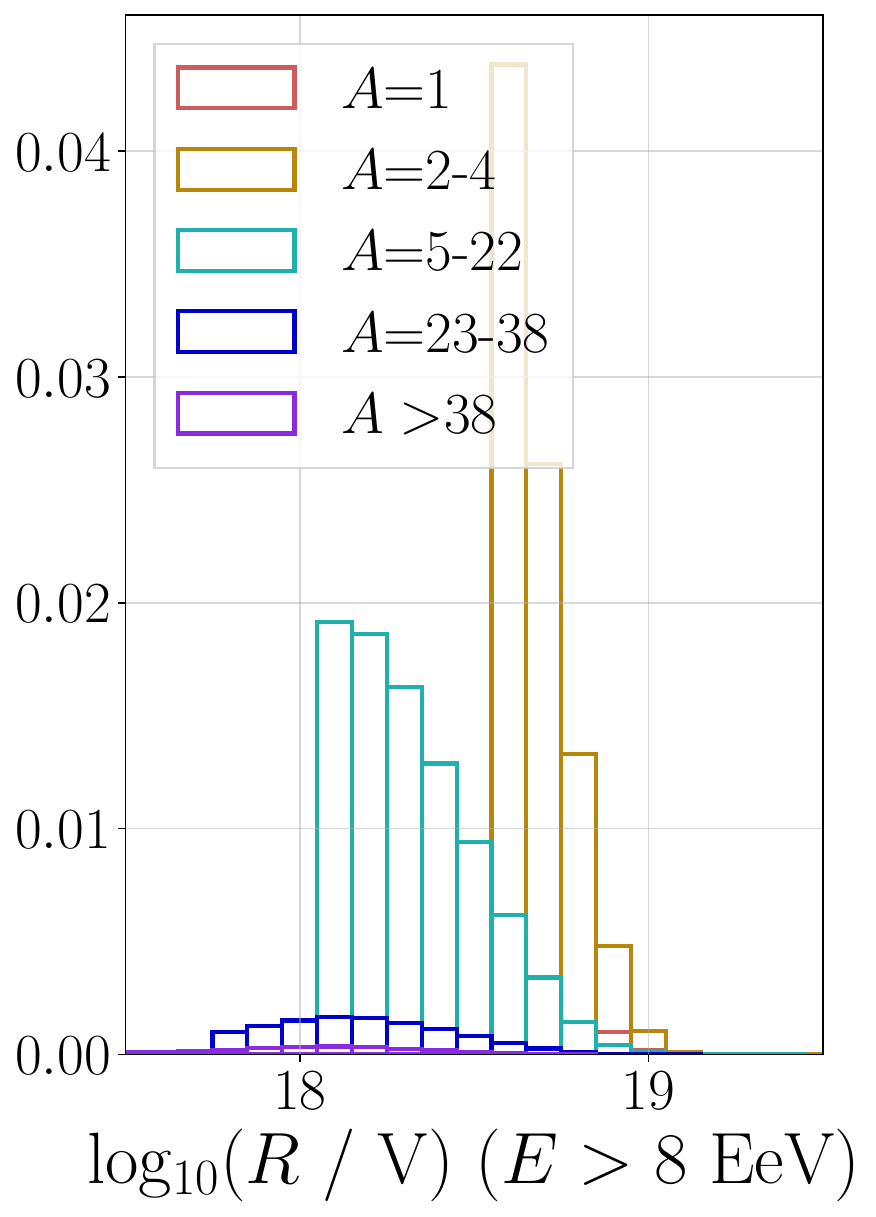}
    \includegraphics[width=0.24\textwidth]{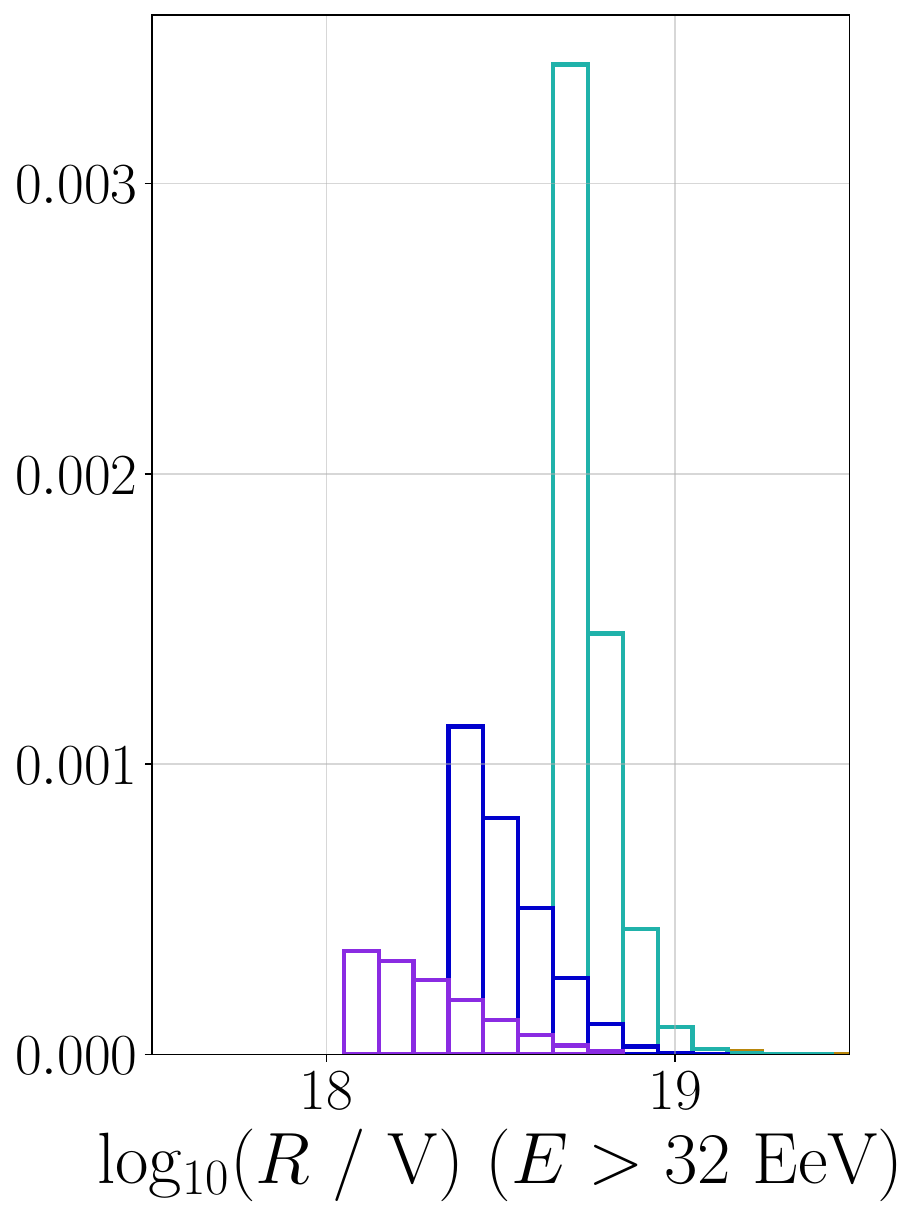}
    \caption{Frequency (a.u.) of arriving rigidities on Earth for the baseline model for two energy thresholds, $E>8$\,EeV (\textit{left}) and $E>32$\,EeV (\textit{right}). The multi-peak structure is due to our coarse binning in mass following Auger.} 
    \label{fig:rig_spec}
\end{figure}

In the \textit{baseline} case (CosmicFlows2 as source proxy, EPOS-LHC as hadronic interaction model, JF12-reg+rand as GMF model, broken exponential at injection), the model describes the measured spectrum and \xmax data at Earth well, as depicted in Fig.~\ref{fig:best_fit_EXmax}. The corresponding best-fit parameters and values of the likelihood are provided in Table~\ref{tab:results} in the Appendix. The best-fit injection parameters are close to recent results (obtained with a different energy threshold) by Auger, using a homogeneous source model~\citep{Auger_CFAD_2023, Auger_CF_2017}.

The best-fit spectrum is quite hard, with spectral index $\gamma=-1.2$, the maximum rigidity is $R_\mathrm{cut}=10^{18.2}$\,V, and the composition is nitrogen-dominated. For the \xmax scale, a best-fit shift %(see Sec.~\ref{sec:fit_method}) 
of $\nu_{\xmax}=-0.9\sigma$ is found. This shift of the data (as displayed in Fig.~\ref{fig:best_fit_EXmax}) leads to a larger likelihood value, and thus a better agreement of model and data, than no shift. It translates to a heavier composition on Earth than without a shift, in agreement with the combined fit by Auger~\citep{Auger_CFAD_2023}, and the comparison of hadronic interaction model predictions of \xmax with Auger data~\citep{Auger_Xmax_Jakub}.

Because the best-fit injected spectrum is very hard, the overlap of different mass groups is small, as visible in Fig.~\ref{fig:best_fit_EXmax}. Consequently, heavier and heavier mass groups dominate with rising energy. This parallel increase of the charge $Z$ with the energy $E$ leads to a relatively constant rigidity $R=E/Z$ that does not evolve with the energy. The rigidity spectra above two energy thresholds, $E>8$\,EeV and $E>32$\,EeV, are depicted in Fig.~\ref{fig:rig_spec}. The mean rigidities are 3.5~EV and 4.5~EV for the respective energy thresholds, and almost all arriving UHECRs have rigidities between 1\,EV and 5\,EV\footnote{For reference, $\log_{10}(5 \times 10^{18}) = 18.7$. For the model using Sibyll instead of EPOS-LHC as hadronic interaction model, (see sec~\ref{sec:model_uncertainties} of the Appendix) the mean rigidities for the two energy thresholds are 3.0~EV and 4.1~EV.}. Note that the multi-peak structure in the rigidity spectrum is because of the coarse binning in mass following Auger. We tested that the effect of this on the analysis is negligible by comparing to a model using the smoother mass distribution from~\citet{Muzio_2019}.

\begin{figure*}[htb]
\centering
\begin{places}{0.96\textwidth}{4}
\place{\textbf{Illumination}}{
  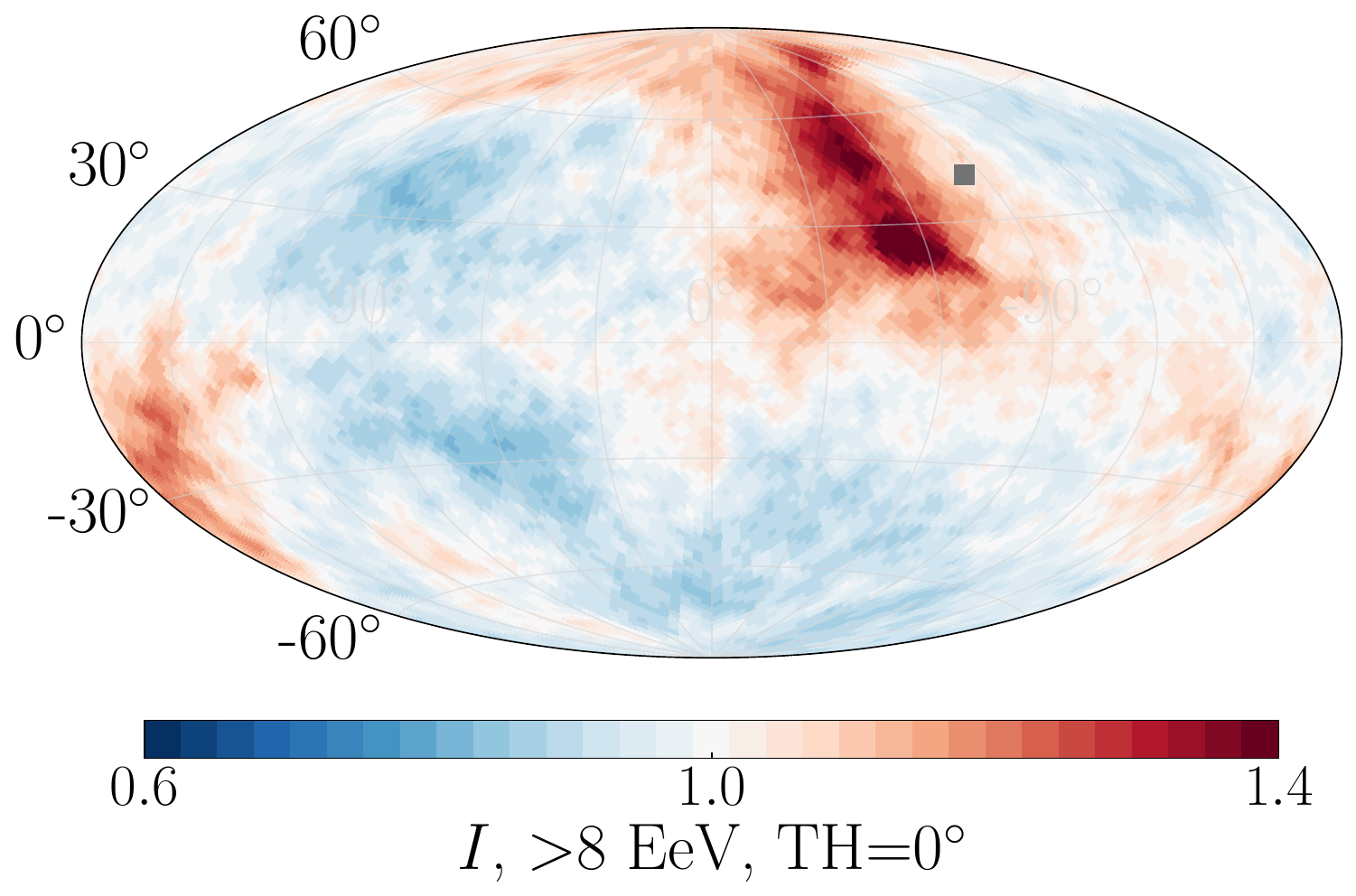,
  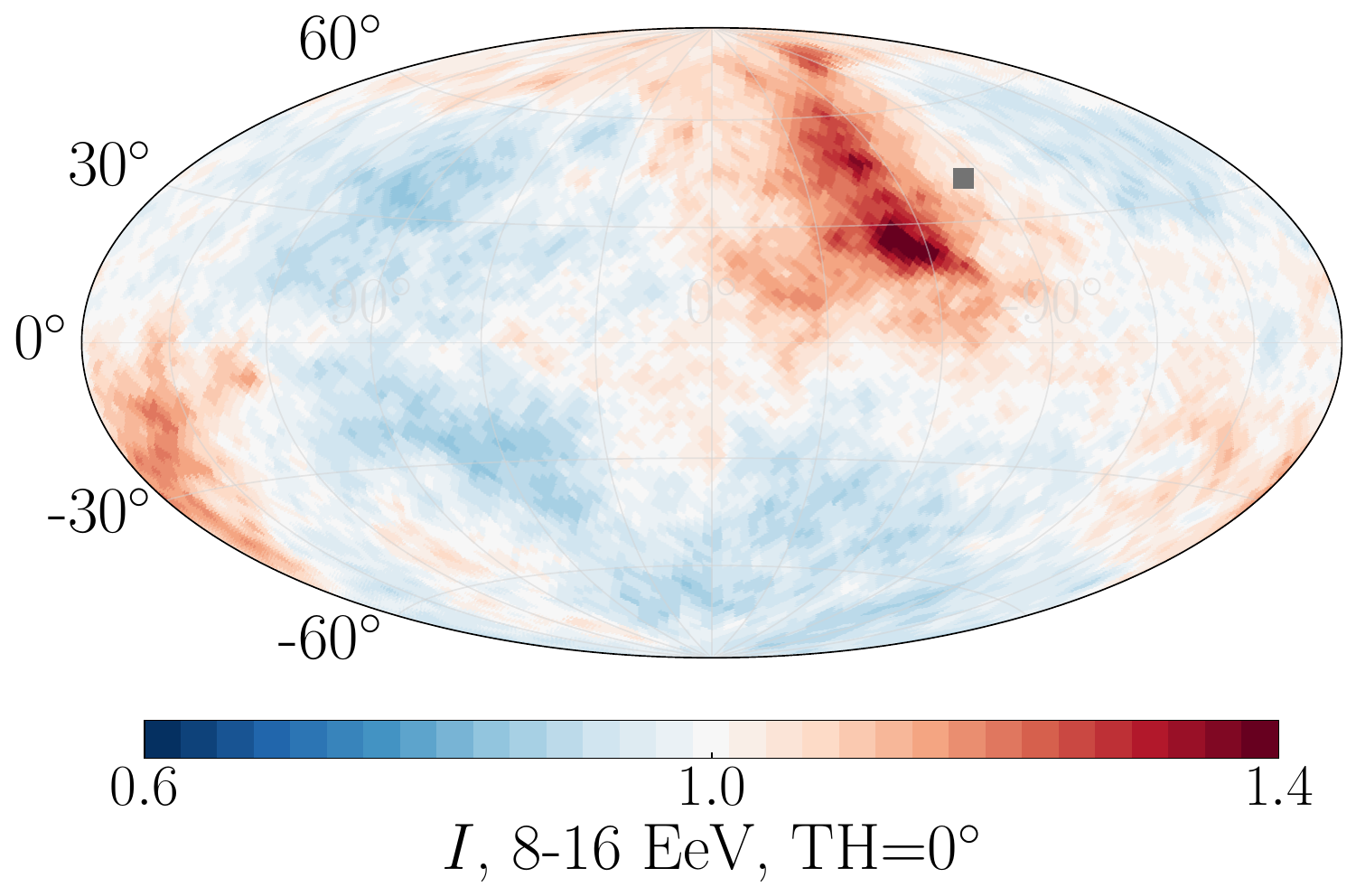,
  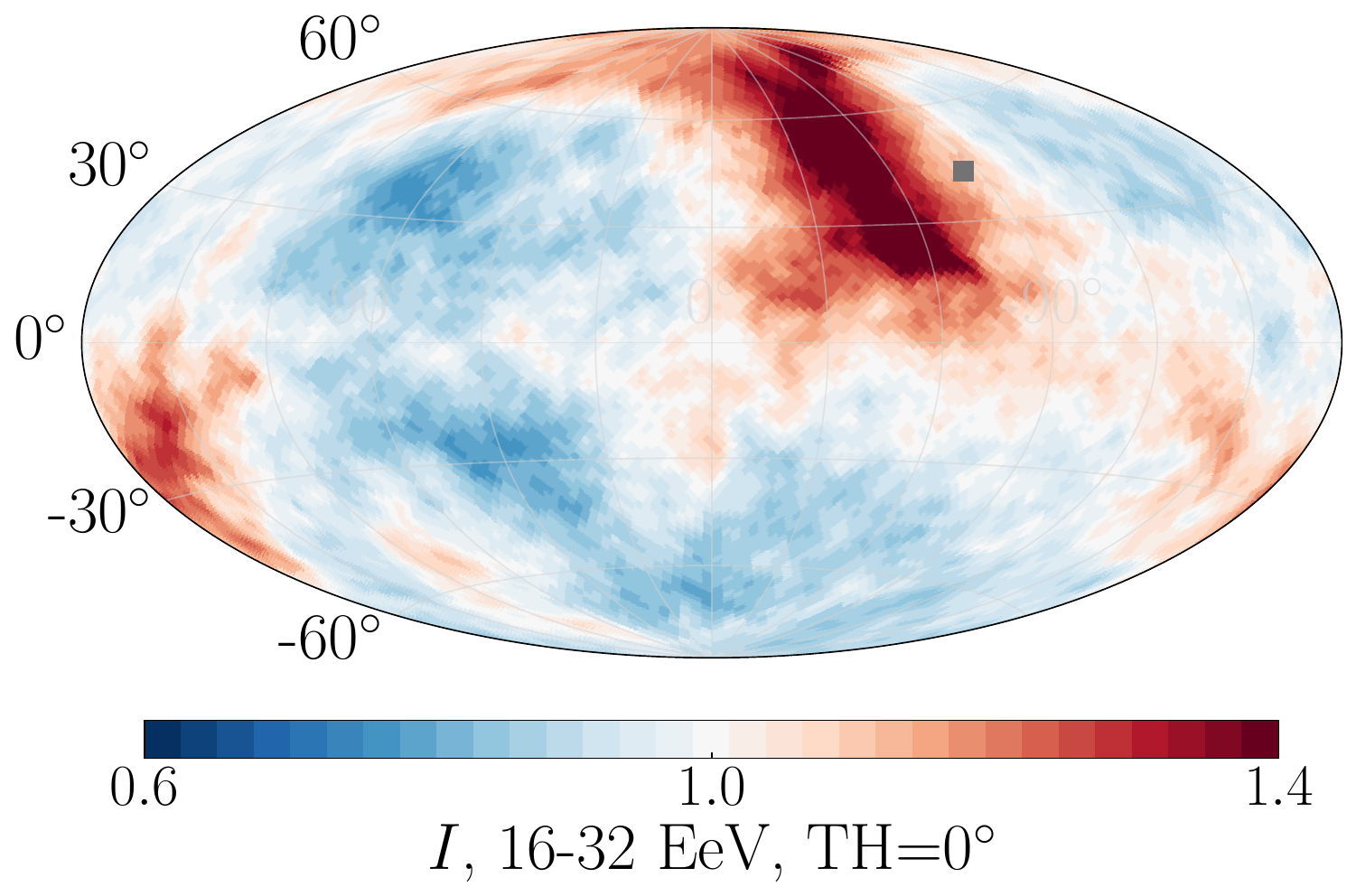,
  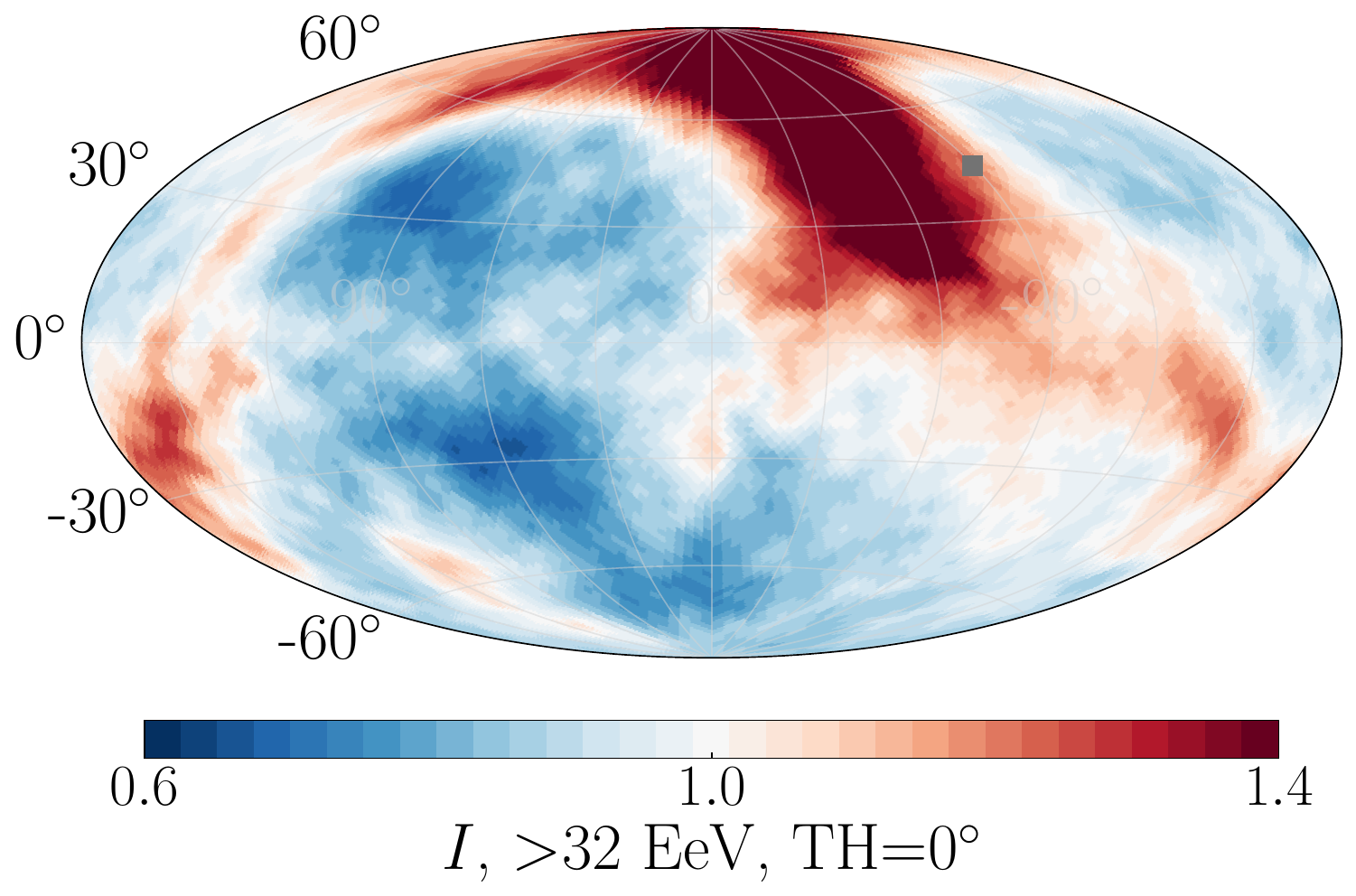,
}
\place{\textbf{flux}, baseline, no TH}{
  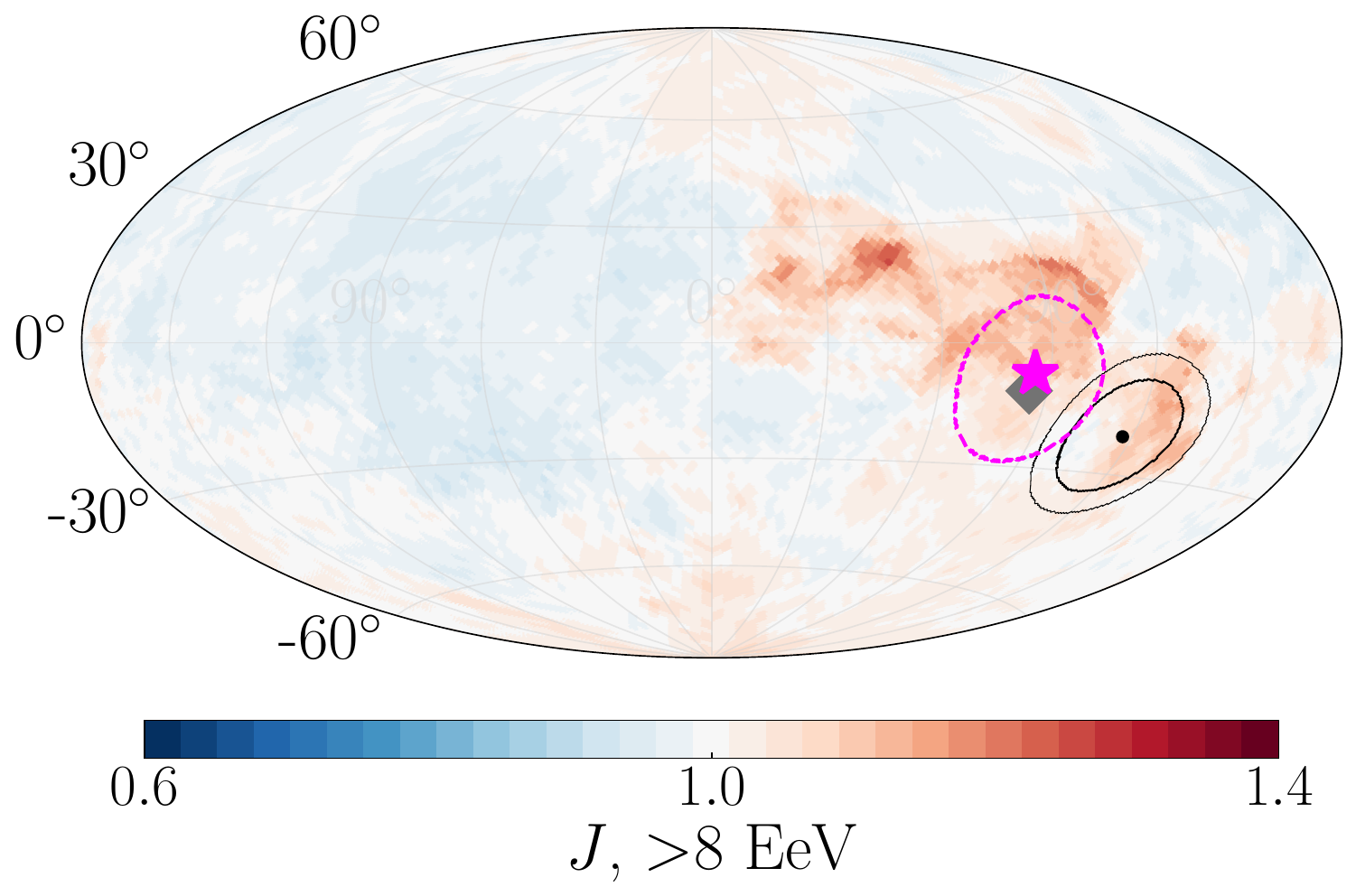,
  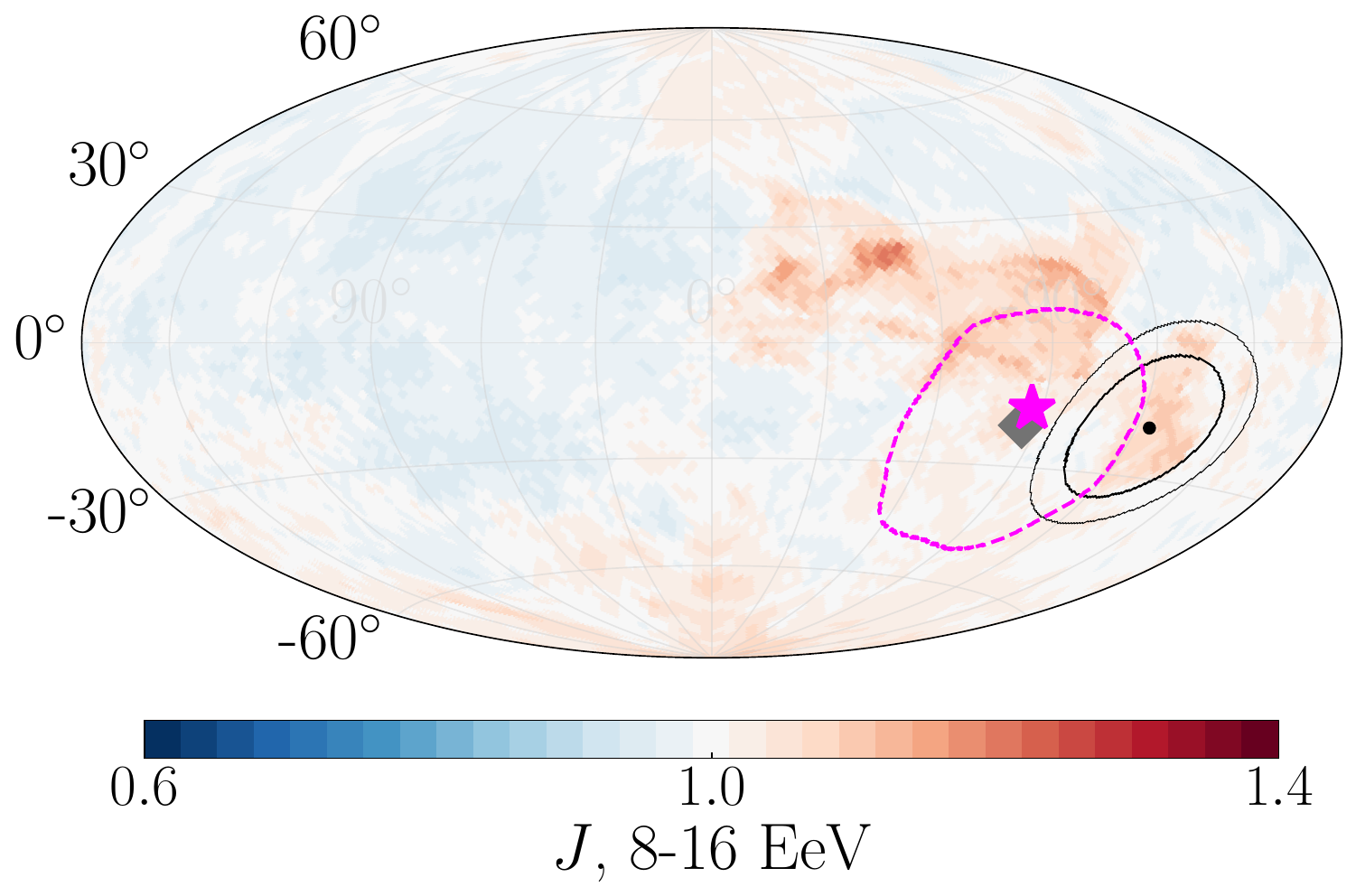,
  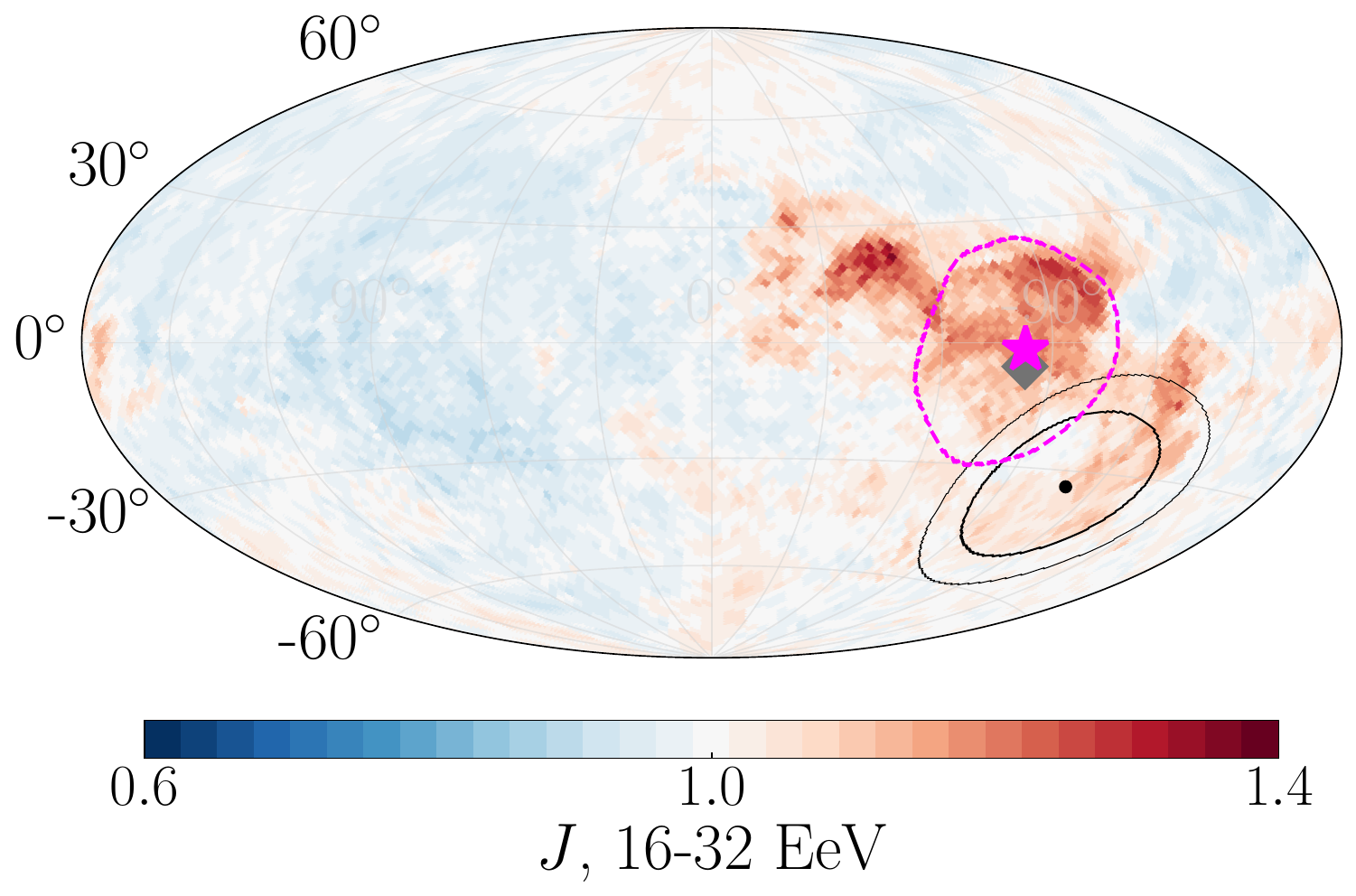,
  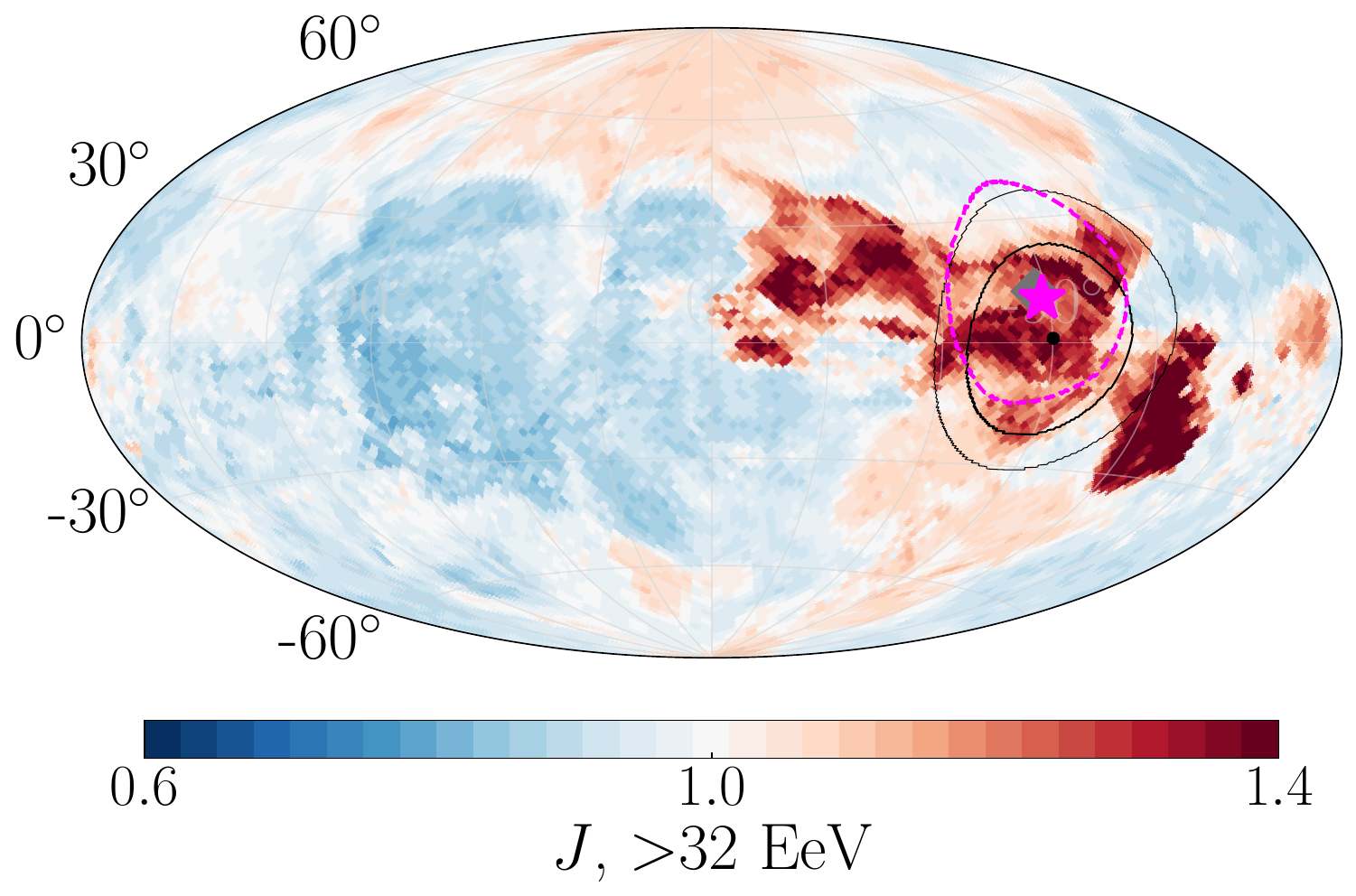,
}
\place{\textbf{flux}, baseline, TH=$45^\circ$}{
  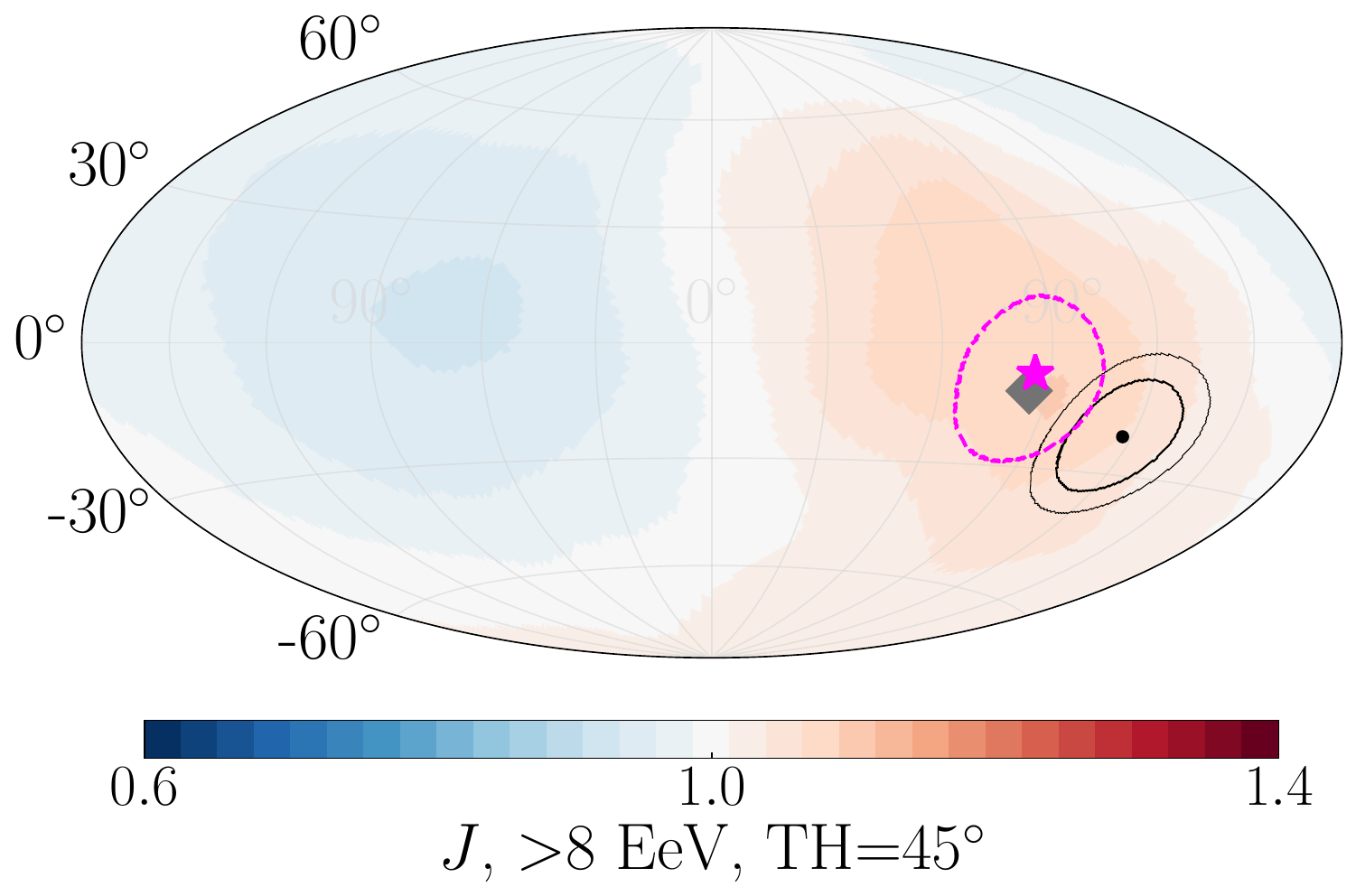,
  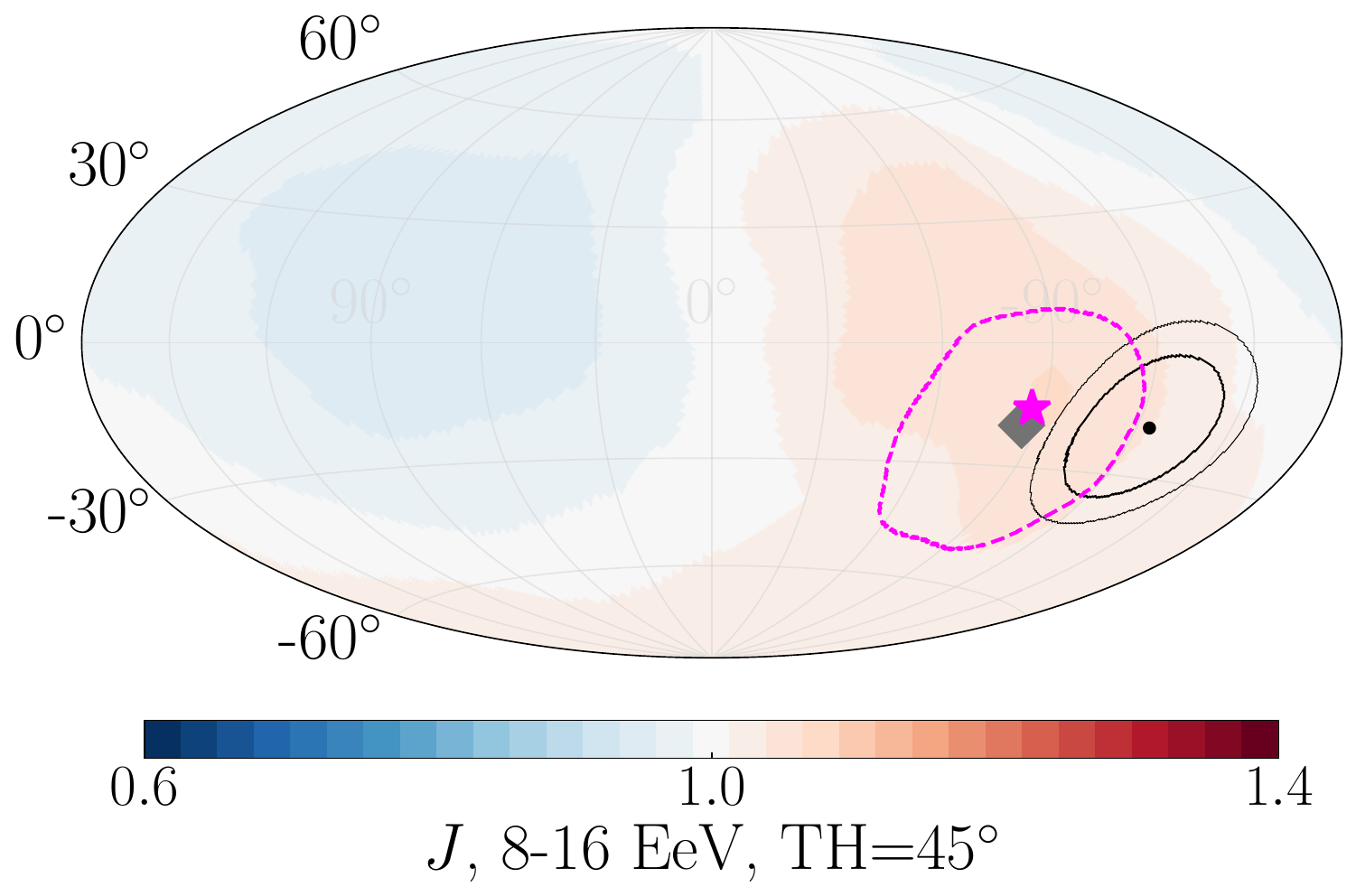,
  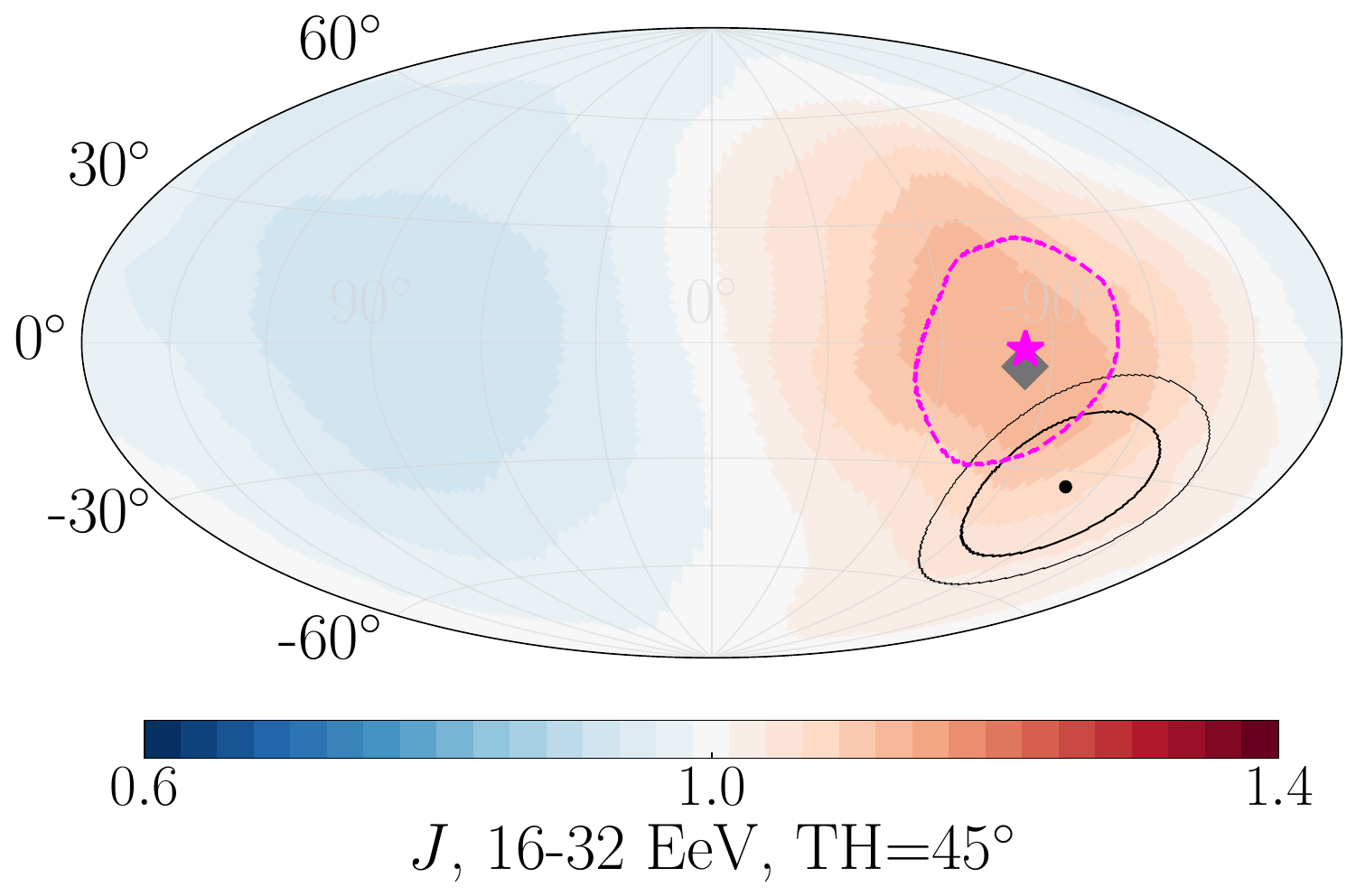,
  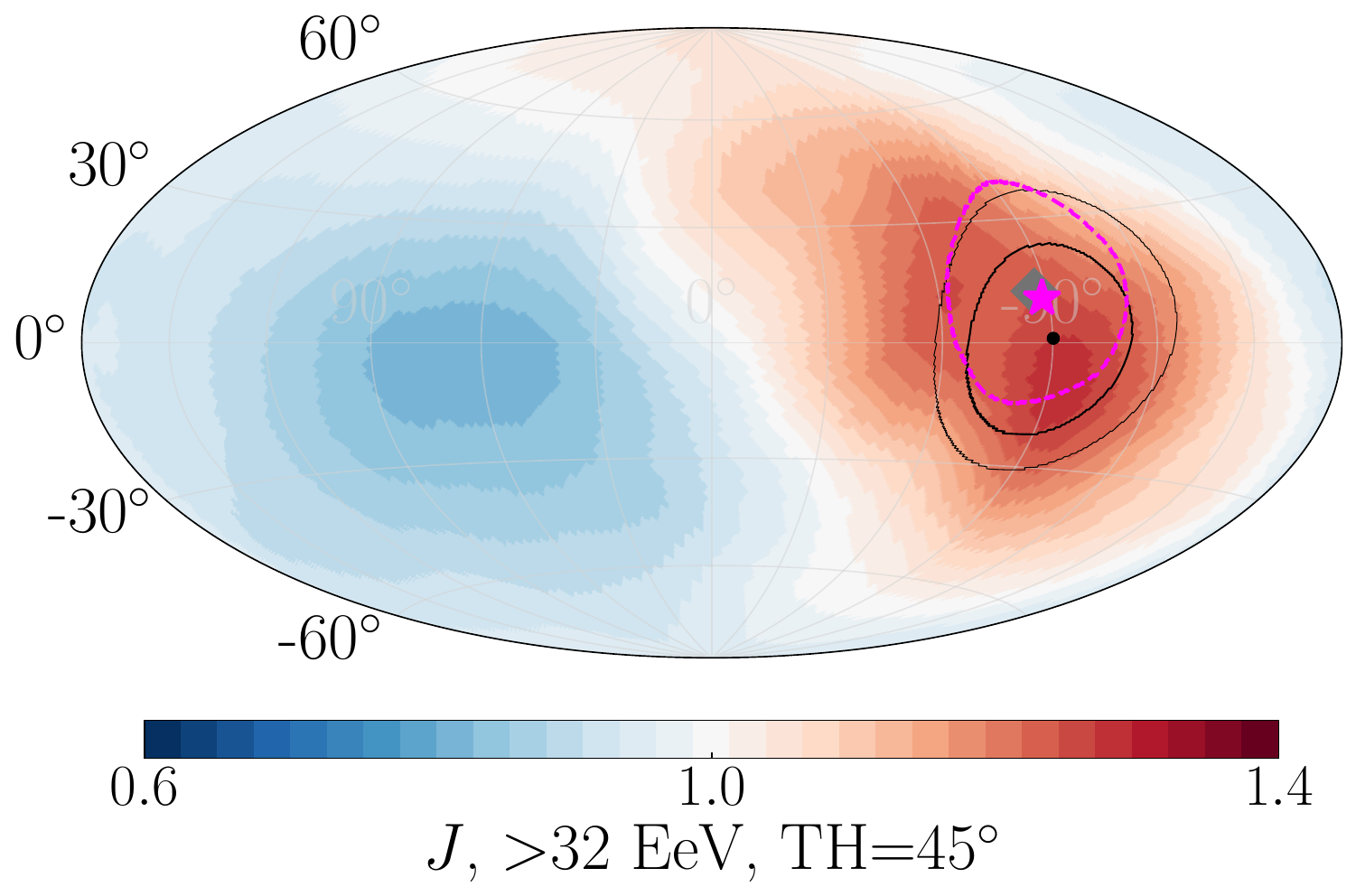,
}
\place{\textbf{flux, JF12 reg}, no TH}{
  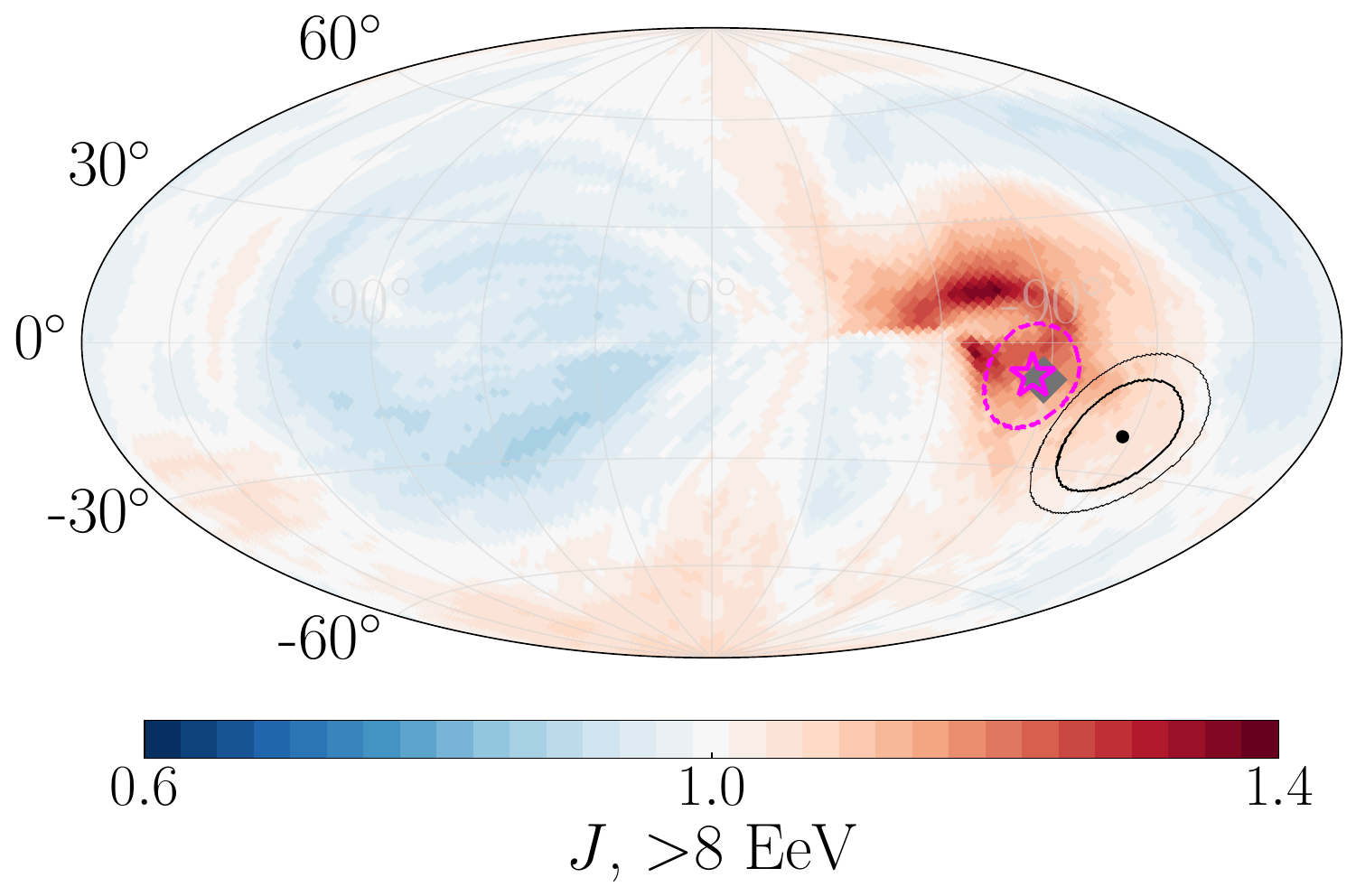,
  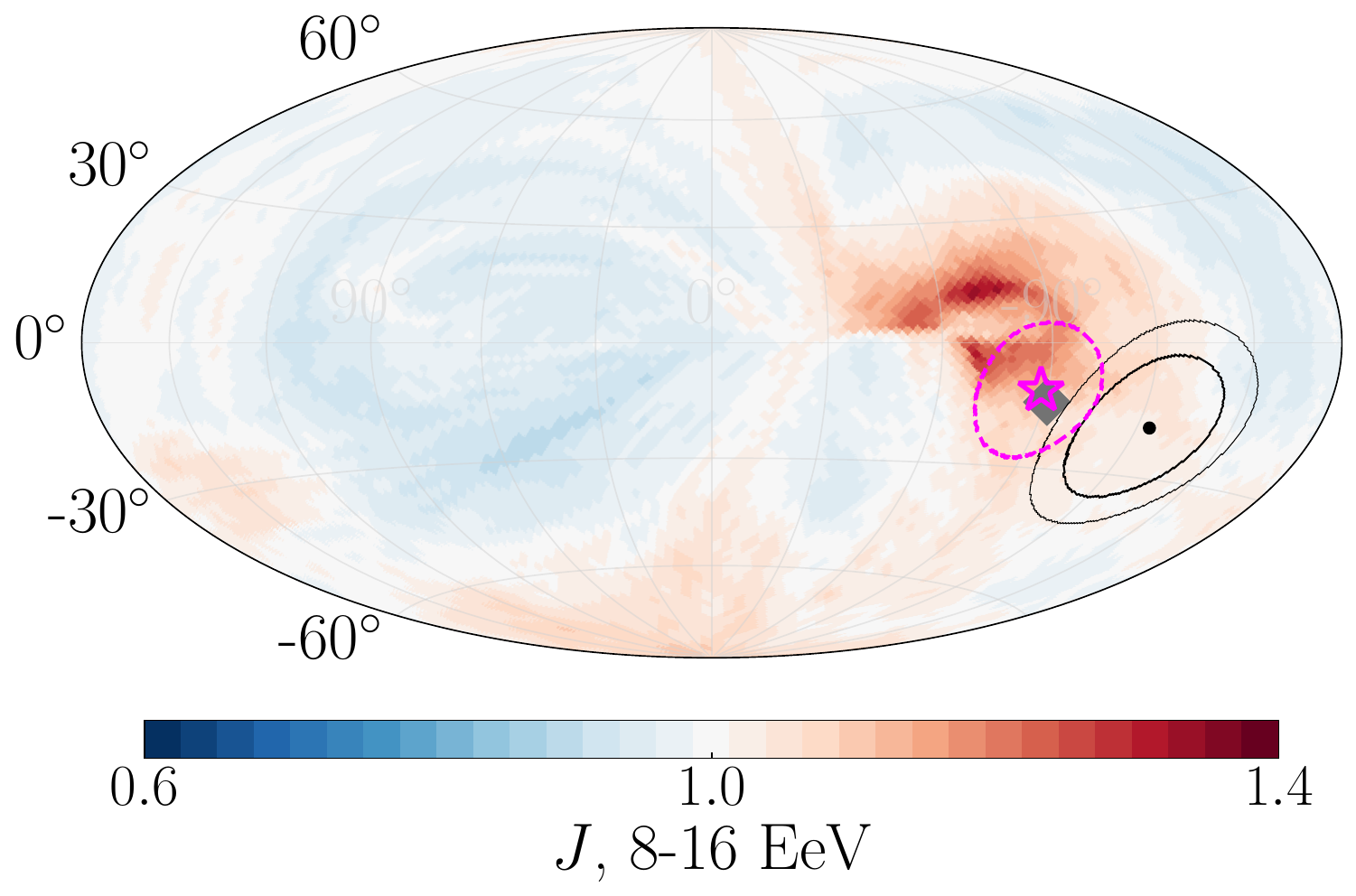,
  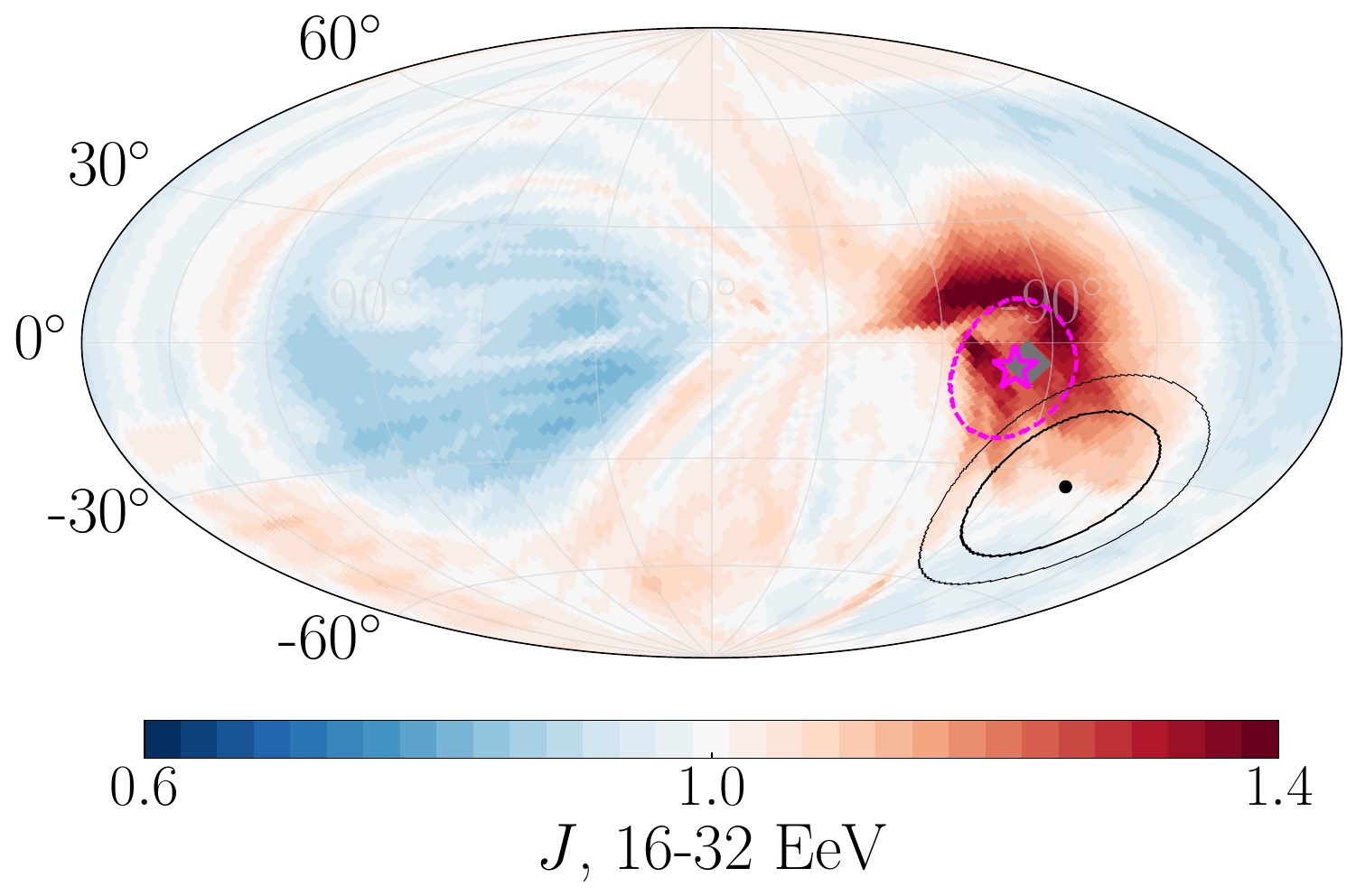,
  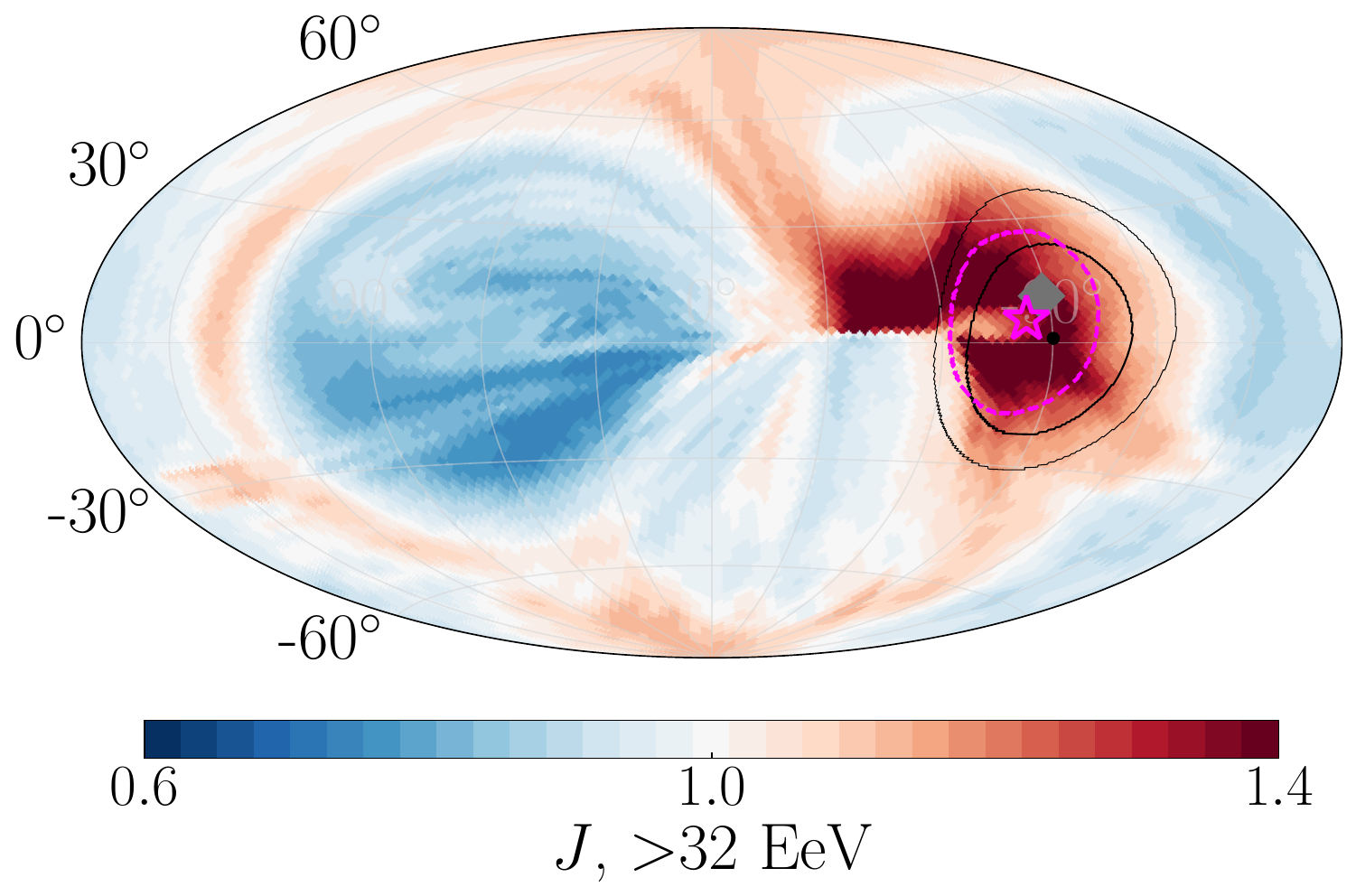,
}
\place{\textbf{flux, JF12 reg}, TH=$45^\circ$}{
  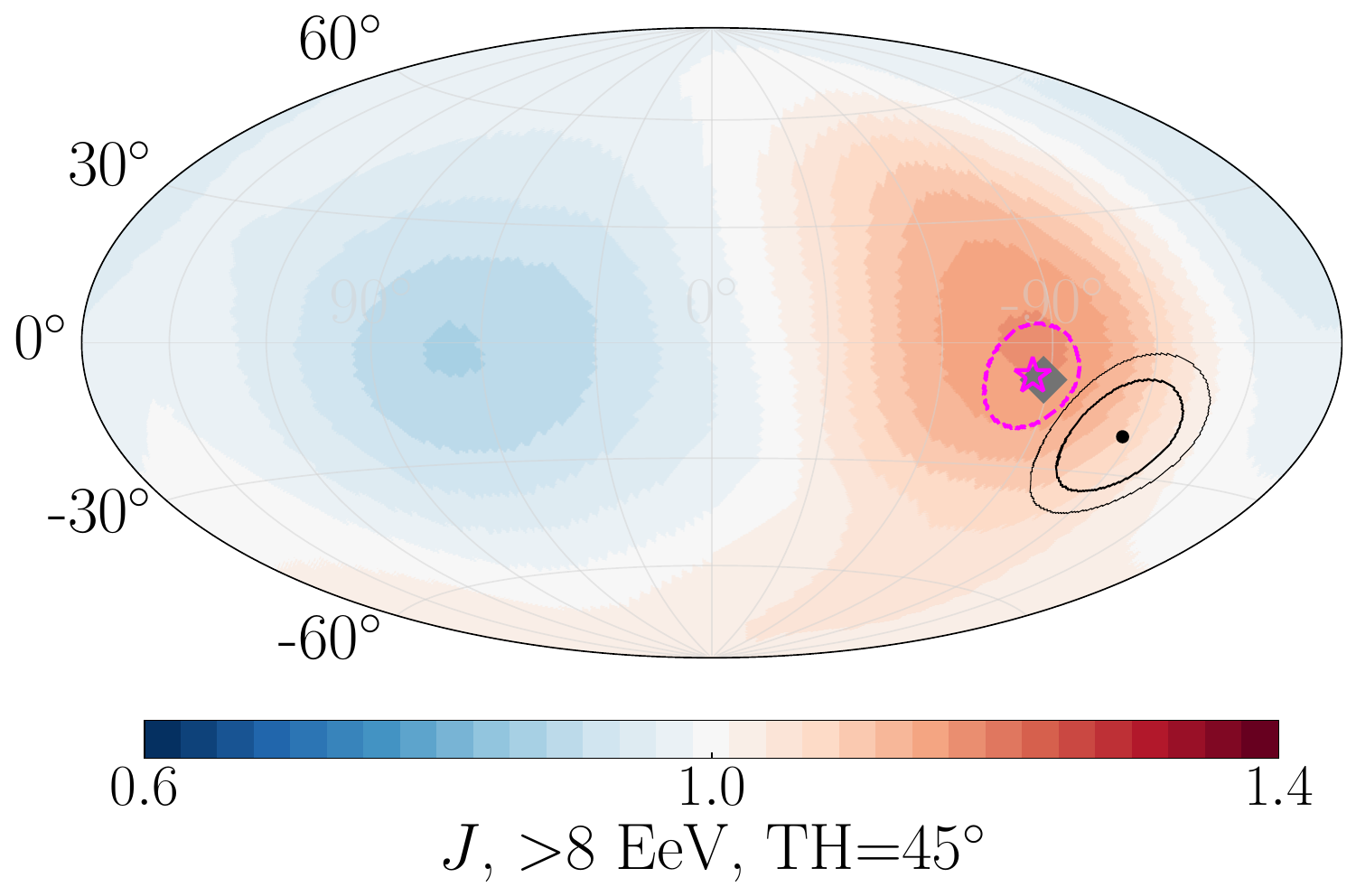,
  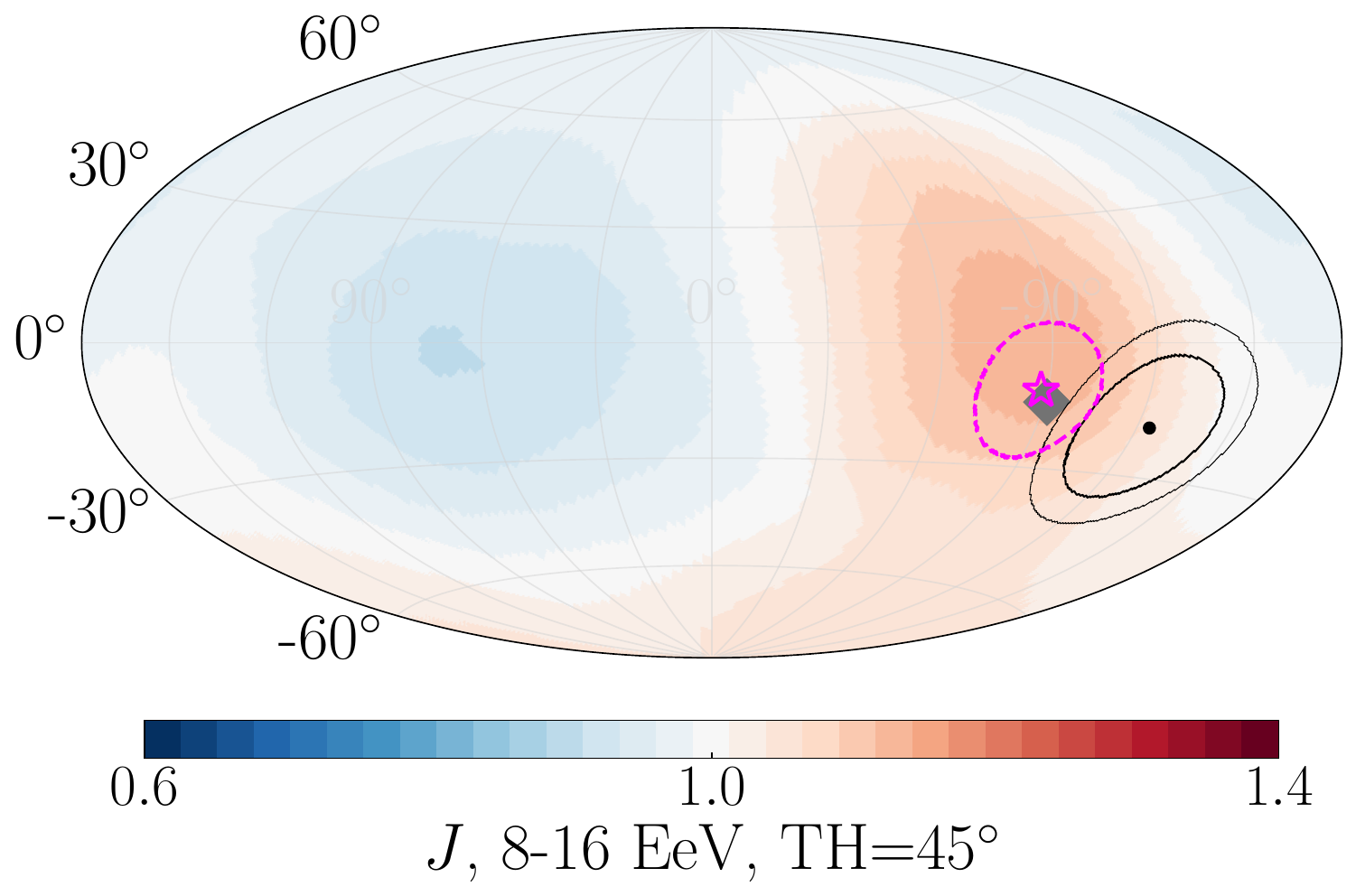,
  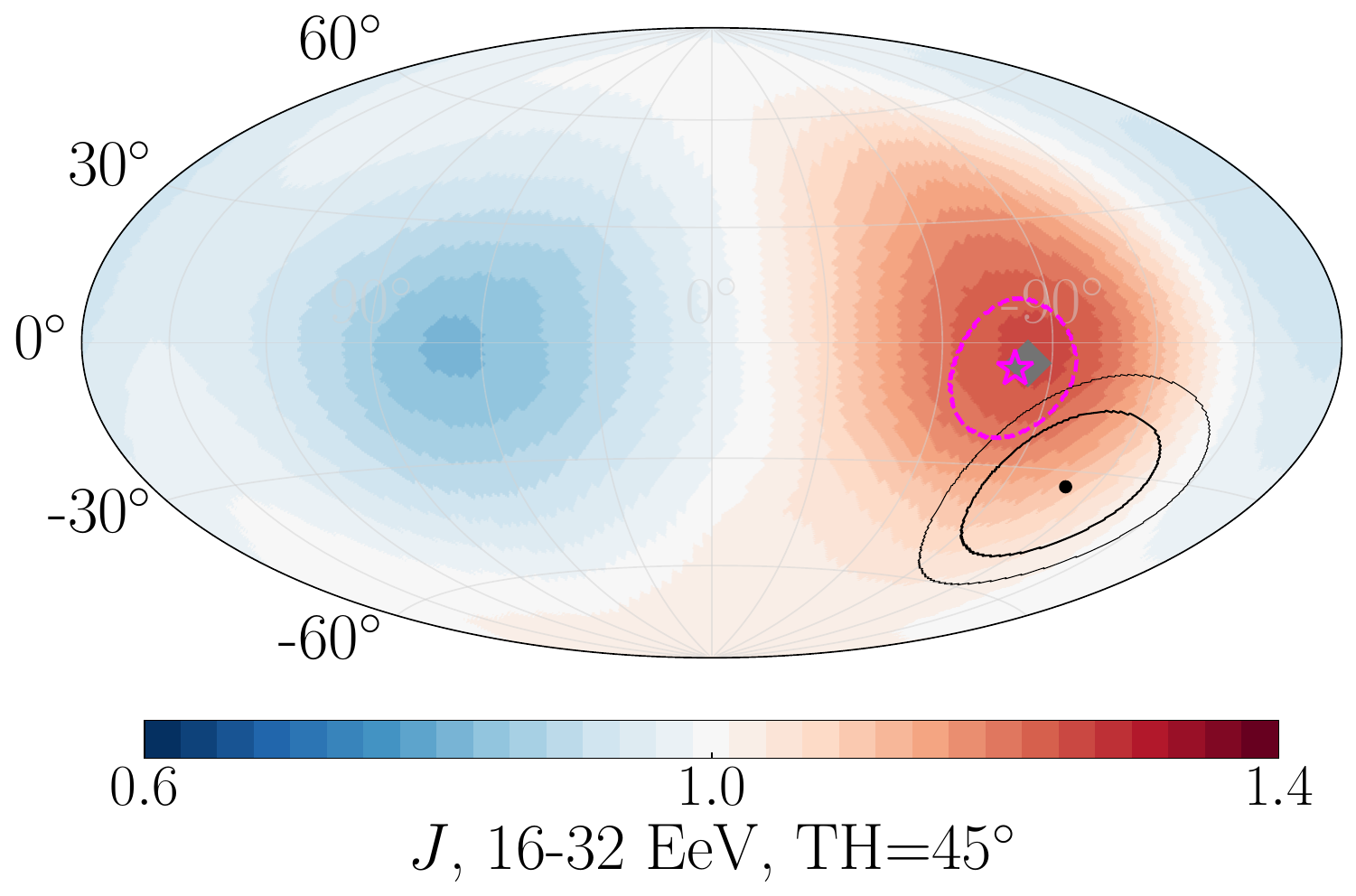,
  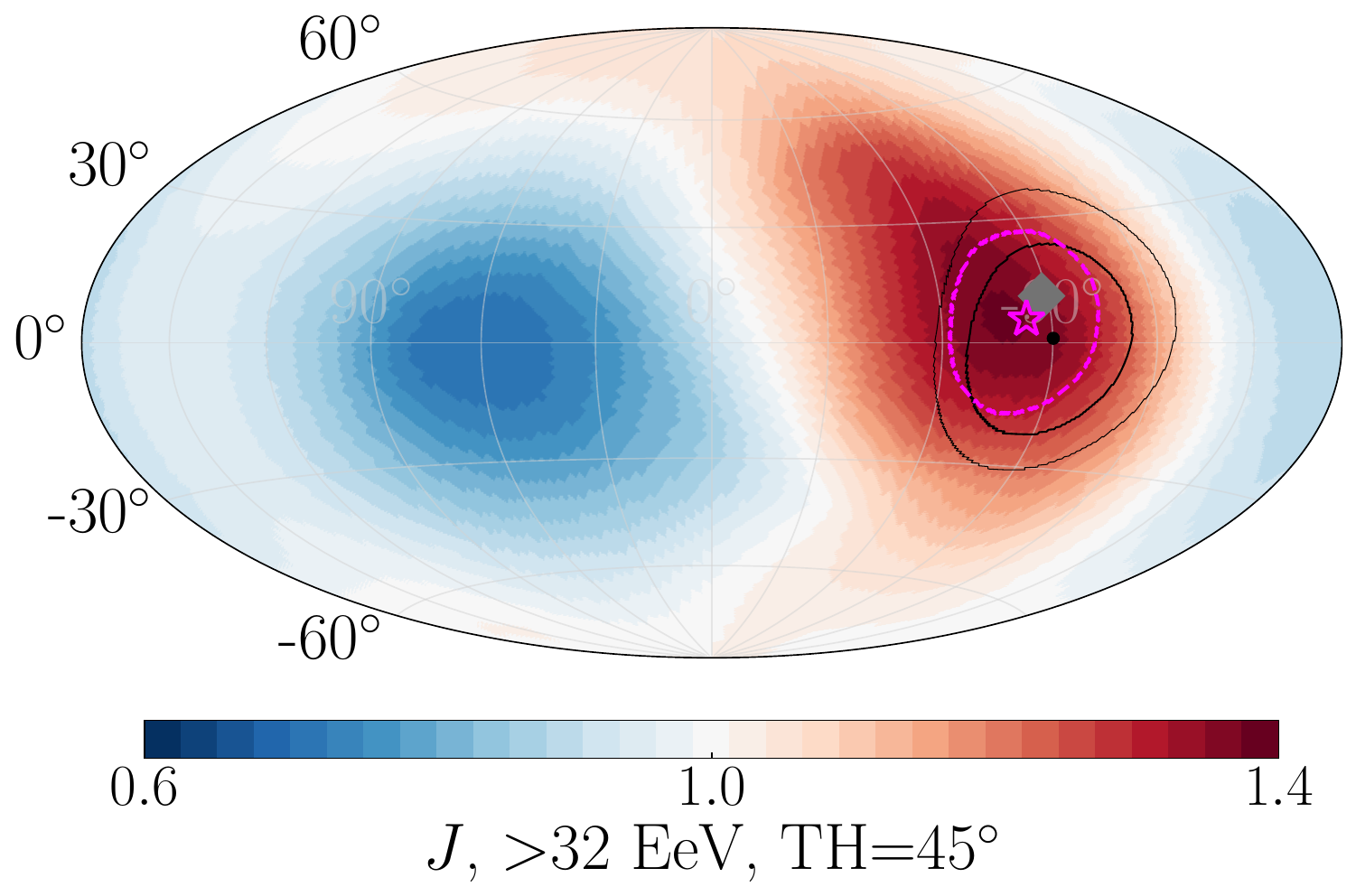,
}
\end{places}
\caption{Baseline model illumination maps $I$ (\textit{upper row}) and arrival maps $J$, without (\textit{second row}) and with (\textit{third row}) $45^\circ$-tophat (TH) smearing. The columns show different energy thresholds or bins; see text below each figure. The last two rows are for the JF12 regular-only GMF model with no random component.  
All maps show the ratio of flux to the map mean, in Galactic coordinates. The black point and circles indicate the measured dipole with $68\%$ and $90\%$ domains from Auger~\citep{Almeida_dipole_2021}. The pink star and circle show the direction and $68\%$ uncertainty of the model dipole, including the Observatory exposure; the grey marker indicates the full-sky dipole direction in the LSS model. The numerical values for the dipole directions and amplitudes are given in the appendix in Table~\ref{tab:dipole_lb}.}
\label{fig:dipole_direc}
\end{figure*}

\subsection{Dipole predictions}
Due to the energy-independence of the rigidity, as seen in Fig.~\ref{fig:rig_spec}, it is not surprising that the observed dipole direction stays relatively constant for different energy bins, both for observations and predictions, as can be seen in the last four rows of Fig.~\ref{fig:dipole_direc}. The Auger dipole is shown in black and the model predictions, including the effect of exposure, in pink. 
At the highest energies, the horizon is relatively close, where sources toward the Galactic North Pole dominate the flux. This moves the dipole direction, even though the rigidity is not much different. This horizon effect can be seen comparing the illumination maps shown in the upper row of Fig.~\ref{fig:dipole_direc} (see also Fig.~\ref{fig:illum}). 

The model prediction of the dipole direction is depicted including its 68\% C.L. statistical uncertainty in Fig.~\ref{fig:dipole_direc}. To estimate this statistical uncertainty, we created $5,000$ realizations of $N_\mathrm{events}=44,000$ events (as in the Auger data set~\citep{Almeida_dipole_2021}), following the model. Earlier studies~\citep{ding_imprint_2021,allard_what_2022} did not provide an estimation of the statistical uncertainties in their dipole direction and amplitude predictions.  
%How Auger created its uncertainty regions is not explained in the papers we found; a jack-knife method is a possible approach when the underlying distribution is not known.  

Note that based on the LSS model, no strong excesses are expected in the Telescope Array field of view, indicating that the possible hotspot~\citep{Kim_ICRC_2023}, if statistically significant, should arise from a local source rather than the LSS. Note also that the speculated correlation of an anisotropy with the Perseus-Pisces supercluster~\citep{Kim_ICRC_2023} is not observed in the UHECR arrival maps in Fig.~\ref{fig:dipole_direc}: even though there is an overdensity in the illumination map from the supercluster (Fig.~\ref{fig:illum}), it is displaced or dissolved completely by the GMF. The small-scale overdensity in the Centaurus region observed by Auger
\citep{Auger_ADs_2022}, however, coincides well with the overdensity predicted by the LSS model at $\geq32$~EeV. The question whether local sources are necessary on top of the LSS to explain the higher-energy small-scale anisotropies observed by Auger and TA will be left for a future publication. Note however that none of these small-scale overdensities is presently statistically significant (especially at the lower energy range that is the focus of this work). In Sec.~\ref{sec:constraints} we use the upper limits on small-scale anisotropies from the power spectrum~\citep{Caccianiga_ICRC2023}.

\begin{figure}[ht]
\includegraphics[width=0.48\textwidth]{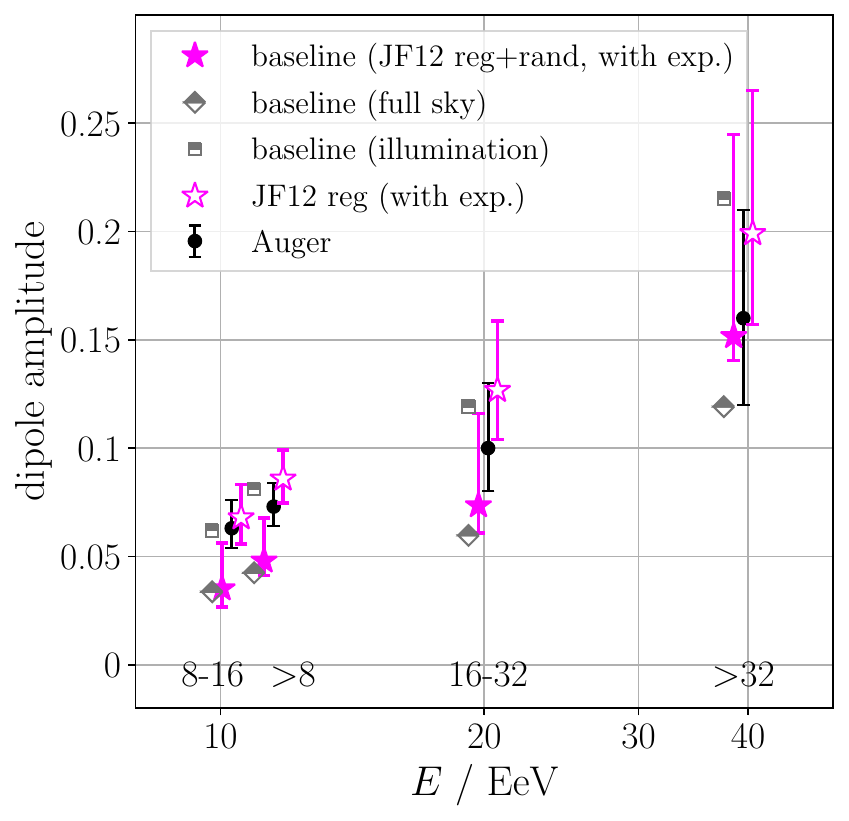}
\caption{Dipole amplitude comparing to the Auger data (black markers, called $d$ in Table 1 of~\citep{Almeida_dipole_2021}). The grey squares are the full-sky dipole moments of the illumination map (from Fig.~\ref{fig:dipole_direc}, upper row) and grey diamonds are the full-sky model prediction. The pink stars are the model predictions including the effect of exposure, with error bars showing the 68\% C.L. statistical uncertainty due to the limited number of events as discussed in the text, filled for the JF12 reg+rand (baseline) model and unfilled for JF12-reg. The data points are positioned at the energy-center of the data set for each energy threshold, with model markers offset on the x-axis for better visibility.}
\label{fig:dipole_amp}
\end{figure}

Figure~\ref{fig:dipole_amp} shows the dipole amplitude. One sees that it increases with the energy. This is due to a combination of the shrinking propagation horizon (inducing greater anisotropy in the illumination maps) and to some extent a decreasing diffusion in the GMF for higher-rigidity particles.  

Both the predicted amplitude and direction of the dipole are very close to the measurements for the highest energy bin \textgreater32\,EeV. For smaller energies, the dipole amplitude is well-bracketed by the baseline model using the JF12 coherent plus random field and the regular-field only model, shown with filled and open stars respectively. However, for both models, the direction is somewhat outside the 2$\sigma$ region. An exciting prospect is that the dipole direction can be used to identify a preferred subset within the new suite of models provided by~\citet{UF23}, designed to encompass the present uncertainties in the GMF.  

\subsection{Model variations}
We tested the influence of the dipole (compared to spectrum and composition) on the overall fit by removing the dipole likelihood part from Eq.~\ref{eq:lik_tot}. In that case, where only energy spectrum and composition are considered, the fit parameters as well as the likelihood values are quite similar to the baseline case. The values are given in Table~\ref{tab:results} in the Appendix. Clearly, the fit is predominantly constrained by the spectrum and composition, and is not distorted in order to fit the dipole components.
This is confirmed by the fact that the fit parameters do not differ much between the JF12-reg+rand (baseline) and the JF12-reg cases, as also seen in Table~\ref{tab:results}.  

The effects of the most important model uncertainties - the shape of the injection cutoff and the uncertainty in the composition due to the hadronic air shower modeling - are investigated in Sec.~\ref{sec:model_uncertainties} of the Appendix, with the respective fit parameters and likelihood values given in Table~\ref{tab:results}. It is evident that the predictions of the model are independent of such systematic uncertainties.  

Thus, in the following, we use the baseline LSS model to obtain further insights into the influence of cosmic magnetic fields, and also the number density and \textit{bias} of sources.

\section{Production of UHECRs in high- and low-matter-density regions} \label{sec:bias}
In this section, we investigate the effect of a possible \textit{bias} in the relation between UHECR sources and the underlying LSS distribution. Such a bias could arise, for example, if UHECR sources are predominantly found only in matter-dense regions, as is the case for example for the most energetic active galactic nuclei (AGNs)~\citep{Hale_2017}, or if UHECRs are accelerated in the large scale shocks of massive galaxy clusters~\citep{blandford+ClusterAccel18}; see also~\citet{Oikonomou:2014zva} for exemplary simulations with varying bias. On the other hand, a scenario where UHECRs are accelerated everywhere, but the stronger magnetic fields inside matter-dense regions inhibit their escape (e.g.,~\citet{condorelli_testing_2023}), would lead to a bias towards low-density regions.

\begin{figure}[htp]
\centering
\begin{places}{0.45\textwidth}{2}
\place{baseline}{
  best_fit_dipole__I_map_9_-1.pdf,
  best_fit_dipole__map_9_-1.pdf
}
\place{cut $\rho < 0.5 \overline{\rho}$}{
  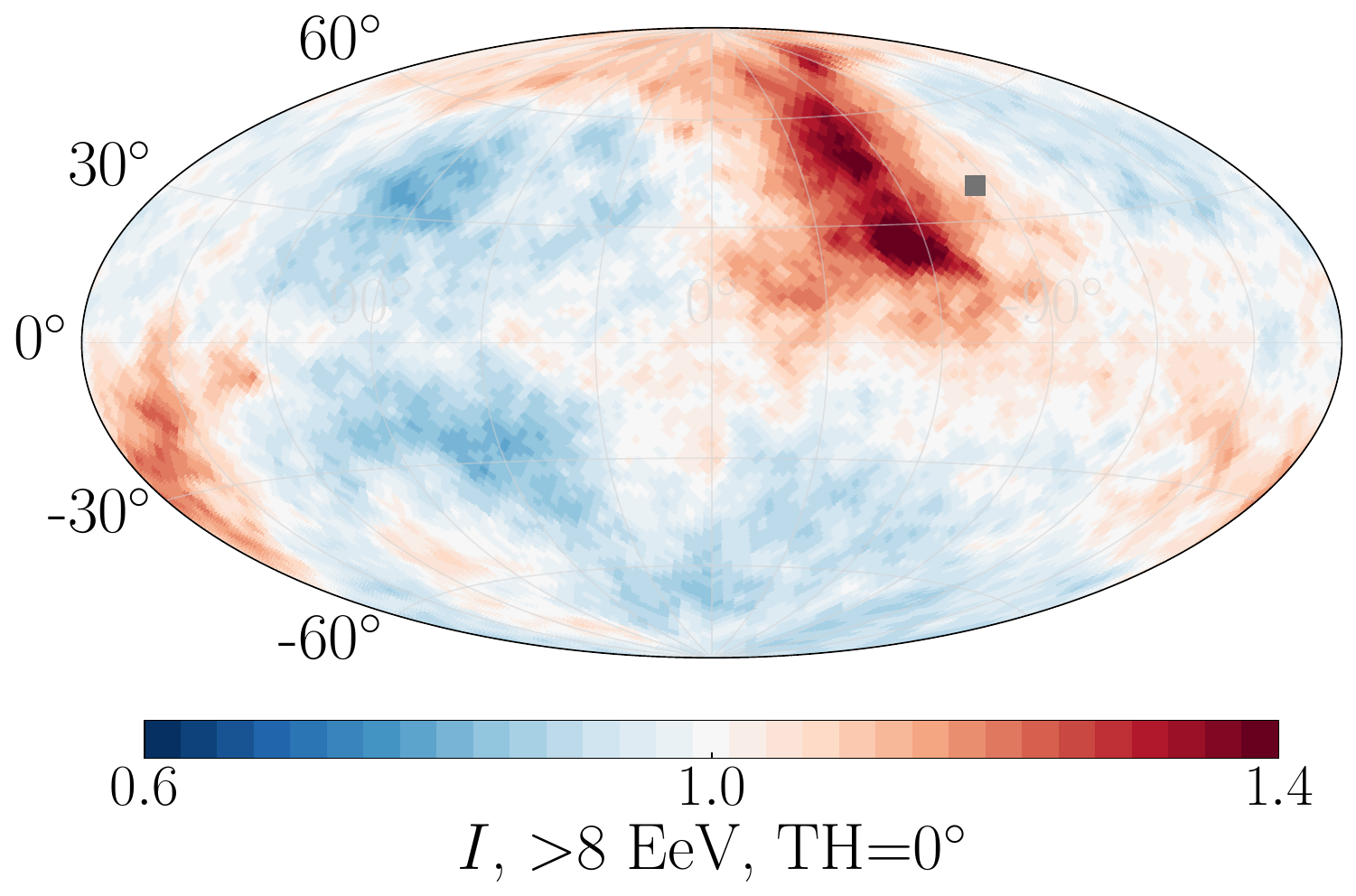,
  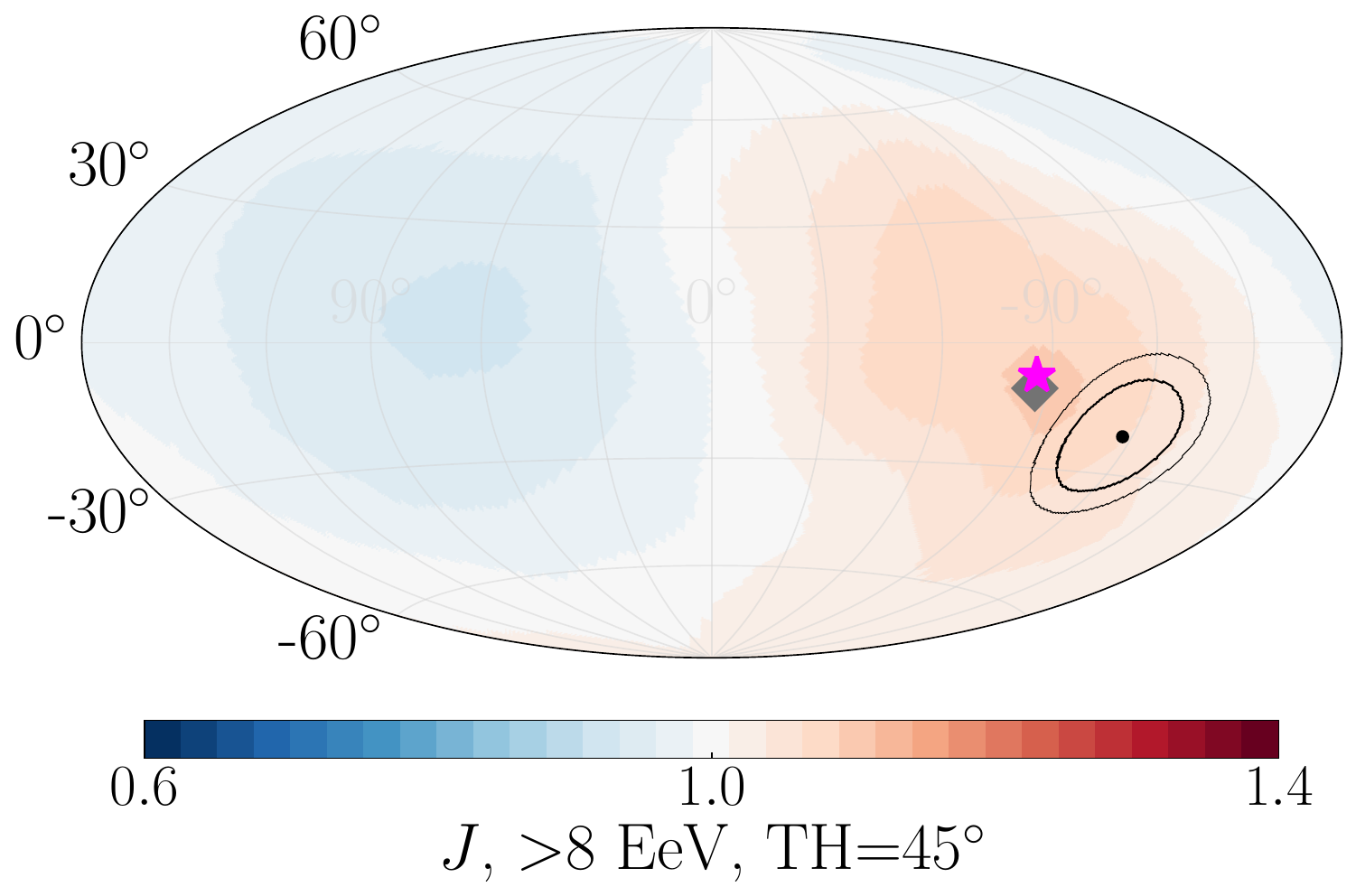
}
\place{cut $\rho < 1.5 \overline{\rho}$}{
  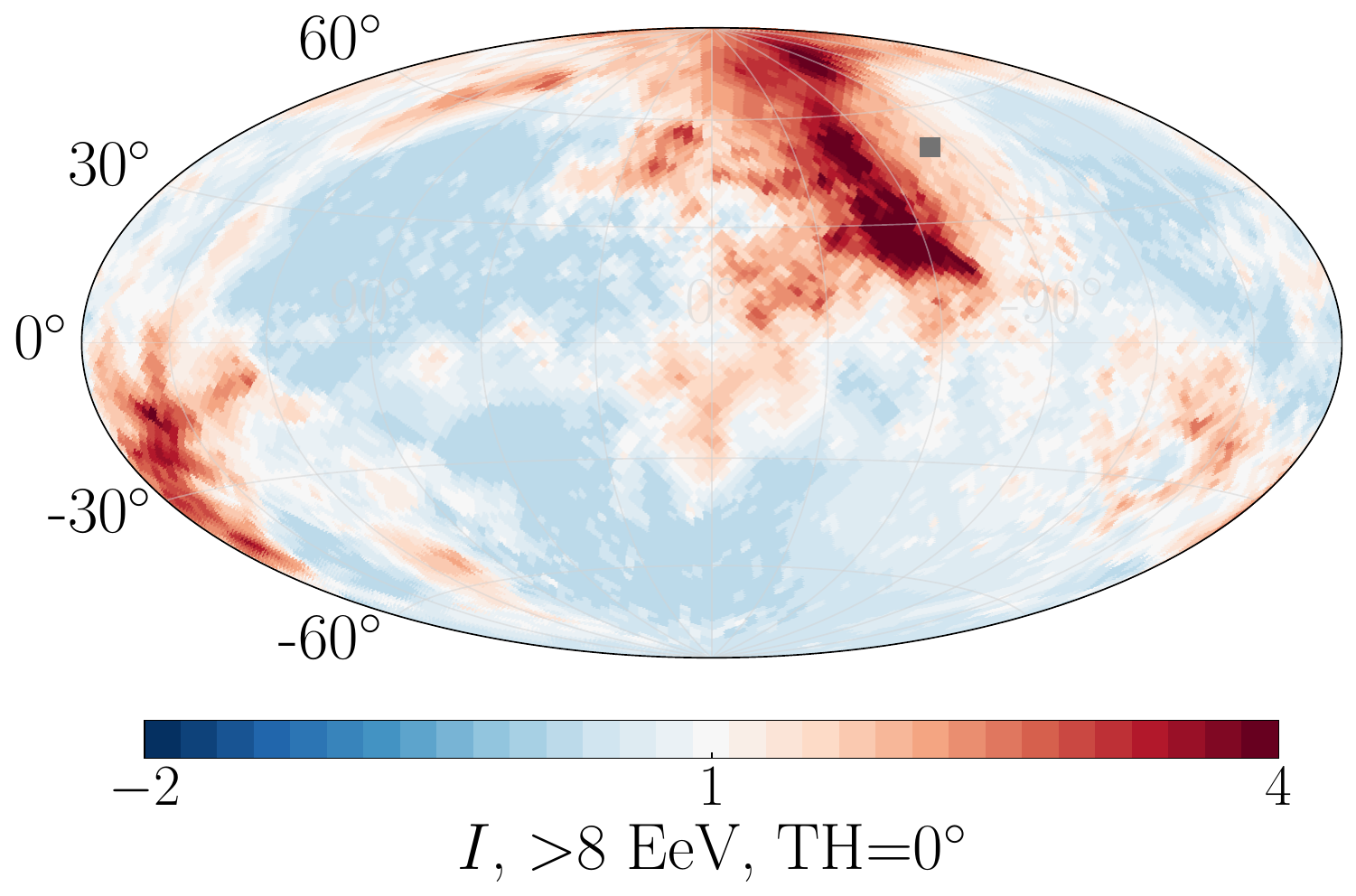,
  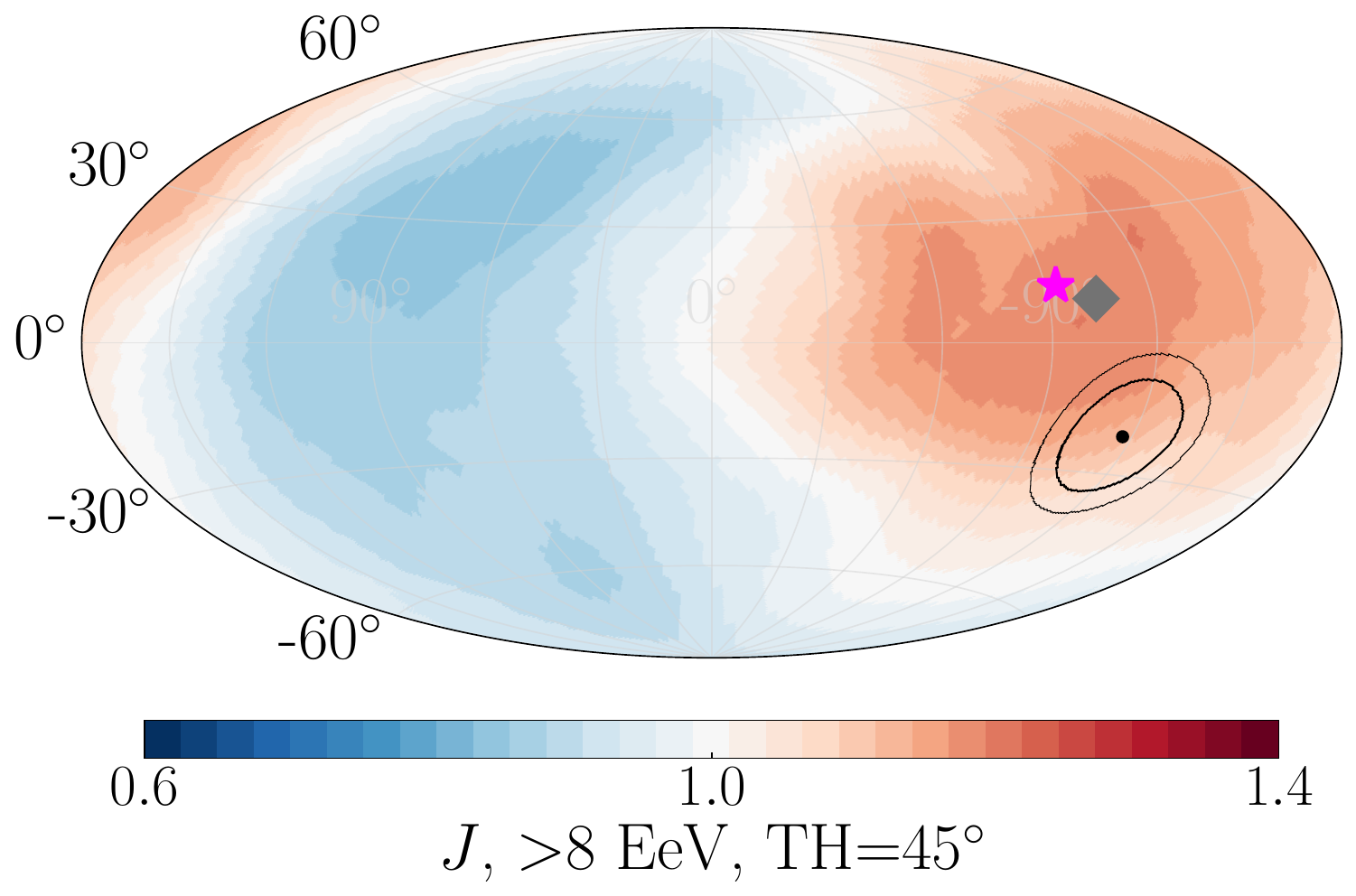
}
\place{cut $\rho > 4 \overline{\rho}$}{
  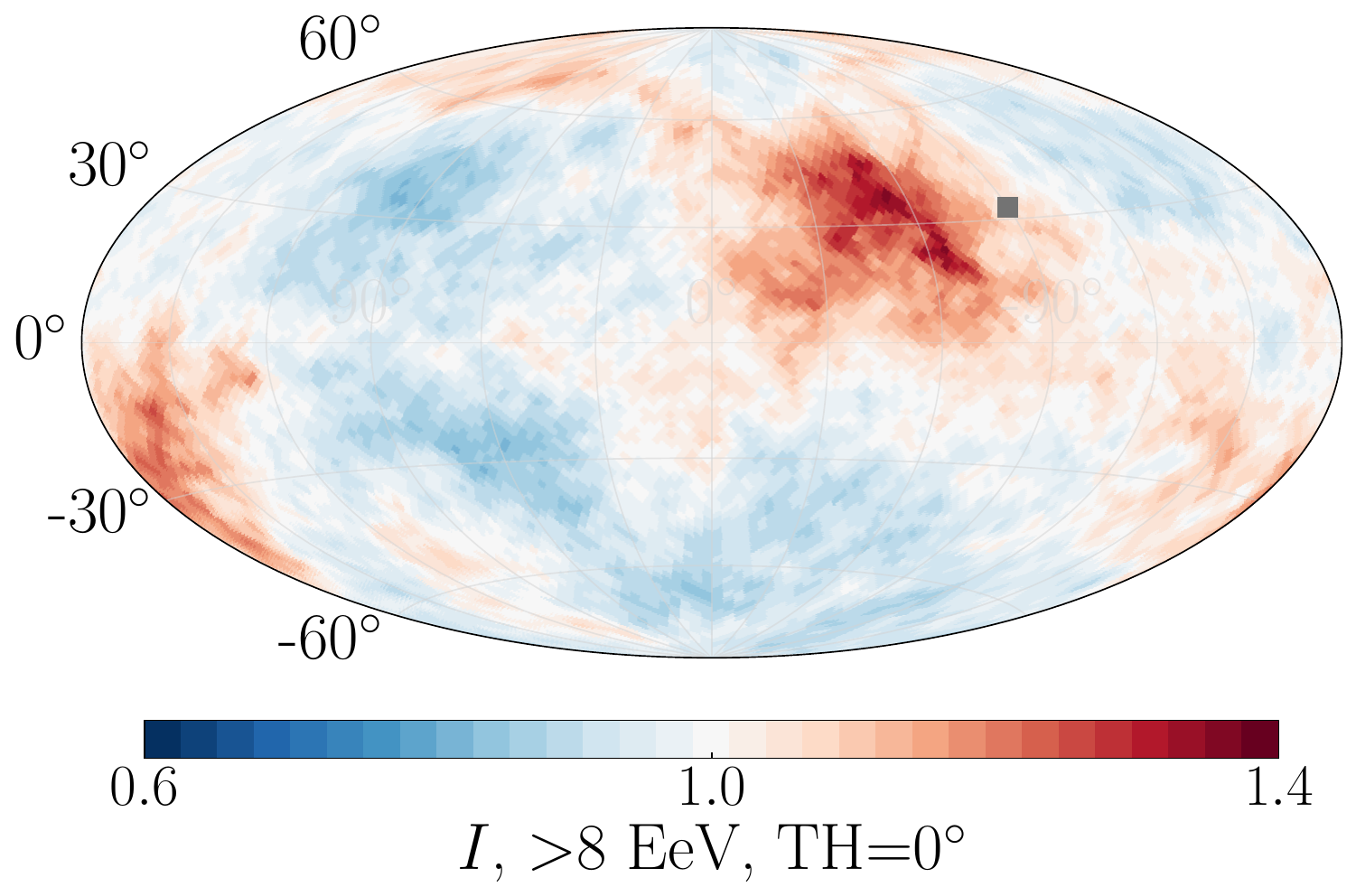,
  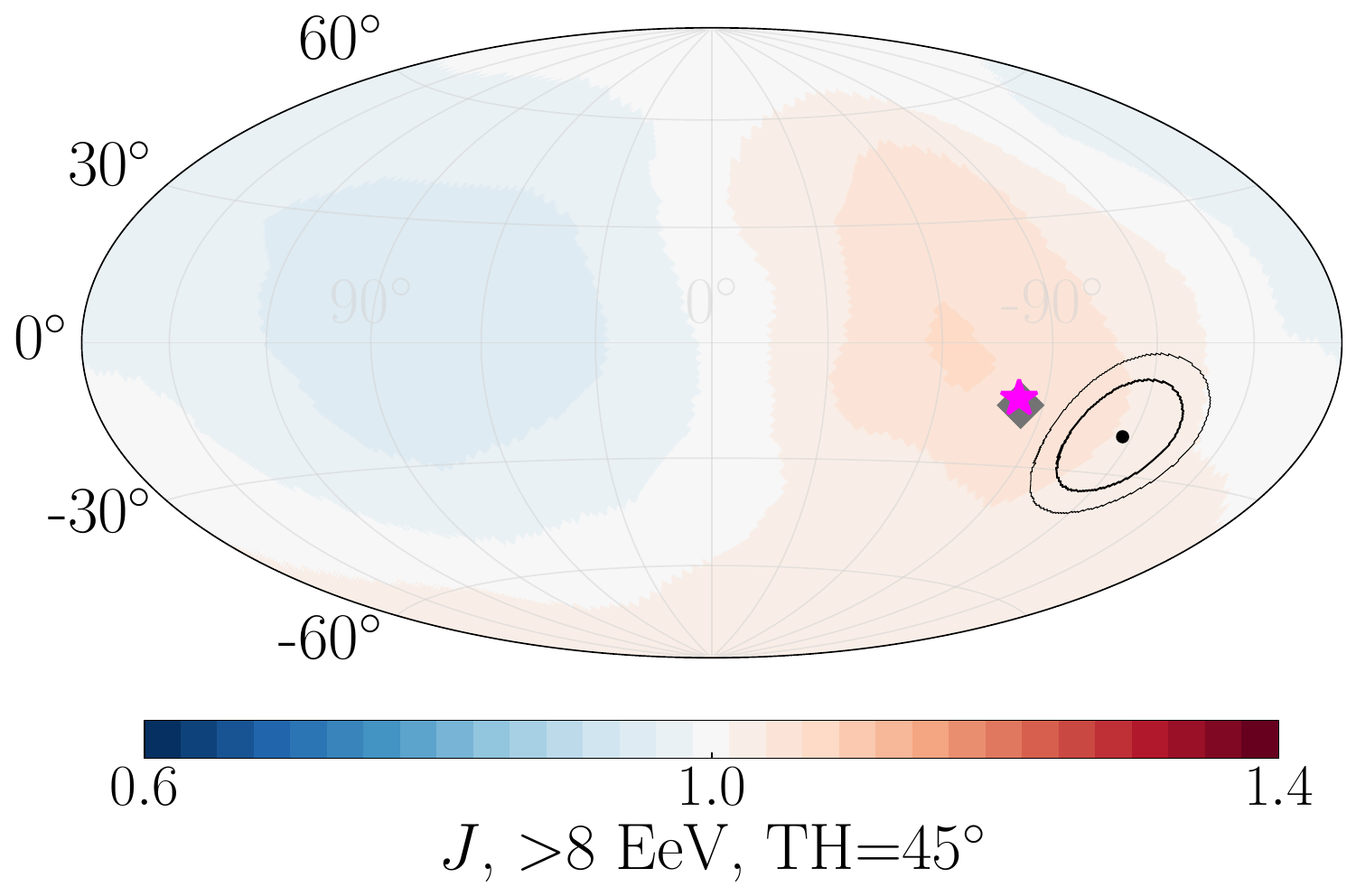,
}
\end{places}
\caption{Example illumination (\textit{left}) and arrival (\textit{right}) maps illustrating the effect of \textit{bias}. \textit{First row}, the baseline case with no bias; \textit{second row}, a small bias away from low-density regions with \{$\rho_\mathrm{min}/\overline{\rho}=0.5$, $\rho_\mathrm{max}/\overline{\rho}=\infty$\};  \textit{third row}, a large bias away from low-density regions with \{$\rho_\mathrm{min}/\overline{\rho}=1.5$, $\rho_\mathrm{max}/\overline{\rho}=\infty$\} (note different color bar normalization!); and \textit{last row} a small bias away from high density regions \{$\rho_\mathrm{min}/\overline{\rho}=0$, $\rho_\mathrm{max}/\overline{\rho}=4.0$\}. In comparison to the bias-free illumination map (\textit{upper row}), it is visible how the cut on high-density regions partly removes the nearby Virgo cluster in the North. The cuts excluding the low-density regions lead to more anisotropy in the illumination and thereby to a larger dipole amplitude. For the extreme cut favoring the high-density regions, the dipole direction is displaced more towards the Galactic North than in the baseline case.}
\label{fig:bias_example}
\end{figure}

We test for both possible directions of bias by excluding regions of the source distribution in the fit, if they either exceed a maximum density $\rho_\mathrm{max}$, or are lower than a minimum density $\rho_\mathrm{min}$ (where the mean density is denoted as $\overline{\rho}$). Example illumination maps and corresponding flux maps are depicted in Fig.~\ref{fig:bias_example}, visualizing the impact of bias. Note that a bias towards high-density regions leads to a shift of the predicted dipole direction towards the Galactic North. This could explain slight differences in the predicted dipole direction between our work and~\citet{allard_what_2022}, who use a volume-limited, inevitably biased 2MRS-based catalog (c.f., discussion in the Introduction).

\begin{figure}[ht]
%\begin{figure}
\includegraphics[width=0.48\textwidth]{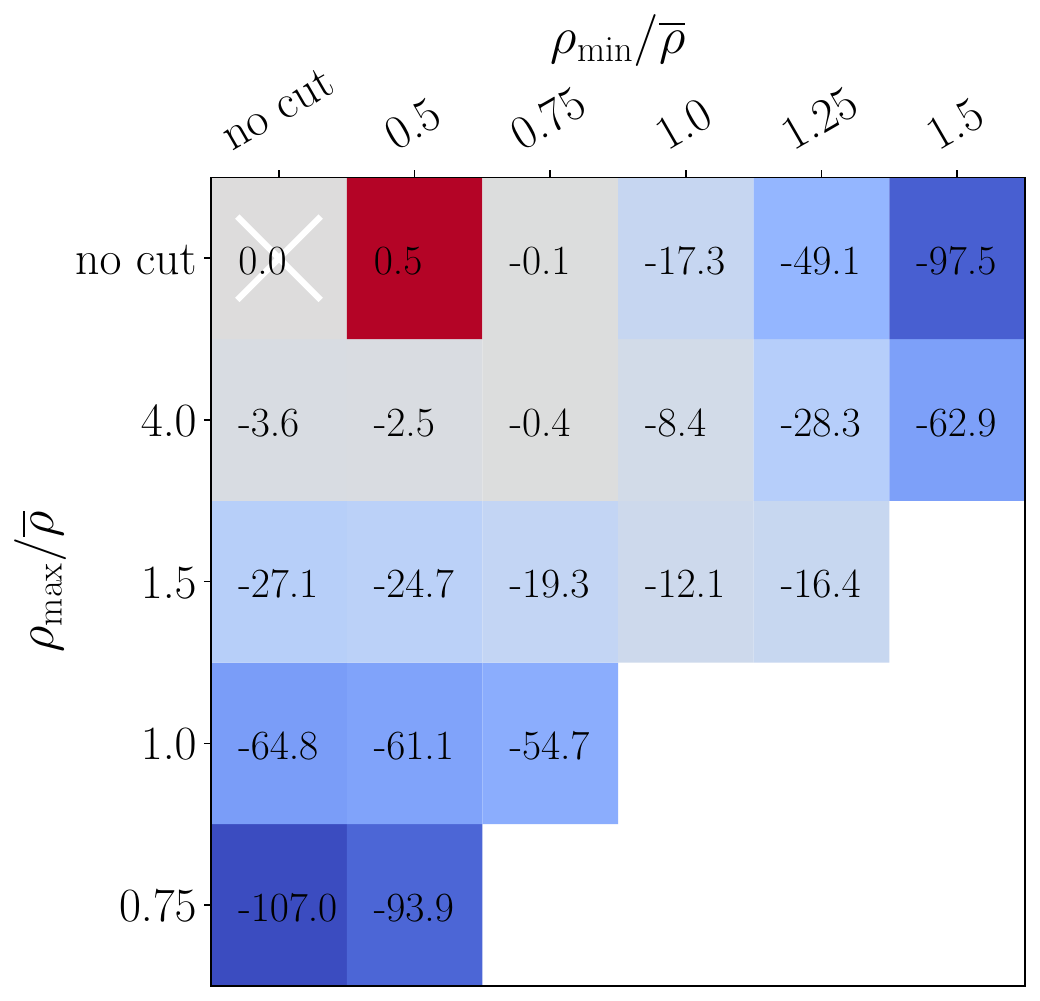}
\caption{Influence of a bias between UHECR source distribution and LSS. Regions with densities below $\rho_\mathrm{min}$ or above $\rho_\mathrm{max}$ are excluded from the UHECR injection during the fit. Colors and numbers indicate the likelihood ratio ($\log \mathcal{L}_\mathrm{tot} - \log \mathcal{L}_\mathrm{tot}(x)$) compared to the baseline model without cuts (marked with white \textit{x}).}
\label{fig:bias}
% \end{figure}
\end{figure}

The result of a scan over bias using multiple combinations of $\rho_\mathrm{min}$ and $\rho_\mathrm{max}$ is shown in Fig.~\ref{fig:bias}.
It can be seen that the likelihood decreases for almost any value of the cuts. Without the high-density regions, the sky becomes more isotropic. Then, the dipole direction is not represented well anymore, especially as even the conservative cut $\rho_\mathrm{max}=4\overline{\rho}$ already removes parts of the nearby Virgo cluster in the Galactic north (see Fig.~\ref{fig:bias_example}, \textit{left}), which is important to reproduce the observed dipole. 

Removing low- and intermediate-density regions with $\rho\lesssim\overline{\rho}$ leads to too large anisotropies, overshooting the dipole amplitude. Only a cut removing the very low-density regions below $0.5\,\overline{\rho}$ leads to a slight, non-significant improvement of the likelihood. This study suggests that UHECR sources reside in both high- and medium-density regions, with the CosmicFlows matter density being a good proxy for the UHECR source distribution. No definite conclusion can at this moment be drawn regarding low-density regions.

\section{Constraining the EGMF and the source number density} \label{sec:constraints}
The observation of the dipole offers a unique opportunity to constrain the intertwined relationship between UHECR sources and cosmic magnetic fields. The dipole amplitude is heavily influenced by several effects, most importantly the anisotropy in the source distribution and also the damping due to turbulent magnetic fields. Within the LSS model, the source anisotropy is most significantly influenced by the source number density and to some extent the bias discussed in the previous section. 

In this section, we investigate the effects of the EGMF and source number density on the dipole. 
We begin by introducing the effect of the extragalactic magnetic field (EGMF) on the dipole amplitude, for the continuous LSS source model. Then, we explore the influence of different possible number densities of UHECR sources on the dipole and higher multipole moments %in sec.~\ref{sec:dens} 
for the case of no EGMF. Finally, we study a combined scenario with both EGMF and varying source density.

\subsection{Constraining the extragalactic magnetic field}  \label{sec:egmf}
We have demonstrated above that the LSS source model with the baseline or JF12-reg models of the Galactic magnetic field, brackets the observed dipole amplitude in the absence of significant extragalactic magnetic field smearing. As described in the Introduction, the EGMF is poorly constrained by observations. It can also have a significant effect on the dipole amplitude, as well as higher order multipoles, through a smoothing of the flux distribution before entering the Milky Way. In this subsection, we derive EGMF constraints for the case of continuous source density.

To investigate the influence of the EGMF on the UHECR arrival directions, we consider the most important effect expected for UHECRs propagating in a turbulent field in the non-resonant scattering regime: a Gaussian beam widening\footnote{We do not consider additional flux suppression due to longer path lengths of low-rigidity particles diffusing in the EGMF, because given the maximum possible values of $B / \mathrm{nG}\,\sqrt{L_c / \mathrm{Mpc}}$ (see Fig.~\ref{fig:egmf_dens}) the increased path length is negligible, from Eq.~\ref{eq:tau_eff}. A more detailed discussion can be found in~\cite{Harari_2015, Mollerach_2019}.}. When a UHECR passes through a turbulent magnetic field and its Larmor radius, $r_\text{g}$, is large compared to the coherence length $L_c$ of the field, its direction of propagation evolves diffusively in a calculable way.  The general relationship for the RMS deflection is clear: $\delta \theta \sim \sqrt{D L_c}/r_\text{g}$, but a review of the literature shows a variety of inconsistent expressions for the coefficient;  this is partly due to imprecision in defining the spectrum of turbulence and what is meant by $L_c$, and also because some authors do not concern themselves with factors of $\mathcal{O}(1)$. We find the analysis of~\citet{Achterberg:1999vr} to be particularly useful and explicit.  We adopt a Kolmogorov spectrum of turbulence, for which the diffusion coefficient is derived at the end of App. A~\citep{Achterberg:1999vr}: $\mathcal{D}_0 = 0.227 L_c / r_\text{g}^2$ where $L_c$ is $ \ell_c $ in the notation of~\citet{Achterberg:1999vr}.  The RMS scattering angle is then
%, derived in~\citep{Waxman_1996} based on analogous considerations for dispersion of X-ray pulses~\citep{AlcockHatchett}:
\begin{equation}
\label{eq:delthet}
    \delta \theta = 2.9^\circ \frac{B}{\mathrm{nG}} \frac{10\,\mathrm{EV}}{E/Z} \frac{\sqrt{D \ L_{c}}}{\mathrm{Mpc}} = 2.9^\circ \beta_{\rm EGMF} \frac{10\,\mathrm{EV}}{E/Z} \sqrt{\frac{\overline{D}}{\mathrm{Mpc}}}, 
\end{equation}
where we have introduced in the second equality the combination $\beta_{\rm EGMF} \equiv B / \mathrm{nG}\,\sqrt{L_c / \mathrm{Mpc}}$ to isolate the quantity our analysis can constrain.

The mean distance $\overline{D}$ is calculated for each energy bin of width $\log_{10}(E/\mathrm{eV})=0.1$ (compare to Fig.~\ref{fig:propagation}). To simulate the influence of the EGMF, we convolve the illumination map for each rigidity $R=E/Z$ with a Gaussian of width $\delta \theta$. It is not necessary to do a refit of the source injection parameters for each value of $\beta_\mathrm{EGMF}$ because, as demonstrated above, the fit is governed by the global energy spectrum and composition observables. This is also supported by the finding that the fit parameters do not really differ for the JF12 reg and JF12 reg+rand (baseline) cases as discussed above.

\begin{figure}[bth]
\includegraphics[width=0.48\textwidth]{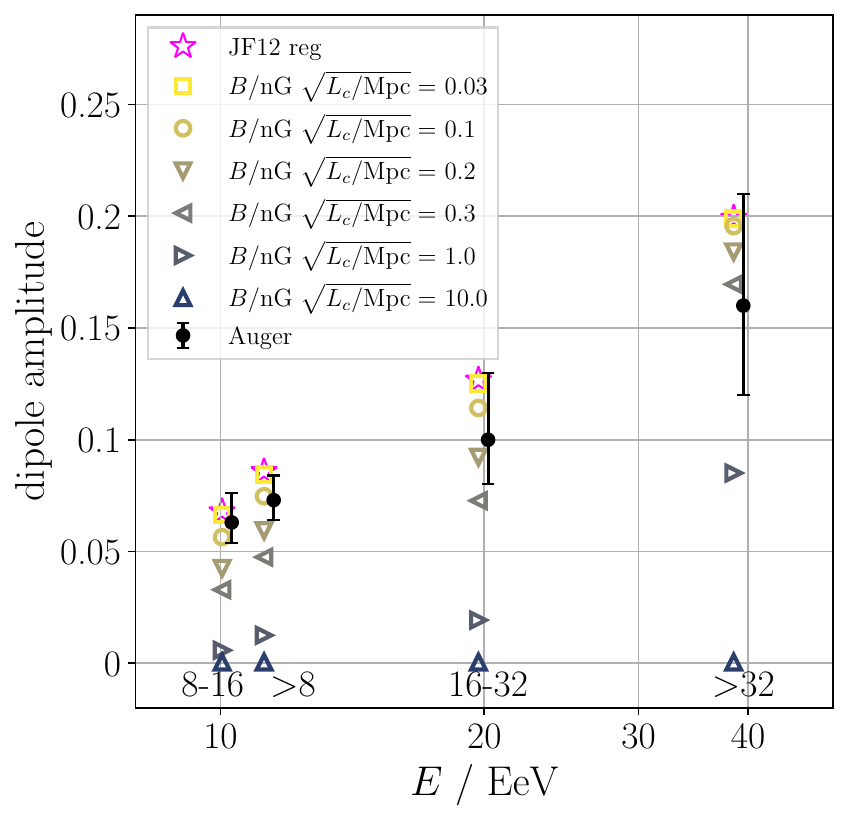}
\caption{Influence of a turbulent EGMF on the dipole amplitude for different values of $\beta_{\rm EGMF} \equiv B / \mathrm{nG}\,\sqrt{L_c / \mathrm{Mpc}}$, compared to the JF12 reg model without turbulent GMF part.}
\label{fig:dipole_amp_EGMF}
\end{figure}
The dipole amplitudes for different values of $\beta_\mathrm{EGMF}$ are shown in Fig.~\ref{fig:dipole_amp_EGMF} for the JF12-reg model, i.e., no turbulent GMF. Using JF12-reg rather than the baseline model minimizes spreading due to the turbulent GMF and hence enables us to place an upper bound on the spreading due to the EGMF in order not to reduce the magnitude of the dipole too much. 
It can be seen in Fig.~\ref{fig:dipole_amp_EGMF} that for values $\beta_\mathrm{EGMF} \gtrsim 0.2$ the dipole amplitude becomes significantly smaller than seen in the Auger data at lower energies. 
Thus, in the limit of continuous sources, the extragalactic magnetic field should respect $\beta_\mathrm{EGMF}\leq 0.2$. Hence for a coherence length of, say, $L_c=1\,\mathrm{Mpc}$, the RMS field strength should not be stronger than $B=0.2$\,nG .

\subsection{Constraining the source number density} ~\label{sec:dens}
In this subsection we place constraints on the source number density, $n$, setting the extragalactic magnetic field to zero. (The lower limit quoted from blazars~\citep{Aharonian_2023} is so much lower than relevant for UHECR smearing we can take it to be zero for this purpose.)  We use both the baseline LSS model and the JF12-reg model, which bracket the amount of random deflections due to the GMF. A value of $n$ which is incompatible with the dipole magnitude and power spectrum for either case can be excluded. When the random component of the GMF is better constrained in the future, the exclusion region will improve. In the next section, Sec.~\ref{sec:egmf_dens}, the more general problem of finding the allowed region in the source density-EGMF plane will be treated.

Reducing the source density from the essentially continuum CosmicFlows model to a finite number density, $n$, produces an additional source of stochasticity in addition to the number of events in the UHECR data set already discussed. This shows up in both the amplitude and direction of the dipole and, importantly, in the rest of the angular power spectrum. Auger~\citep{Almeida_dipole_2021}, as well as the combined Auger+TA working group~\citep{Caccianiga_ICRC2023}, has found that apart from the dipole, all higher multipoles are consistent with being in the 99\% range of variation found in isotropic data sets. This proves to be highly constraining. 

A virtue of using the power spectrum rather than the dipole direction to constrain the number density is the insensitivity of the power spectrum to the exact coherent GMF configuration~\citep{Tinyakov_2015}. This means that we obtain robust constraints in spite of the non-perfect reproduction of the dipole direction at lower energies. How the source number density affects the dipole direction, as well as the energy dependence of the dipole, is reported in Appendix~\ref{sec:density_dipole}.

To model the effect of different source densities, we draw -- for each $n$ studied -- 1000 explicit catalogs of sources from the continuous source distribution. Concretely, for speed reasons, for each catalog the closest $10,000$ sources are drawn and the continuous distribution is used from the distance of the most distant explicit source outwards. We verified that this procedure does not affect the analysis results.
From each source catalog, we draw $N_\mathrm{event}=44,000$ events from those sources, that being the event statistic in the Auger analysis~\citep{Almeida_dipole_2021}, as we did for the continuum model analysis in Sec.~\ref{sec:fit_results}. Additionally, we draw 6000 events following the TA exposure and statistics as in the combined data set used by the Auger+TA working group~\citep{Caccianiga_ICRC2023} in order to have full-sky coverage to calculate the power spectrum.

For each realization, we ask: 
\\ 
1) 
\emph{Are all the higher multipoles within the 99\% isotropic expectation, for all $C_{l>1}$ and energy ranges?} \\ 
2) 
\emph{Is the dipole amplitude large enough?} 

We use the dipole from Auger, rather than the combined Auger-TA dipole, because even though having full-sky coverage in principle would give a more accurate result, in practice the systematic uncertainty in the relative calibration between observatories means that the Auger-only dipole is more precise~\cite{Caccianiga_ICRC2023}, as described in sec.~\ref{sec:fit_method}. A too-large dipole amplitude for a given source density can be mitigated by EGMF smoothing, so we only require that the dipole amplitude be above a minimum, taken conservatively to be $d_\mathrm{8\,EeV}>5\%$ for $E>8\,\mathrm{EeV}$. This value is around $2.5\sigma$ below the Auger value (see Fig.~\ref{fig:dipole_amp}) so it constitutes approximately a $99\%$ C.L. lower limit on the dipole amplitude. This requirement is easily fulfilled by many samples, especially for lower densities.

Table~\ref{tab:dens} reports the number of cases, out of the 1000 realizations for each source density, fulfilling the two criteria. The combination of both criteria excludes densities $n\leq10^{-3.5}\, {\rm Mpc}^{-3}$ at more than 99\% C.L., for the case of negligible EGMF being treated in this section.

\begin{table}[ht]
    \centering
    \begin{tabular}{l | l l l  l  l  l  l}
         $n / {\rm Mpc}^{-3}$ & $10^{-5}$ & $10^{-4}$  & $10^{-3.5}$  & $10^{-3}$  & $10^{-2}$  & $10^{-1}$  \\
         \hline \hline
         $d_\mathrm{8\,EeV}$ >5$\%$ & 906 & 802 & 717 & 587 & 416 & 288\\
         $C_{l>1}$ in $99\%$ iso  & 0 & 0 & 3 & 12 & 273 & 569 \\
         both &  \textbf{0} & \textbf{0} & \textbf{0} & 5 & 100 & 152\\
    \end{tabular}
    \caption{The entries in the table give the number of realizations\, out of 1000, which satisfy the criterion in the first column, for the $\beta_\mathrm{EGMF}=0$ case. 
    "$C_{l>1}\,99\%$ iso" means that apart from the dipole, all multipoles are within the 99\% isotropic expectation for all energy bins and thresholds, % ($8-16$ EeV, $16-32$ EeV, $>8$ EeV, and $>32$ EeV) 
    as observed for the data (c.f., Fig.~\ref{fig:power_spectrum}). }
    \label{tab:dens}
\end{table}

\begin{figure}[ht]
\includegraphics[width=0.49\textwidth]{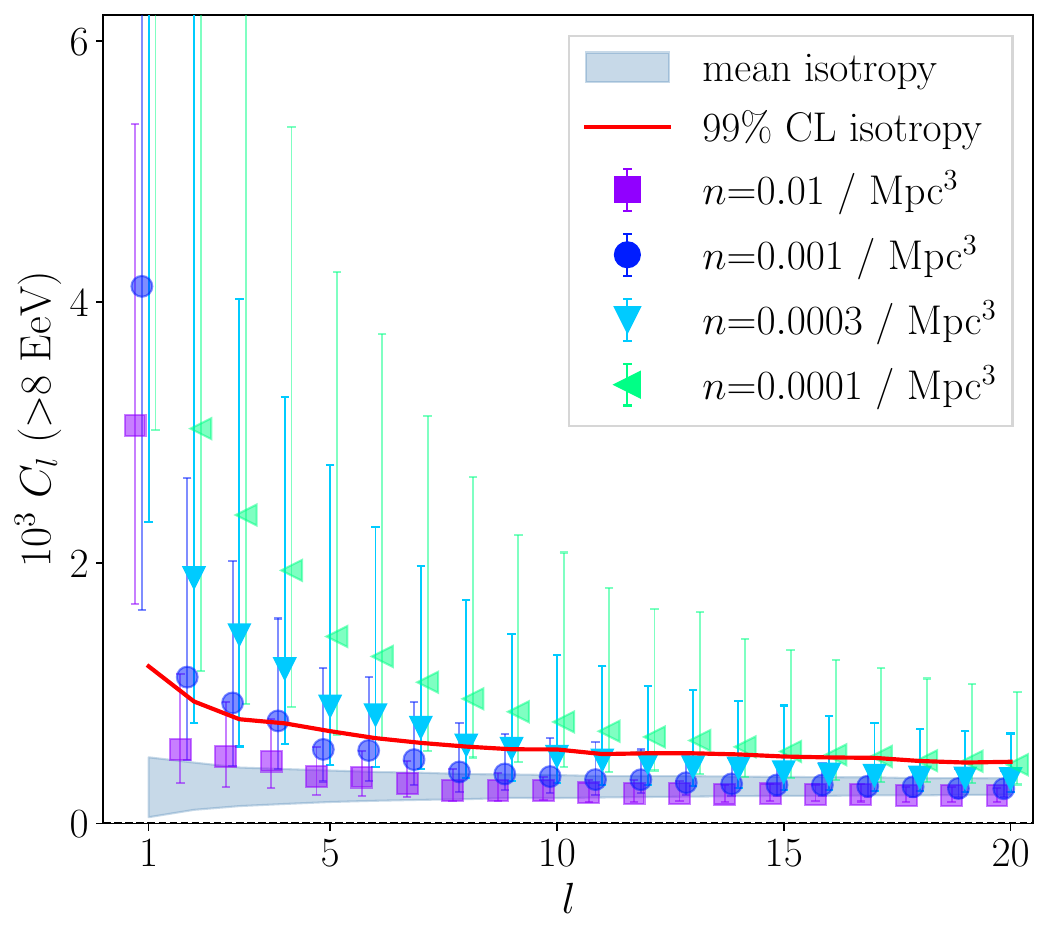}
\includegraphics[width=0.49\textwidth]{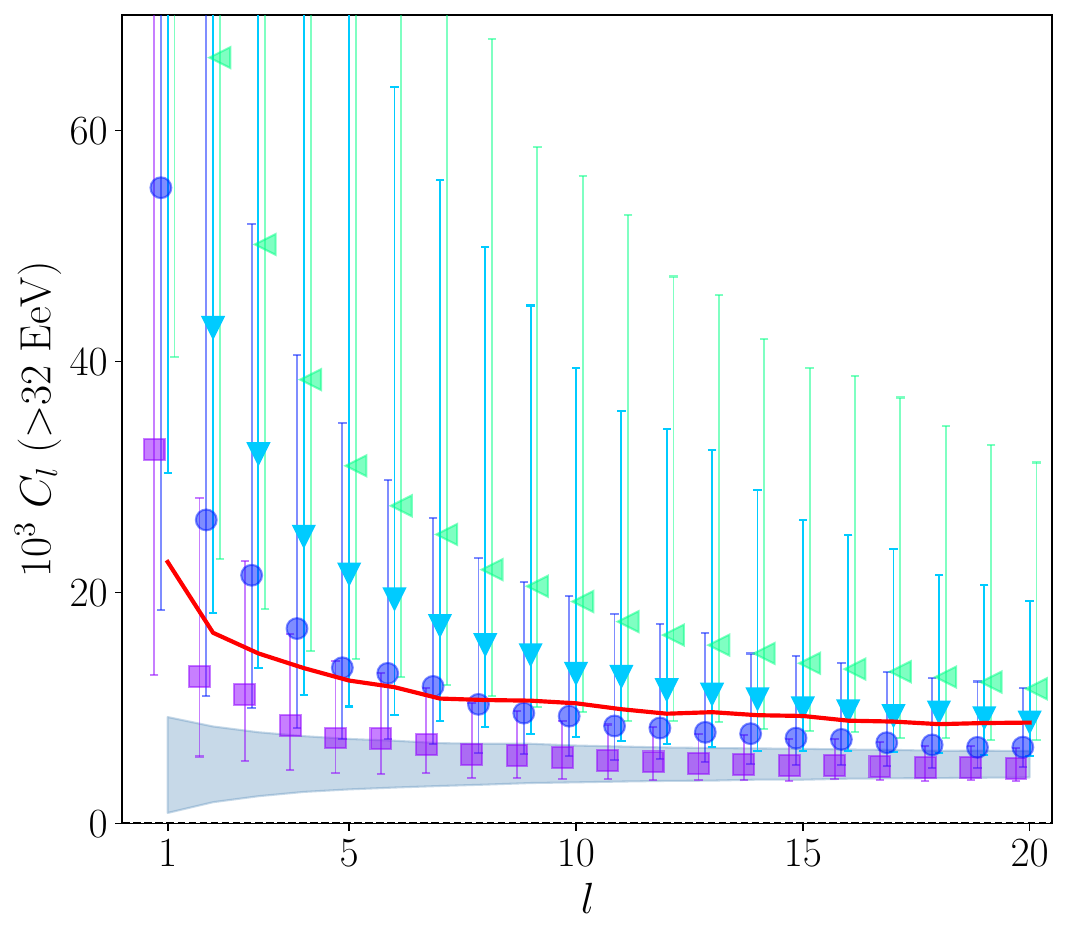}
\caption{Angular power spectrum in the baseline model, for different source number densities $n$, for $E>8\,\text{EeV}$ (\textit{top}) and $E>32\,\text{EeV}$ (\textit{bottom}). The markers show the median value for the given density and the error bars indicate the range encompassing $68\%$ of the variations from statistical fluctuations in source distribution and cosmic ray samples. The 99\% expectation from isotropic arrival directions is shown as a red line, with the mean $\pm$ standard deviation shown as a grey band, as in~\citet{Caccianiga_ICRC2023}. Note that for Auger and combined Auger+TA data, all $C_{l>1}$ are within isotropic expectations, i.e., below the red line, apart from the dipole. The dipole amplitudes for different densities, and for the data, are shown in Fig.~\ref{fig:dipole_amp}.}
\label{fig:power_spectrum}
\end{figure}

For reference, the power spectrum predicted in the baseline model is displayed in Fig.~\ref{fig:power_spectrum} for the two cumulative energy bins, $>$8 EeV and $>$32 EeV, and four source number densities. For good comparability, the same style is used to mark the expectations from isotropic simulations as used by the combined working group between Auger and TA~\citep{Caccianiga_ICRC2023}. One sees from Fig.~\ref{fig:power_spectrum} that the level of anisotropy in the arrival flux increases significantly with decreasing source number density. This is as expected: the smaller the number of sources, the greater the impact of the few local sources on the total flux.

The power spectrum for the JF12-reg model is displayed in Appendix~\ref{sec:JF12-reg}. There are sizeable quadrupole and octupole moments for all densities. That means that if the magnetic field is solely coherent without any turbulence from GMF or EGMF, it is not possible to obtain a sizeable dipole moment without significant higher multipole moments. In order to reproduce the data, some level of turbulence, from either GMF or EGMF, is hence absolutely necessary.

\subsection{Combined constraints on the EGMF\\ and source number density} ~\label{sec:egmf_dens}
As was demonstrated in the previous subsection, the source number density has a strong impact on the anisotropy level of UHECRs, specifically the power spectrum and dipole amplitude. These observables are also impacted by random fields along the UHECRs' trajectories as we saw in Sec.~\ref{sec:egmf}. Therefore, in this section, we perform a combined analysis considering a range of values for $\beta_\mathrm{EGMF} = B / \mathrm{nG}\,\sqrt{L_c / \mathrm{Mpc}}$ as well as the source number density. We again create 1000 realizations of each scenario based on the baseline and JF12-reg models and determine how many of them fulfill the two criteria outlined in Sec.~\ref{sec:dens}: sufficient dipole amplitude but higher multipoles not exceeding the 99\% isotropic C.L.. It is important to note that the dipole amplitude is affected less by turbulent magnetic fields than higher multipoles~\citep{deoliveira2023revisiting}, so that we do not expect a linear dependency between the allowed values of $\beta_\mathrm{EGMF}$ and $n$.

\begin{figure}[ht]
\includegraphics[width=0.49\textwidth]{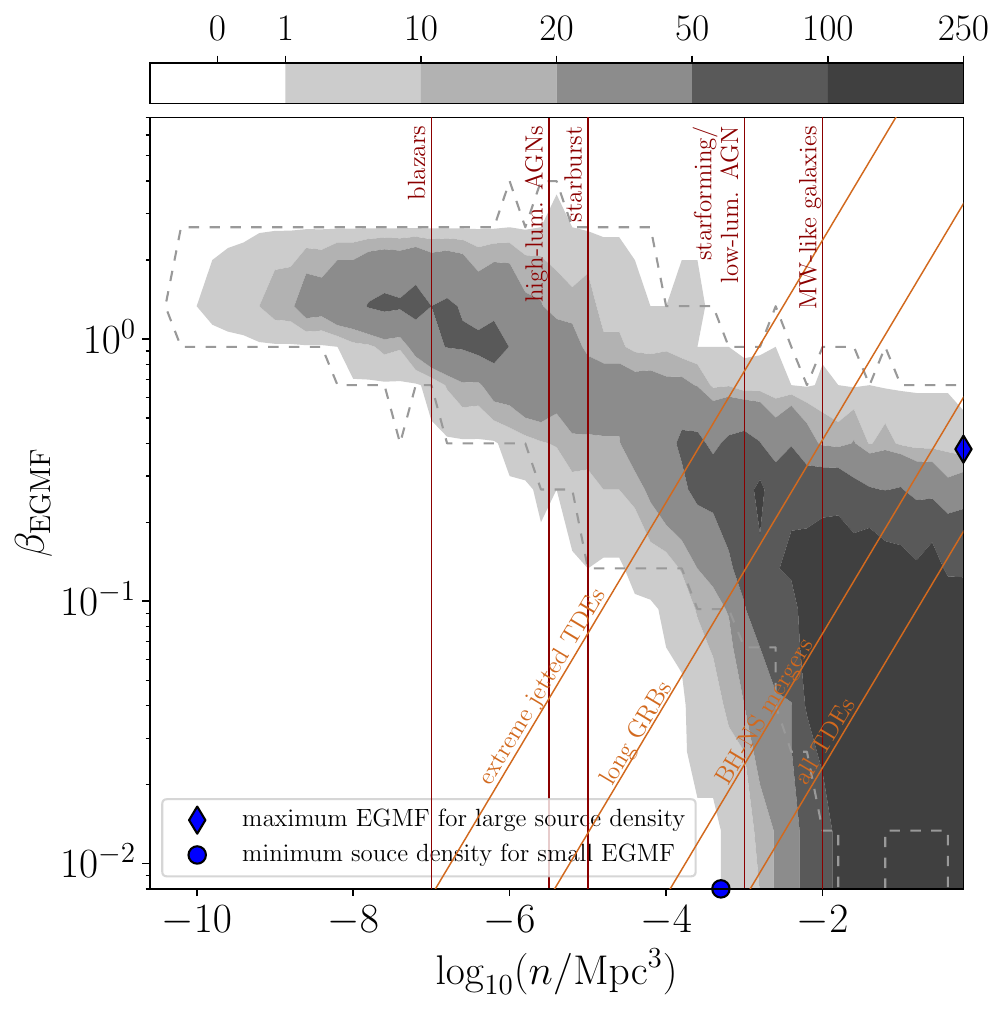}
\caption{Combined constraints on the the source number density $n$ and EGMF parameter $\beta_\mathrm{EGMF} \equiv B / \mathrm{nG}\,\sqrt{L_c / \mathrm{Mpc}}$; dark grey regions are favored. The intensity bar at the top shows the number of simulations out of 1000 total for the baseline model that have both a sufficiently large dipole and higher multipole moments small enough to be compatible with the 99\% isotropic expectations as found for the data. The white region is excluded at more than 99\% C.L.. The dashed region indicates the excluded region with no Galactic random field.
%The allowed region is correlated: for weak EGMF, the number density has to be large while for strong EGMF, the number density should be relatively small. 
Two extreme cases discussed in the text are depicted by blue markers. Characteristic estimates of the number densities of some steady source candidates are shown with red lines, and indicative locii of transient source candidates with orange lines; see Sec.~\ref{sec:AstImp} for discussion and references. %Note that the effect of beaming due to a limited source opening angle (relevant e.g. for blazars or GRBs) is not taken into account.
}
\label{fig:egmf_dens}
\end{figure}

In Fig.~\ref{fig:egmf_dens}, the number of simulations that fulfill both criteria combined is depicted graphically. The white area is excluded at more than 99\% C.L. as there are no simulations out of 1000 that fulfill both criteria at the same time. The lower left part of the phase space is excluded by too-large multipole moments, and the upper right part by a too-small dipole amplitude.
A clear correlation of the allowed area is visible: if the EGMF is small with $\beta_\mathrm{EGMF}\lesssim0.1$, the source density has to be large with $n\gtrsim 10^{-4}\, {\rm Mpc}^{-3}$. This part of the phase space also includes the case of no EGMF discussed in Sec.~\ref{sec:dens}, indicated with a circle marker in Fig.~\ref{fig:egmf_dens}. If, instead, the EGMF is sizeable -- $\beta_\mathrm{EGMF}\gtrsim0.3$ -- both large source densities $n\gtrsim10^{-3}\, {\rm Mpc}^{-3}$ and very small densities $n\lesssim10^{-9}\, {\rm Mpc}^{-3}$ are excluded at 99\% C.L., producing too small dipole amplitude or too large multipoles in the respective cases. An EGMF stronger than $\beta_\mathrm{EGMF}\geq3$ leads to too small dipole moments for any source number density and is thus excluded at 99\% C.L.. Note that this is a competitive upper limit on the EGMF~\citep{Durrer_2013}. It is in the same range as the 95\% C.L. upper limit of 1\,nG for 1\,Mpc coherence length derived in \citet{Pshirkov:2015tua}. 
As the EGMF $\beta_\mathrm{EGMF}$ and/or UHECR source density become better constrained by other means, the constraints from UHECR anisotropies embodied in Fig.~\ref{fig:egmf_dens} will further restrict the other. For an example of the impact of EGMF smearing for a particular source model, see~\citet{van_Vliet_2021}.

Another scenario that can be excluded based on the combination of sufficient dipole magnitude and absence of higher multipoles is a homogeneous source distribution in the case of negligible EGMF. For a sufficiently low source density, the required dipole magnitude can result from the local inhomogeneity of sources arising randomly in an otherwise homogeneous distribution~\citep{Harari_2015}. However, our analysis shows that the required small source density leads to too-large higher multipole moments, and there is no density for which both criteria can be fulfilled simultaneously. This is shown in Sec.~\ref{sec:homogeneous} of the Appendix.
%Thus, the dipole being a fluctuation of a homogeneous source distribution can be excluded at 99\% C.L. from the dipole amplitude and power spectrum. 
For non-negligible EGMF with $\beta_\mathrm{EGMF}\simeq1$, it is in principle possible to reproduce the dipole amplitude while keeping all higher multipoles within the 99\% isotropic C.L. (see Fig.~\ref{fig:egmf_dens_iso} in the Appendix) but much less likely for both criteria to be satisfied at the same time than in the LSS model. Additionally, the direction is obviously only very rarely coincident with the data for a random fluctuation of a homogeneous source distribution.

\subsection{Astrophysical Implications} \label{sec:AstImp}
The obtained limits on the source number density enable conclusions on the possible classes of UHECR sources. The densities of different classes of objects can be measured by astronomical observations. For reference, the number of Milky Way-like galaxies is $\mathcal{O}(10^{-2})$ per cubic Mpc (e.g.,~\citet{Conselice_2016}).
The number density of AGN is a strong function of their luminosity\footnote{The number density of AGN evolves strongly with the redshift; the values quoted here are at $z=0$, as appropriate to UHECRs in our energy range which cannot propagate over cosmological distances. See~\citet{zcf19} for a complete, homogeneous catalog of local AGN.}. Low-luminosity %and FR-I 
AGN 
are quite common, with 20-40\% of all galaxies in SDSS at low redshift being broad-line AGN or satisfying the Kauffman criterion for narrow-line AGN~\citep{zcf19}; see%$\mathcal{O}(10^{-2}-10^{-3})\, {\rm Mpc}^{-3}$
~\citep{Ho_2008, Danforth_2016} for earlier work. At high-luminosity, the density is $\mathcal{O}(10^{-5}-10^{-6})\, {\rm Mpc}^{-3}$~\citep{Gruppioni_2013, Best_2012}; blazars have been estimated to have a density as sparse as $\sim\mathcal{O}(10^{-7})\, {\rm Mpc}^{-3}$~\citep{Ajello_2013}.
Starforming galaxies also containing an AGN are very common with density $\mathcal{O}(10^{-3}-10^{-2})\, {\rm Mpc}^{-3}$~\citep{Gruppioni_2013}, however the highest luminosity starburst galaxies (also known as ULIRGs, for Ultra-Luminous Infrared Galaxies) are much rarer with density around $\mathcal{O}(10^{-5}-10^{-4})\, {\rm Mpc}^{-3}$~\citep{Gruppioni_2013, Murase_2019, alves_batista_open_2019}.

Thus if the EGMF smearing is insignificant, i.e., $\beta_\mathrm{EGMF} \lesssim 0.1$, our limit on the source density of $n \geq 10^{-3.5}\, {\rm Mpc}^{-3}$, rules out very sparse object classes such as high-luminosity AGNs and ULIRGs. In this case, UHECRs must come from more common source types such as starforming / normal galaxies or low-luminosity AGNs, or else transient sources (see below). If, however, the EGMF produces more smearing, so $\beta_\mathrm{EGMF} \gtrsim 0.3$, sources have to be sparse in order to produce a sizeable enough dipole. With EGMF strengths in the range of $\beta_\mathrm{EGMF}\simeq 0.5-1$, the sources could be e.g. starburst galaxies, high-luminosity AGNs, or blazars. This is also in agreement with the finding of \citet{Eichmann:2022ias} that radio galaxies could explain the observed dipole for $\beta_\mathrm{EGMF}\simeq1$. Our constraints imply that $\beta_\mathrm{EGMF} \gtrsim 3$ is highly disfavored, except in the context of a specially crafted scenario.

However, another possibility is that UHECRs are produced in transient events, for instance giant AGN flares, tidal disruption events (TDEs)~\citep{fg08}, collapsars (long gamma-ray bursts) (LGRBs)~\citep{Waxman_1996} or compact binary mergers (Neutron Star (NS) - Black Hole (BH) or binary NS). %Strong limits on astrophysical neutrino production by GRBs~\citep{ICgrblim22} seem to exclude GRBs as a primary source of UHECRs.

The estimated $z=0$ rate of long GRBs is $\Gamma_\mathrm{long\,GRBs}=1.3\,\mathrm{Gpc}^{-3}\,\mathrm{yr}^{-1}$~\citep{grbRatePiran10}. However, given the lack of neutrino-GRB correlations~\citep{ICgrblim22} and also the fact that the total power of long GRBs in UHECRs would have to exceed their gamma-ray luminosities by more than an order of magnitude~\citep{fg08}, at present long GRBs are disfavored as UHECR sources.

Meanwhile, TDEs have become an even more attractive option thanks to discoveries of coincidences between TDEs and high energy neutrinos~\citep{stein+TDENature20} and theoretical developments~\citep{fp14,piran+UHECR23}; AGN flares as a possible source class have received little follow-up.  And with recent evidence of jet formation in NS-BH and BNS mergers, these are also potential accelerators. %with NS-BH especially attractive on account of having less baryon loading.

Due to deflections in the EGMF during propagation, the UHECRs produced in a given transient event arrive to Earth over an extended time period we denote as $\tau_\mathrm{eff}$. Therefore the effective number density of sources contributing at any given epoch is
\begin{equation}
    n_{\rm eff} \approx \Gamma \, \tau_{\rm eff}~,
\end{equation}
where $\Gamma$ is the rate of the transient events which produce the UHECRs.  From~\citep{Achterberg:1999vr} the time-spreading is $ \mathcal{D}_0 D^2/(3 c)$, giving 
\begin{equation} \label{eq:tau_eff}
\tau_{\rm eff} = 0.14 \left( \frac{D}{\mathrm{Mpc}} \frac{\mathrm{EV}}{R} \beta_\mathrm{EGMF} \right)^2 \! \mathrm{Myr}.   
\end{equation}
When evaluated for rigidity $R = 4.5\,\mathrm{EV}$ and distance $D \simeq 70$ Mpc appropriate to the more constraining $E \geq 32$ EeV dataset (see Fig.~\ref{fig:propagation} and \ref{fig:rig_spec}),  $\tau_{\rm eff} = 34 \, \beta_\mathrm{EGMF}^2 \, \mathrm{Myr}$.  

The total rate of TDEs is estimated to be $\Gamma_\mathrm{all\,TDE}=10^{2-3}\,\mathrm{Gpc}^{-3}\,\mathrm{yr}^{-1}$ \citep{vVfTDErate14,teboul2023loss, Yao_2023}, but not all TDEs may be capable of producing UHECRs. The rate of jetted TDEs is difficult to determine robustly because the radio emission signaling the presence of a jet may only emerge many years after the event~\citep{piran+UHECR23}.  With this in mind the estimate of $\Gamma_\mathrm{jetted\,TDE}=0.03\,\mathrm{Gpc}^{-3}\,\mathrm{yr}^{-1}$~\citep{Andreoni_2022} can be considered a lower bound on the jetted TDE rate. The LIGO-Virgo-Kagra collaboration estimates the rate of NS-BH mergers as (7.8-140)$\,\mathrm{Gpc}^{-3}\,\mathrm{yr}^{-1}$, with the BNS rate possibly greater than an order of magnitude larger~\citep{LVK_binaryMergerRate23}.

In Fig.~\ref{fig:egmf_dens}, we show with orange lines the relation of $\beta_\mathrm{EGMF}$ and $n_\mathrm{eff}=\Gamma \tau_\mathrm{eff} = 34 \, \Gamma \beta_\mathrm{EGMF}^2$ following Eq.~\ref{eq:tau_eff} for the different transient candidates discussed above and using $
30\,\mathrm{Gpc}^{-3}\,\mathrm{yr}^{-1}$ as a representative choice for NS-BH mergers. It is visible that all of them cross the allowed region, so none can be excluded as sources from this study.  In general, transient source candidates with the lowest rates require stronger EGMF time-averaging in order to have a sufficient number density of sources contributing at any given epoch.

\section{Summary and Conclusions} \label{sec:conclusions}
In this work \textbf{we explored the implications of the hypothesis that UHECR sources follow the Large Scale Structure of the local universe}, improving upon the earlier study by~\citet{ding_imprint_2021} by implementing a self-consistent treatment of propagation effects and a simultaneous fit to the UHECR spectrum, composition, and dipole data. The best-fit parameters of the source injection are in agreement with previous works using a homogeneous source distribution~\citep{Auger_CF_2023, Auger_CFAD_2023}, with a hard spectral index and predominantly nitrogen injection. We verified that the results are almost completely independent of systematic uncertainties from the hadronic interaction model and variations of the source injection.  We investigated the sensitivity of various features of the data to uncertainties in the strength of the Galactic random field.
\\
$\bullet$
We demonstrated that \textbf{the \emph{dipole amplitude} and its energy evolution are well reproduced by the LSS model}. The measured amplitude is bracketed by our two tested GMF models, chosen to encompass the uncertainty in the random component of the GMF: the baseline JF12  model~\citep{jansson_galactic_2012, jansson_new_2012} including a turbulent component with 30~kpc coherence length, and ``JF12-reg" with only the coherent field part. %This is in agreement with the finding that the turbulent component of the original JF12 field is probably overestimated~\citep{fComptRend14,Planck_2016}.
\\
$\bullet$
We demonstrated that \textbf{the \emph{dipole direction} is independent of the turbulent part of the GMF}. It is, however, strongly influenced by the regular component of the GMF and not perfectly reproduced using JF12. \textbf{In the future, the dipole direction can help select among different coherent magnetic field models}~\citep{UF23},   %,Unger_ICRC_2023, 
largely independently of the turbulent field part.

\textbf{We investigated the possibility of a \emph{bias} between UHECR sources and the LSS}. Based on the ability to reproduce the observed dipole, we find that \textbf{the LSS matter distribution is a good, bias-free proxy for the UHECR distribution}. In particular, UHECR sources reside in both high- and intermediate-density regions; we came to no definite conclusion regarding their presence in low-density regions. 

\textbf{We used the amplitude of the dipole and upper limits on higher multipoles to constrain the extragalactic magnetic field and the number density of UHECR sources.}  The arrival direction smearing due to the EGMF depends on the combination of rms field strength, $B$, and its coherence length $L_c$, through the combination $\beta_\mathrm{EGMF} \equiv B \sqrt{L_c}$.
\\
$\bullet$
In the ``continuum limit" of many low-luminosity UHECR sources, we place the 99\% C.L. upper limit $\beta_\mathrm{EGMF} \lesssim 0.2 \,\, \mathrm{ nG \, Mpc^{1/2}}$.   If the extragalactic smearing were larger than this, the strength of the dipole amplitude could not be as large as observed, within the LSS framework. 
\\
$\bullet$
If the arrival direction smearing by the EGMF is negligible, the number density of UHECR sources cannot be too small:  $n>10^{-3.5}\, {\rm Mpc}^{-3}$ at 99\% C.L..  If the sources are more rare than this, the power spectrum would have significant quadrupole and octupole moments, contrary to observation. 
Compared to the previous best limit of $n\gtrsim \mathcal{O}(10^{-4})\, {\rm Mpc}^{-3}$~\citep{Auger_dens_2013}, our limit is not only more constraining but also more accurate in that it takes into account deflections in the Galactic magnetic field and consistently treats the mixed composition. The greatly enhanced event statistics accumulated in the decade since that work call for use of the more realistic LSS source distribution as done here.
\\
$\bullet$
We set combined constraints in the parameter space of the source density $n$ and the EGMF smearing parameter $\beta_\mathrm{EGMF} $, shown in Fig.~\ref{fig:egmf_dens}. A correlation between the two quantities is visible.  $\beta_\mathrm{EGMF}$ up to $ \approx 1 \sqrt{\mathrm{Mpc}} \ \mathrm{nG}$ is possible, but requires the sources to be rare with density $n$ in the range $\approx (10^{-9}-10^{-4})\, {\rm Mpc}^{-3}$.
\\
$\bullet$ We investigated the idea that the dipole can arise as a statistical fluctuation of a homogeneous source distribution~\citep{Harari_2015}. In general that scenario is in conflict with the absence of higher multipoles, but it can work for a sufficiently rare source population $n \lesssim 10^{-5}\, {\rm Mpc}^{-3}$ and relatively strong EGMF $\beta_\mathrm{EGMF} \simeq (0.1-2)$, albeit in a fairly small fraction of trials (see Fig.~\ref{fig:egmf_dens_iso}). The direction of the dipole is of course totally random and not predictable or interpretable in that scenario, unlike for the LSS model.
\\
$\bullet$
An EGMF smearing parameter $\beta_\mathrm{EGMF} \geq 3 \sqrt{\mathrm{Mpc}} \ \mathrm{nG}$ is excluded at 99\% C.L. for all source densities, for both LSS and homogeneous sources. This places a competitive upper limit on the EGMF~\citep{Durrer_2013} which will be investigated in more depth in the future.

\textbf{We placed limits on the effective source density, which restrict the possible classes of UHECR sources}. If the EGMF smearing is small, UHECR sources must be common -- for example low-luminosity AGNs, star forming galaxies, or normal galaxies.  Rare, powerful sources like high-luminosity AGNs, blazars, or luminous starburst galaxies can be excluded as the exclusive source of UHECRs unless there is sufficient EGMF smearing (but that is limited by the magnitude of the dipole -- see above).  Transient sources can also be constrained by our bounds, if their rate is accurately enough known, since their effective number density increases as the square of the spreading parameter $\beta_\mathrm{EGMF}$ times the rate.  %For these sparser candidates to be the only source of UHECRs, the EGMF has to be $B\,L_c \approx \mathcal{O}(0.1-3) \sqrt{\mathrm{Mpc}} \ \mathrm{nG}$.

Finally, \textbf{we investigated correlated anisotropies in flux, mean rigidity, and composition} (see Appendix).  The effects are too small to be statistically significant at present, but may be detectable in the not-too-distant future thanks to ongoing data taking and upgrades for better mass determination by Auger and TA. 

%\vspace{-0.1in}

% \begin{acknowledgments}
\section*{Acknowledgements}
We are grateful for fruitful discussions with many colleagues including  Eric Mayotte, Marco Muzio, Foteini Oikonomou, Michael Unger, and members of the Pierre Auger Collaboration. Also, we thank Chen Ding for sharing insights into his work, Noemie Globus for sharing the CosmicFlows 2 density distribution, and Esteban Roulet and Silvia Mollerach for comments on the manuscript. The work of T.B. is supported by a Radboud Excellence fellowship from Radboud University in Nijmegen, Netherlands and that of G.R.F. is supported by National Science Foundation Grant No.~PHY-2013199. T.B. thanks the Center for Cosmology and Particle Physics of New York University for its kind hospitality, including support from the Alfred P. Sloan Foundation facililtating this research and the production of this paper.  
% \end{acknowledgments}

\software{CRPropa3 \citep{alves_batista_crpropa_2016},
          healpy \citep{healpy},  
          numpy \citep{numpy}, 
          scipy \citep{scipy},
          astrotools (\href{https://git.rwth-aachen.de/astro/astrotools}{Bister et al.})
          }

\newpage
\appendix

\section{Influence of model uncertainties on the results}  \label{sec:model_uncertainties}
In this section, the influence of the most important model uncertainties on the fit results is investigated.

\subsection{Fitted model parameters}
The fitted parameters for the different tested models are given in Table~\ref{tab:results}. The predicted dipole directions (Fig.~\ref{fig:dipole_direc}) and amplitudes (Fig.~\ref{fig:dipole_amp}) for the baseline and JF12-reg models are given in Table~\ref{tab:dipole_lb}.

\begin{table*}[ht]
% \scriptsize
\footnotesize
\centering
\begin{tabular}{l  |l  |l  |l  |l  |l  |l  |l  |l }
 % model & \textbf{baseline} & \textbf{Sibyll} & \textbf{sech1} & \textbf{sech2} & \textbf{sech3} & \textbf{isotropic} & \textbf{no} $\boldsymbol{\xmax}$ \textbf{shift} & \textbf{no} $\boldsymbol{\ln} \boldsymbol{\mathcal{L}_d}$ & \textbf{only} $\boldsymbol{\ln} \boldsymbol{\mathcal{L}_d}$ \\
& \textbf{baseline} & \textbf{JF12 reg} & \textbf{homogeneous} & \textbf{no} \boldmath $\ln \mathcal{L}_d$ \unboldmath & \textbf{Sibyll} & \textbf{sec1} & \textbf{sec2} & \textbf{sec3} \\
\hline
sources & LSS & LSS & homogenous & LSS & LSS & LSS  & LSS & LSS \\
HIM & EPOS-LHC & EPOS-LHC & EPOS-LHC & EPOS-LHC & Sibyll2.3d & EPOS-LHC & EPOS-LHC & EPOS-LHC \\
cutoff $f_\mathrm{cut}$ & b.e. & b.e. & b.e. & b.e. & b.e. & sec($x$) & sec($x^2$) & sec($x^3$) \\
GMF & reg+rand & reg only & reg+rand & reg+rand & reg+rand & reg+rand & reg+rand & reg+rand \\ \hline
$\gamma$ & $-1.17$ & $-1.23$ & $-1.34$ & $-1.12$ & $-0.73$ & $-1.10$ & \phantom{+}$0.74$ & \phantom{+}$1.08$ \\
$\log_{10}R_\mathrm{cut}$ & \phantom{+}18.2 & \phantom{+}18.2 & \phantom{+}18.2 & \phantom{+}18.2 & \phantom{+}18.2 & \phantom{+}18.2 & \phantom{+}18.8 & \phantom{+}18.9 \\
$I_\mathrm{H}$ & \phantom{+}0.01 & \phantom{+}0.02 & \phantom{+}0.02 & \phantom{+}0.01 & \phantom{+}0.00 & \phantom{+}0.00 & \phantom{+}0.06 & \phantom{+}0.07 \\
$I_\mathrm{He}$ & \phantom{+}0.27 & \phantom{+}0.27 & \phantom{+}0.24 & \phantom{+}0.27 & \phantom{+}0.27 & \phantom{+}0.28 & \phantom{+}0.11 & \phantom{+}0.00 \\
$I_\mathrm{N}$ & \phantom{+}0.57 & \phantom{+}0.56 & \phantom{+}0.58 & \phantom{+}0.57 & \phantom{+}0.55 & \phantom{+}0.57 & \phantom{+}0.69 & \phantom{+}0.80 \\
$I_\mathrm{Si}$ & \phantom{+}0.12 & \phantom{+}0.12 & \phantom{+}0.13 & \phantom{+}0.12 & \phantom{+}0.14 & \phantom{+}0.11 & \phantom{+}0.12 & \phantom{+}0.07 \\
$I_\mathrm{Fe}$ & \phantom{+}0.01 & \phantom{+}0.04 & \phantom{+}0.04 & \phantom{+}0.03 & \phantom{+}0.04 & \phantom{+}0.03 & \phantom{+}0.03 & \phantom{+}0.06 \\
$\nu_{\xmax} / \sigma$ & $-0.88$ & $-0.93$ & $-0.95$ & $-0.91$ & $-0.84$ & $-0.84$ & $-1.26$ & $-1.50$ \\ \hline
$\ln \mathcal{L}_E$ & $-92.3$ & $-91.1$ & $-92.4$ & $-92.3$ & $-91.4$ & $-92.2$ & $-94.2$ & $-99.5$ \\
$\ln \mathcal{L}_{\xmax}$ & $-228.7$ & $-229.2$ & $-229.3$ & $-228.7$ & $-229.1$ & $-229.0$ & $-230.4$ & $-230.0$ \\
$\ln \mathcal{L}_d$ & \phantom{+}12.1 & \phantom{+}$11.8$ & $-8.5$ & \phantom{+}\textit{12.0}* & \phantom{+}11.2 & \phantom{+}12.1 & \phantom{+}10.6 & \phantom{+}$9.5$ \\
$\ln \mathcal{L}_\mathrm{syst}$ & $-0.4$ & $-0.4$ & $-0.5$ & $-0.4$ & $-0.4$ & $-0.4$ & $-0.8$ & $-1.1$ \\
$\ln \mathcal{L}_\mathrm{sum}$ & $-309.3$ & $ -308.9$ & $-330.6$ & $-321.4$ & $-309.6$ & $-309.4$ & $-314.9$ & $-321.1$
\end{tabular}
\caption{Best-fit source parameters and likelihood values for the multiple tested models. The differences between models are indicated in the top rows (see also text), with HIM relating to the hadronic interaction model and \textit{b.e.} to the broken exponential cutoff. The baseline model results and interpretation are presented in Sec.~\ref{sec:fit_results}. Some of the other tested models are discussed in Appendix~\ref{sec:model_uncertainties}. (* not included in $\ln \mathcal{L}_\mathrm{sum}$)}
\label{tab:results}
\end{table*}

\begin{table*}[]
\movetableright=-0.4in
\footnotesize
\begin{tabular}{l|l|l|l|l|l|l |l|l|l}
&  \multicolumn{3}{c|}{illumination} & \multicolumn{3}{c|}{at Earth (baseline)} & \multicolumn{3}{c}{at Earth (Jf12 reg)}\\
% \hline
 & $l/^\circ$ & $b/^\circ$ & $d$ & $l/^\circ$ & $b/^\circ$ & $d$ & $l/^\circ$ & $b/^\circ$ & $d$ \\
 \hline
$E>8$~EeV & $-84.2$ & \phantom{+}$41.2$ & \phantom{+}$0.081$ & $-85.7$ & $-7.4$ & \phantom{+}$0.048$ & $-85.1$ & $-7.9$ & \phantom{+}$0.086$ \\
$E=(8-16)$~EeV & $-82.8$ & \phantom{+}$40.8$ & \phantom{+}$0.062$ & $-87.0$ & $-15.7$ & \phantom{+}$0.035$ & $-88.2$ & $-11.4$ & \phantom{+}$0.068$ \\ 
$E=(16-32)$ EeV & $-84.7$ & \phantom{+}$42.5$ & \phantom{+}$0.12$ & $-82.5$ & $-1.6$ & \phantom{+}$0.074$ & $-80.0$ & $-6.2$ & \phantom{+}$0.13$ \\ 
$E>32$ EeV & $-89.3$ & \phantom{+}$43.7$ & \phantom{+}$0.22$ & $-88.2$ & \phantom{+}$10.6$ & \phantom{+}$0.15$ & $-89.1$ & \phantom{+}$43.6$ & \phantom{+}$0.20$\\
    \end{tabular}
    \caption{Predicted Galactic longitudes $l/^\circ$, latitudes $b/^\circ$, and amplitudes $d$ of the dipole for the baseline illumination, and the baseline and JF12-reg models at Earth. Compare to Fig.~\ref{fig:dipole_direc} and Fig.~\ref{fig:dipole_amp}.}
    \label{tab:dipole_lb}
\end{table*}

\subsection{Uncertainty from the air shower model} \label{sec:EAS}
% For testing the influence of the systematic uncertainty from the air shower measurement, we include a shift of the \xmax scale as described above. In the best-fit case, that shift is determined to $\nu_{\xmax}=-0.9\sigma$, making the composition on Earth heavier. Similar results were also found in~\citet{Auger_CFAD_2023}. To determine the influence of the shift, we compare the baseline model to a model without shift (third column of Table~\ref{tab:results}). It is visible that the shift improves the total likelihood by around $\Delta \ln \mathcal{L}_\mathrm{tot} \approx 10$, while softening the injected spectrum. The composition at the sources is relatively independent of the shift, as well as the dipole amplitude and directions. This is also visible by the almost constant value of $\ln \mathcal{L}_d$.

\begin{figure*}[ht]
\includegraphics[width=0.33\textwidth]{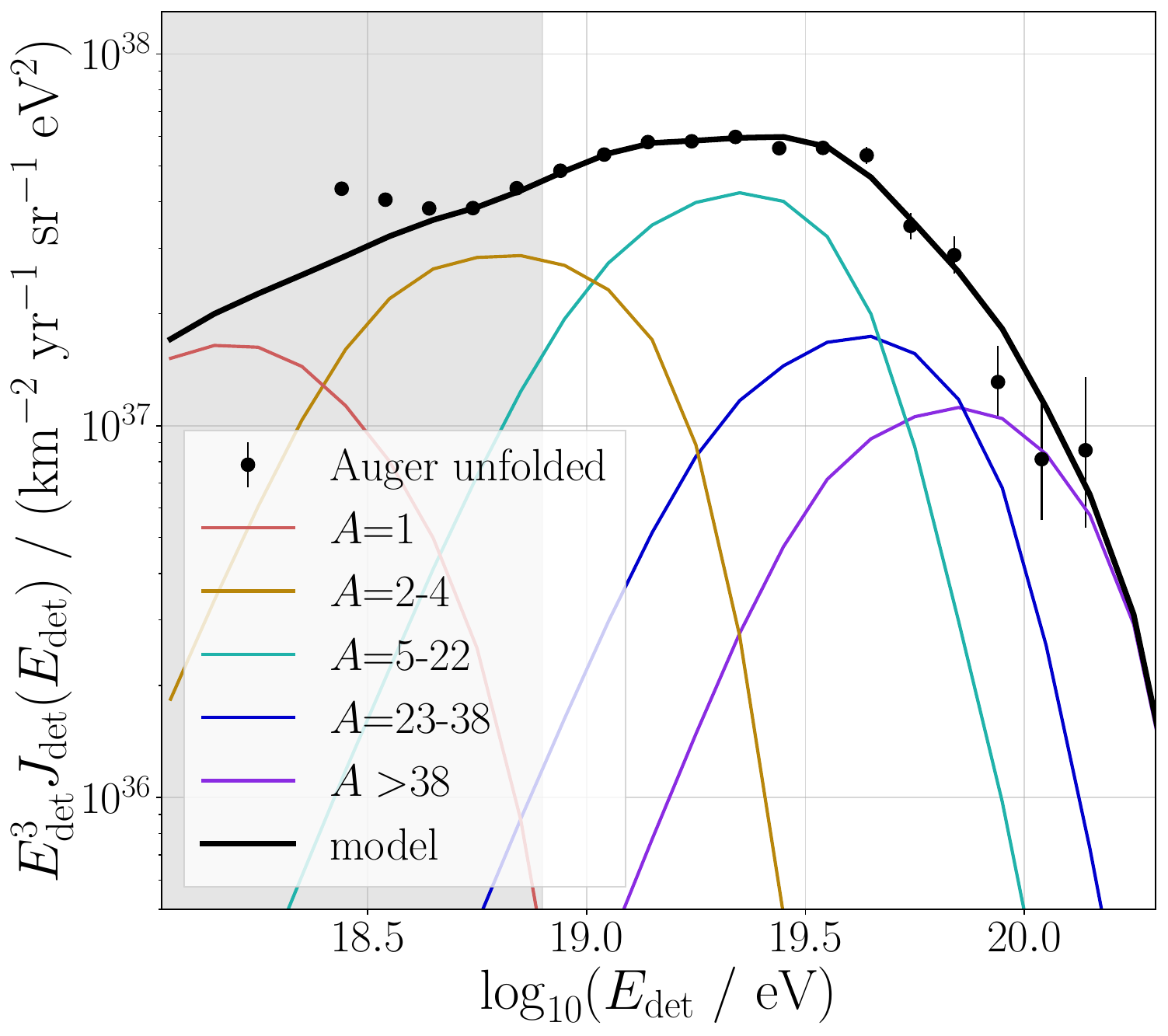}
\includegraphics[width=0.35\textwidth]{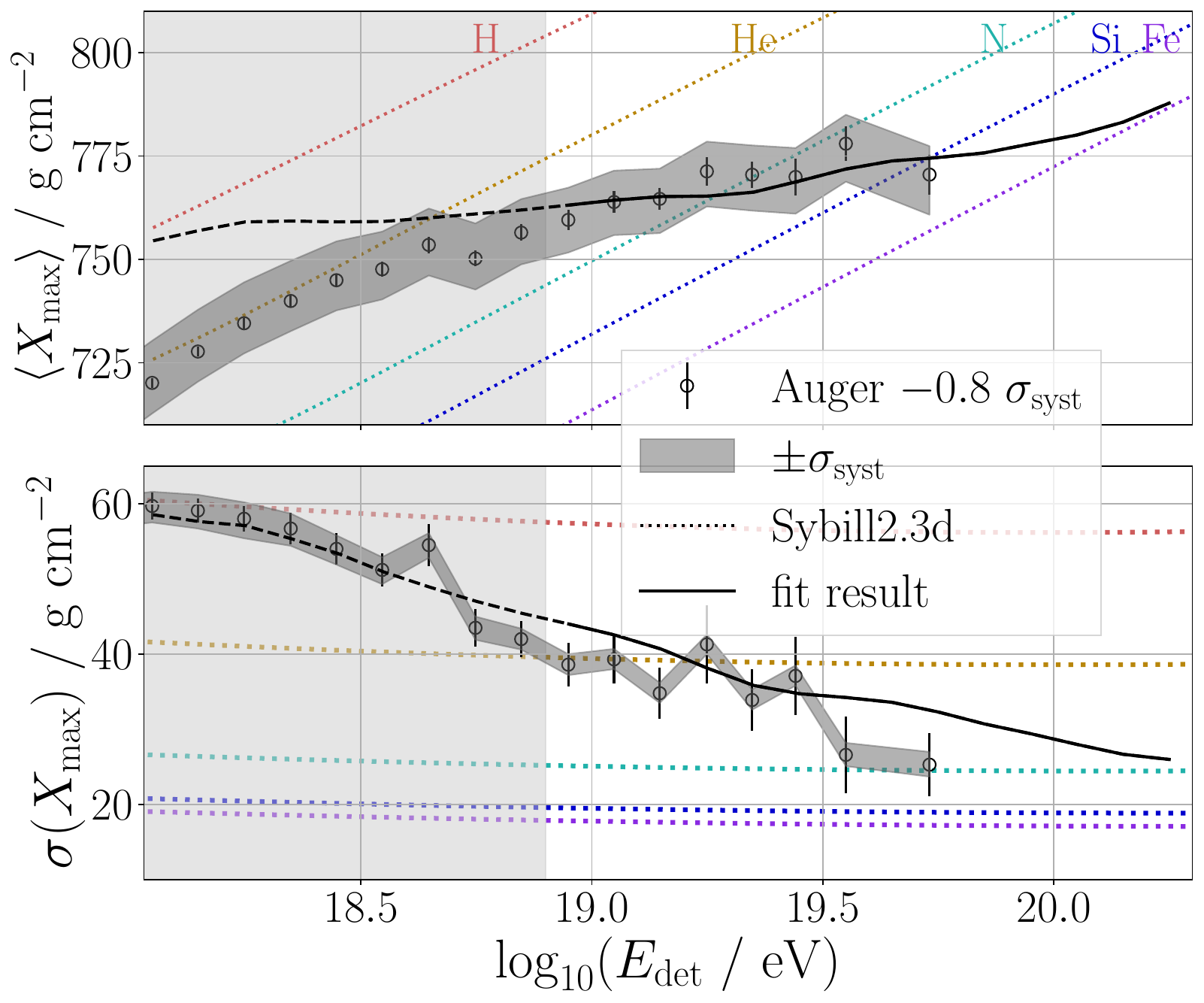}
\includegraphics[width=0.31\textwidth]{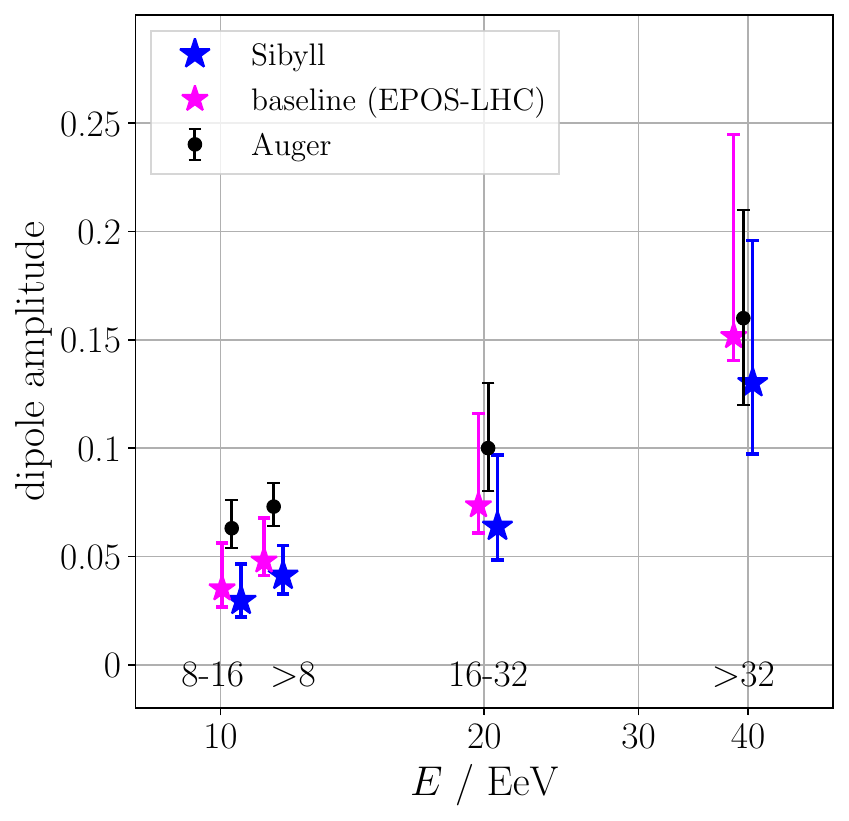}
\caption{Best-fit energy spectrum (left), first two moments of the \xmax distributions (middle), and dipole amplitudes (right) for the \textbf{Sibyll} model (see best-fit parameters in Table~\ref{tab:results}), using Sibyll 2.3d as hadronic interaction model.}
\label{fig:best_fit_EXmax_syb}
\end{figure*}

To test the influence of the hadronic interaction model on the result, we compare a model using Sibyll2.3d~\citep{Sybill} to the baseline one using EPOS-LHC. The best-fit parameter values for that case are given in Table~\ref{tab:results}. The fit parameters as well as the modeled observables (see Fig.~\ref{fig:best_fit_EXmax_syb}), are almost independent of the hadronic interaction model. Only the spectral index at injection is softer in the case of Sibyll, as was also found in~\citet{Auger_CF_2023}. The likelihood value of the Sibyll model is almost the same as the baseline one. This is not as found in~\citet{Auger_CF_2023}, where the fit using Sibyll is significantly worse. Note, however, that compared to~\citep{Auger_CF_2023} we use CRPropa instead of Simprop, and also a higher energy threshold of 8\,EeV.

As the dipole amplitude and direction are almost not influenced by the hadronic interaction model, the conclusions in Sec.~\ref{sec:constraints}, Sec.~\ref{sec:dens}, and Appendix~\ref{sec:predictions} can be made independently of these uncertainties.

\subsection{Influence of the cutoff function shape at injection}
To investigate the influence of the choice of cutoff function on the fit result, we vary the functional form of $f_\mathrm{cut}$ as described in Sec.~\ref{sec:injection}. The resulting fit parameters are given in Table~\ref{tab:results}.
This is important because of the small baseline best-fit maximum source rigidity of $R_\mathrm{cut}=10^{18.2}$\,V - hence the cutoff of the dominant nitrogen contribution already happens at $E_\mathrm{cut}^N= 7 \times 10^{18.2} \, \text{eV} \approx 10 \, \text{EeV}$. This means that the source injection in the energy range $E>8$\,EeV studied in this work is dominated by the cutoff region instead of the power-law with spectral index $\gamma$.

For the soft cutoff with $f_\mathrm{cut}(x)=\text{sec}(x)$, almost no difference can be seen in the fit parameters and likelihood values compared to the broken exponential cutoff. 
When the steeper cutoff functions with $\text{sec}(x^2)$ and $\text{sec}(x^3)$ are used, the injected spectrum becomes significantly softer, leading to positive values of $\gamma$ more in agreement with expectations of shock acceleration. The same is also observed in~\citet{Gonzalez_ICRC2023}. The softening of the injected spectrum happens because the pronounced features of the measured spectrum at Earth are now modeled by the steep cutoff instead of a hard injection spectrum. As the spectral index is correlated with the maximum rigidity, this also leads to higher values of $R_\mathrm{cut}$. The integral element contributions are relatively independent of the cutoff shape.

The value of the log-likelihood decreases for the steeper cutoffs, indicating a worse description of the data. This is mainly due to the energy spectrum, which cannot be described as well with the softer injected spectra. Additionally, the dipole likelihood $\mathcal{L}_d$ decreases with the steepening of the injection cutoff. The reason for this is the larger mixing of elements induced by the softer spectral index. This leads to stronger contributions of heavier elements at smaller energies as in the baseline case, which diffuse stronger. This induces a further decrease of the amplitude of the modeled dipole in the lower energy bins compared to the baseline model.
Based on these results, we only use the broken exponential injection for the main part of this work.

\section{Source number density influence on dipole amplitude and direction} \label{sec:density_dipole}
\begin{figure*}[ht!]
\includegraphics[width=0.48\textwidth]{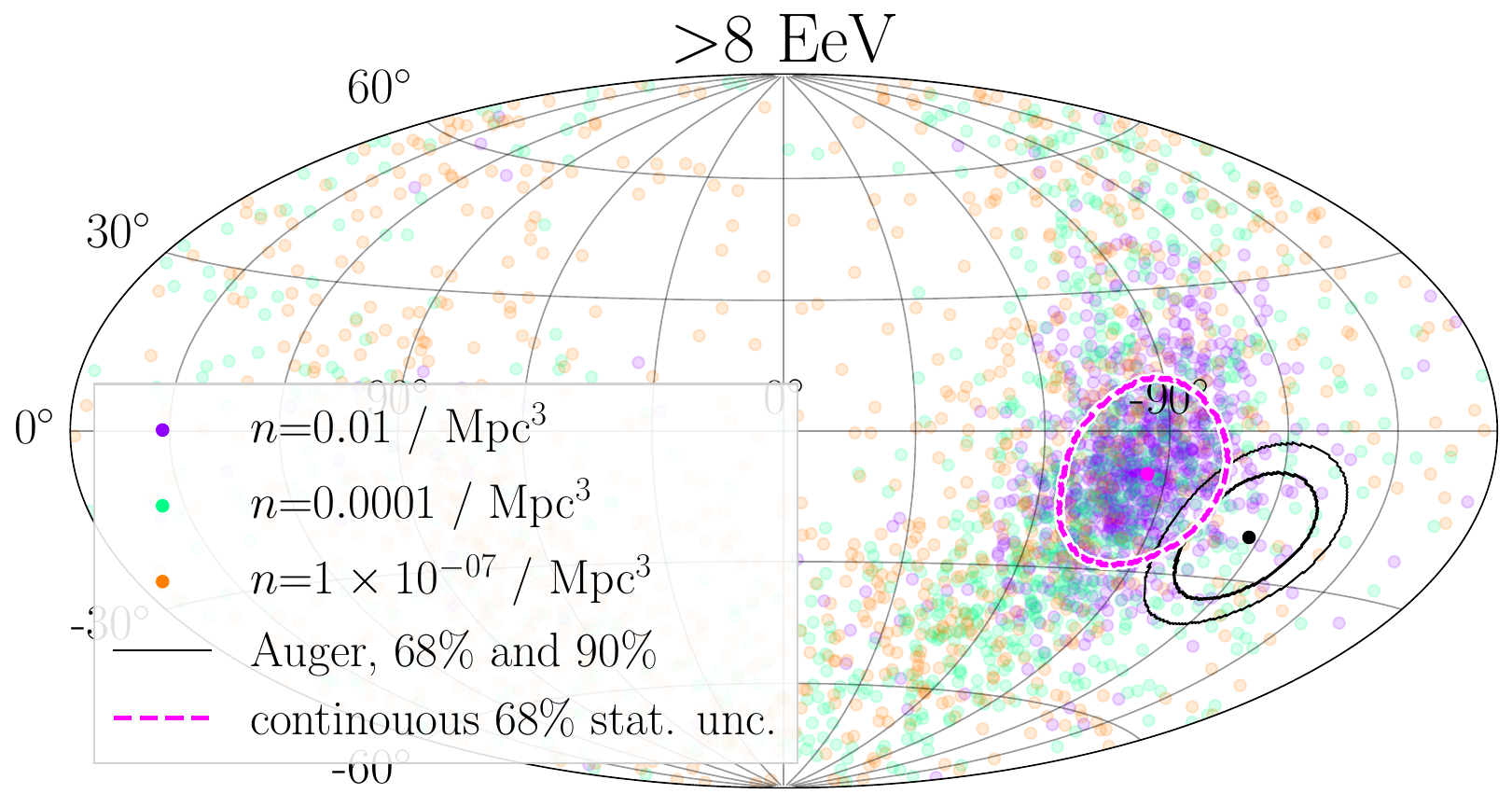}
\includegraphics[width=0.48\textwidth]{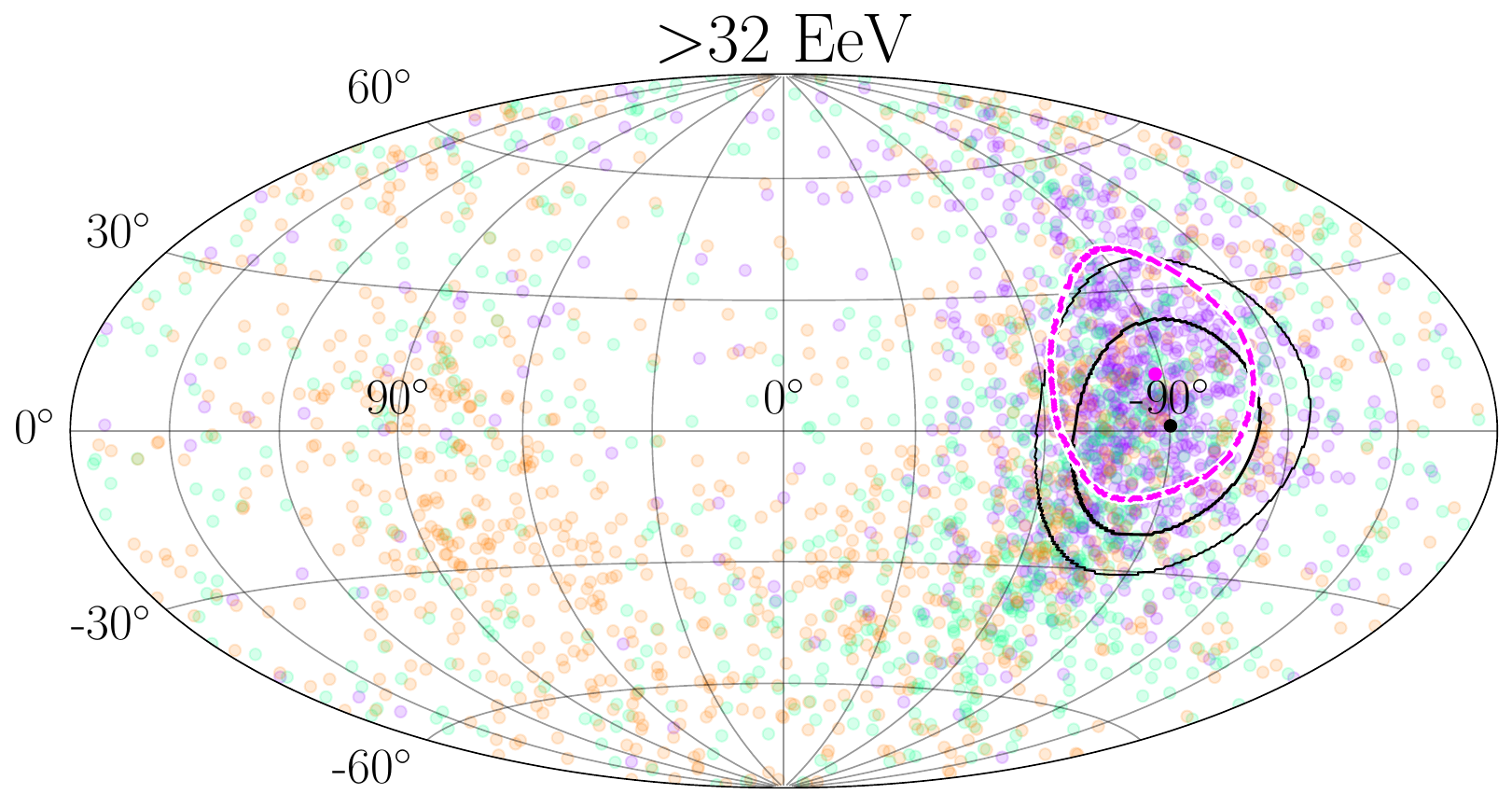}
\includegraphics[width=0.48\textwidth]{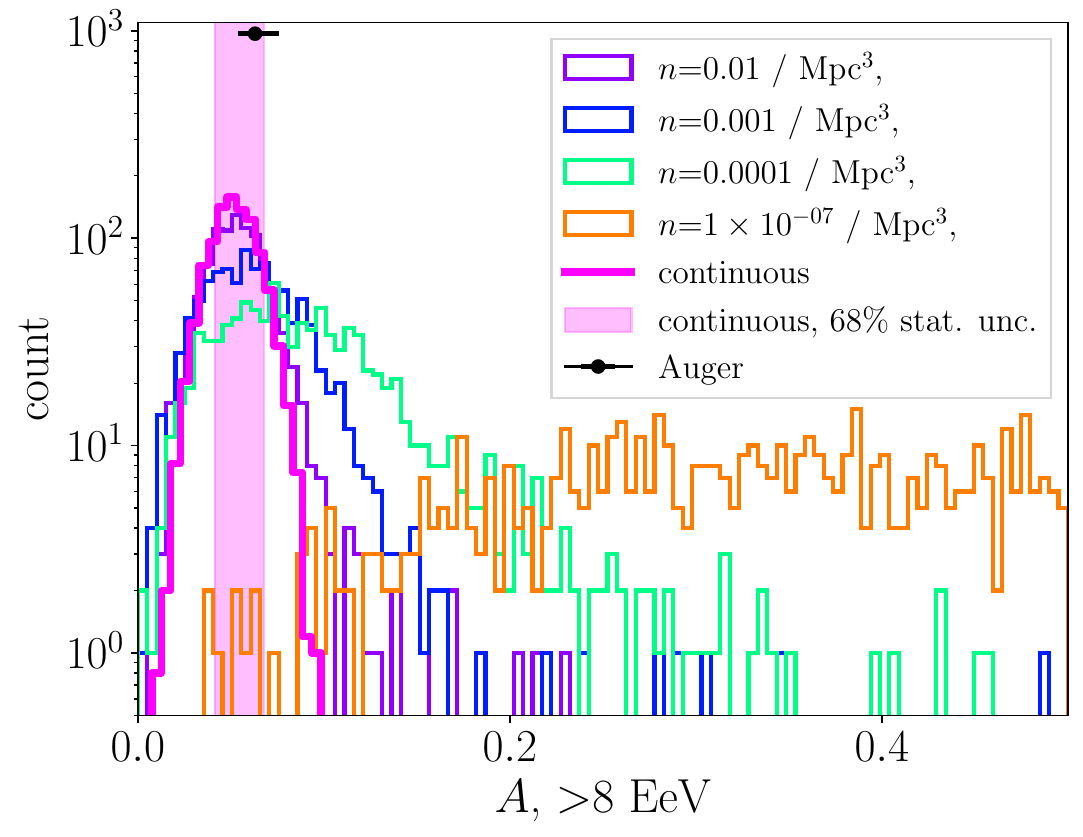}
\includegraphics[width=0.48\textwidth]{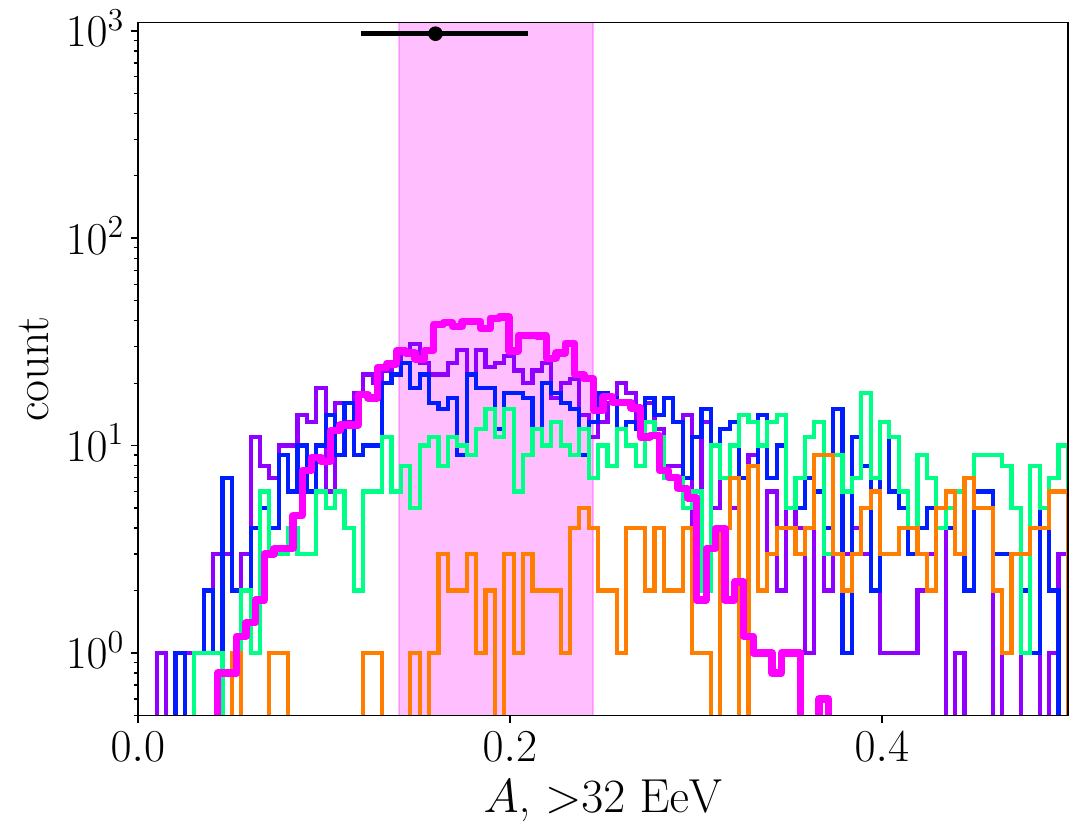}
\caption{Spread of dipole direction (\textit{upper row}) and dipole amplitude $A$ (\textit{lower row}) for various source number densities $n$, for two different energy thresholds. The pink contours in the skymaps and filled region in the histogram indicate the statistical variations due to the limited number of events for the continuous (baseline) model, as in Fig.~\ref{fig:dipole_direc}.}
\label{fig:dens_A_skymaps}
\end{figure*}

Here we investigate how well the observed dipole amplitude and direction are reproduced for different source number densities for the baseline GMF model, without EGMF, using the same setup as in Sec.~\ref{sec:dens}.  %Note that these results can depend on the Galactic magnetic field model (see Sec.~\ref{sec:gmf}) and the extragalactic magnetic field (see Sec.~\ref{sec:egmf}).
The variation of the dipole direction and amplitude are shown in Fig.~\ref{fig:dens_A_skymaps}. For small densities, $n\lesssim 10^{-4}\, {\rm Mpc}^{-3}$, only a few of the sparse sources make significant contributions to the total flux at Earth, so that the variations between realizations are sizeable. This leads to a big spread of dipole directions and generally too-large dipole amplitudes.
For large densities, $n\gtrsim 10^{-3} \text{Mpc}^{-3}$, the variations due to the number of events dominate over the variations in the source distribution, so that the dipole amplitude and directions mostly agree with the continuous case.

Because even in the continuous case, the agreement between the predicted and measured dipole direction is not perfect (especially at lower energies, see Fig.~\ref{fig:dipole_direc}), it is currently not useful to characterize the source density by measuring the number of realizations that reproduce the observed dipole direction.  
So, we choose to outline the intended method that could be used in future works with updated GMF, where the continuous model gives an accurate description of the dipole at all energies: instead of characterizing the fraction inside the \textit{measured} 68\% region, here, the fraction inside the \textit{modeled} statistical uncertainty will be used.

\begin{figure*}[ht!]
\includegraphics[width=0.33\textwidth]{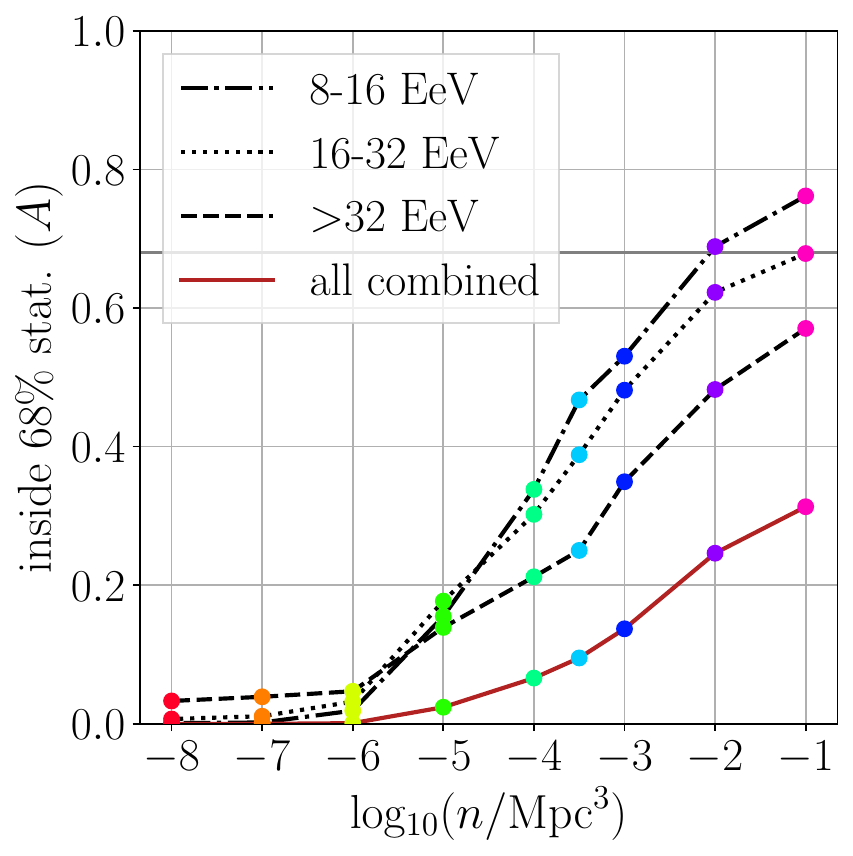}
\includegraphics[width=0.33\textwidth]{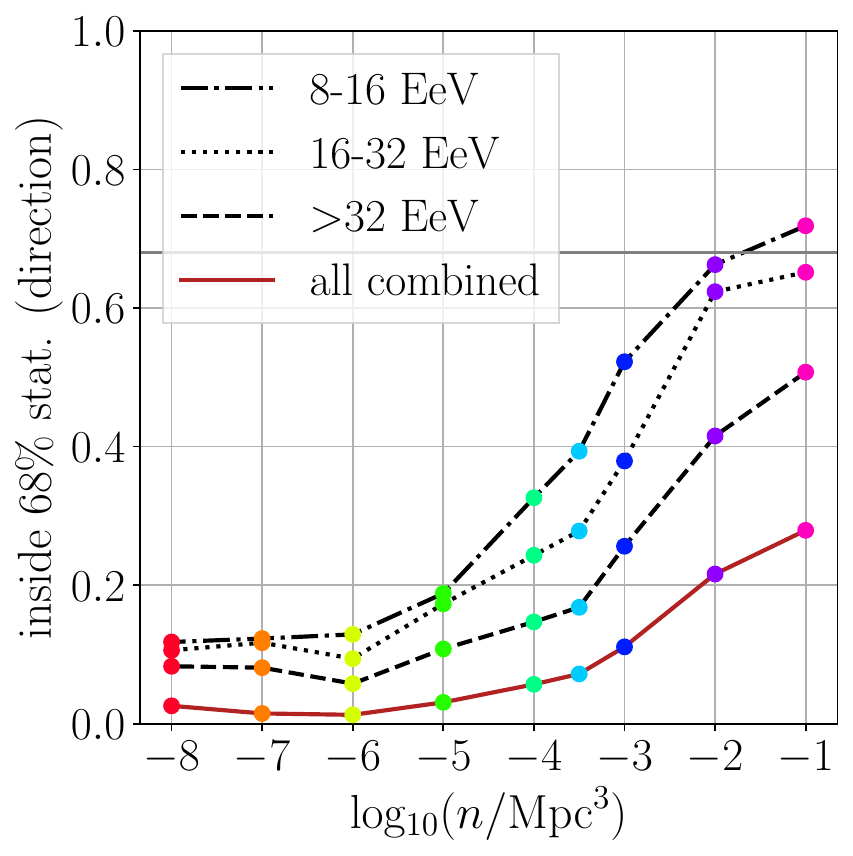}
\includegraphics[width=0.33\textwidth]{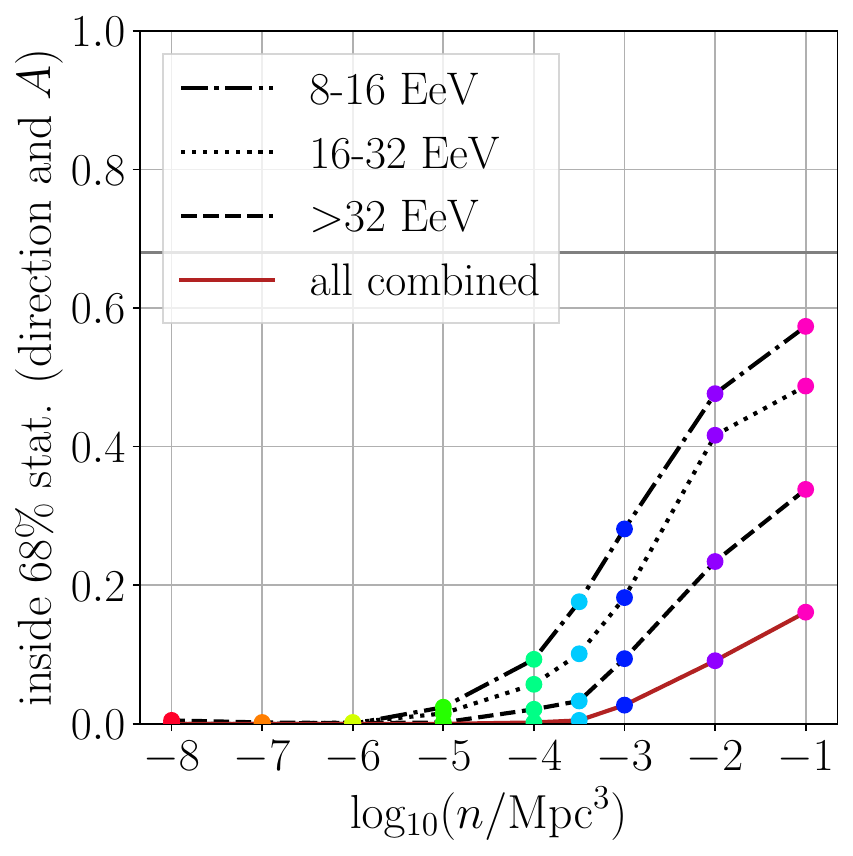}
\caption{Fraction of realizations where dipole amplitude (\textit{left}) / direction (\textit{middle}) / both at the same time (\textit{right}) are within the 68\% uncertainty of the continuous model depending on the source number density $n$. Three separate energy bins are shown, as well as a red line indicating the combination of all three at the same time. The black line at 68\% indicates the expected value for infinite density for each separate energy bin.  The colors of the points correspond to the different densities for better visual comparability to Fig.~\ref{fig:power_spectrum} and Fig.~\ref{fig:dens_A_skymaps}.}
\label{fig:inside_fraction}
\end{figure*}

In Fig.~\ref{fig:inside_fraction}, the fraction of realizations where the dipole amplitude and/or direction is within the $68\%$ statistical uncertainty of the continuous model (from the measured number of events, same as Fig.~\ref{fig:dipole_direc} and~\ref{fig:dipole_amp}) is depicted. 
It is evident that for small densities $n\lesssim 10^{-4}\, {\rm Mpc}^{-3}$, the amplitudes become too large, and the dipole directions too random as was also visible in Fig.~\ref{fig:dens_A_skymaps}. For $n=10^{-4}\, {\rm Mpc}^{-3}$, only 1 out of the 1000 realizations reproduces both measures at the same time for all energy bins. Even smaller densities are excluded completely, mostly due to the too-strong dipole amplitudes.

\section{Power spectra without turbulent GMF component}  \label{sec:JF12-reg}
\begin{figure*}[ht]
\includegraphics[width=0.45\textwidth]{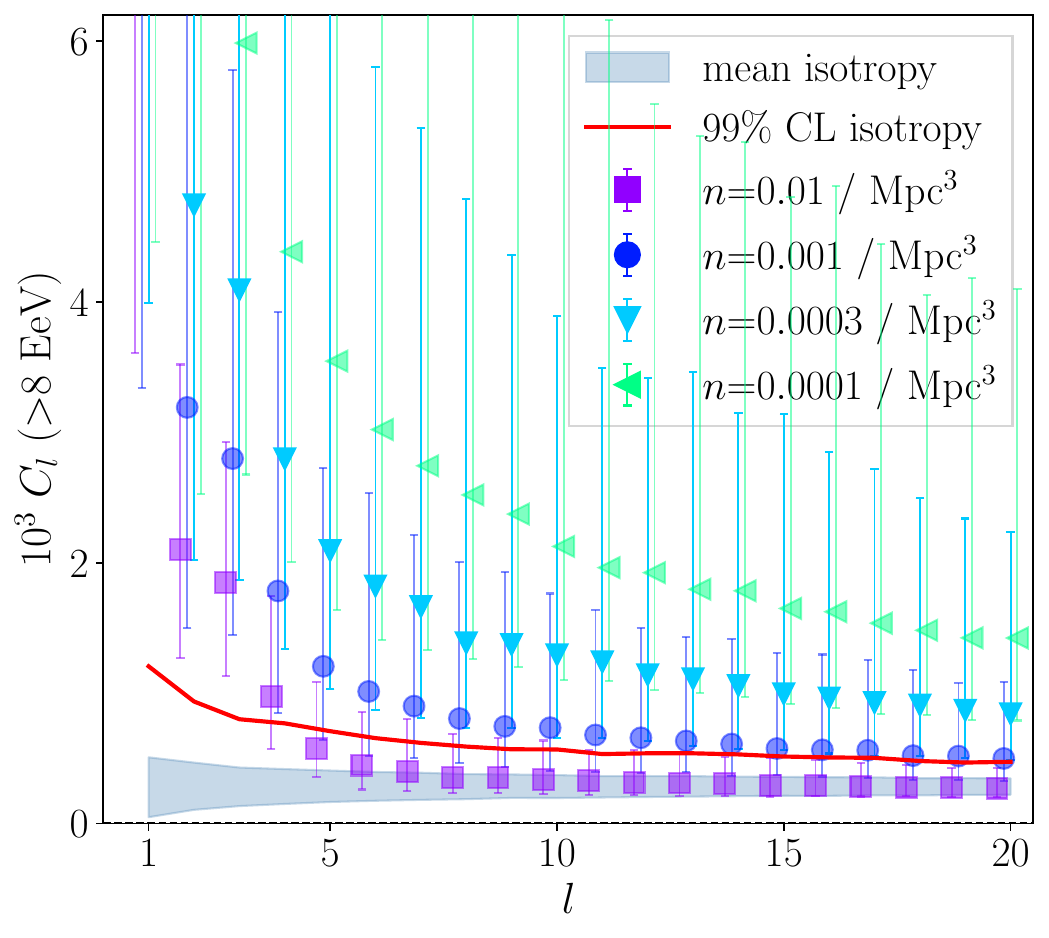}
\includegraphics[width=0.46\textwidth]{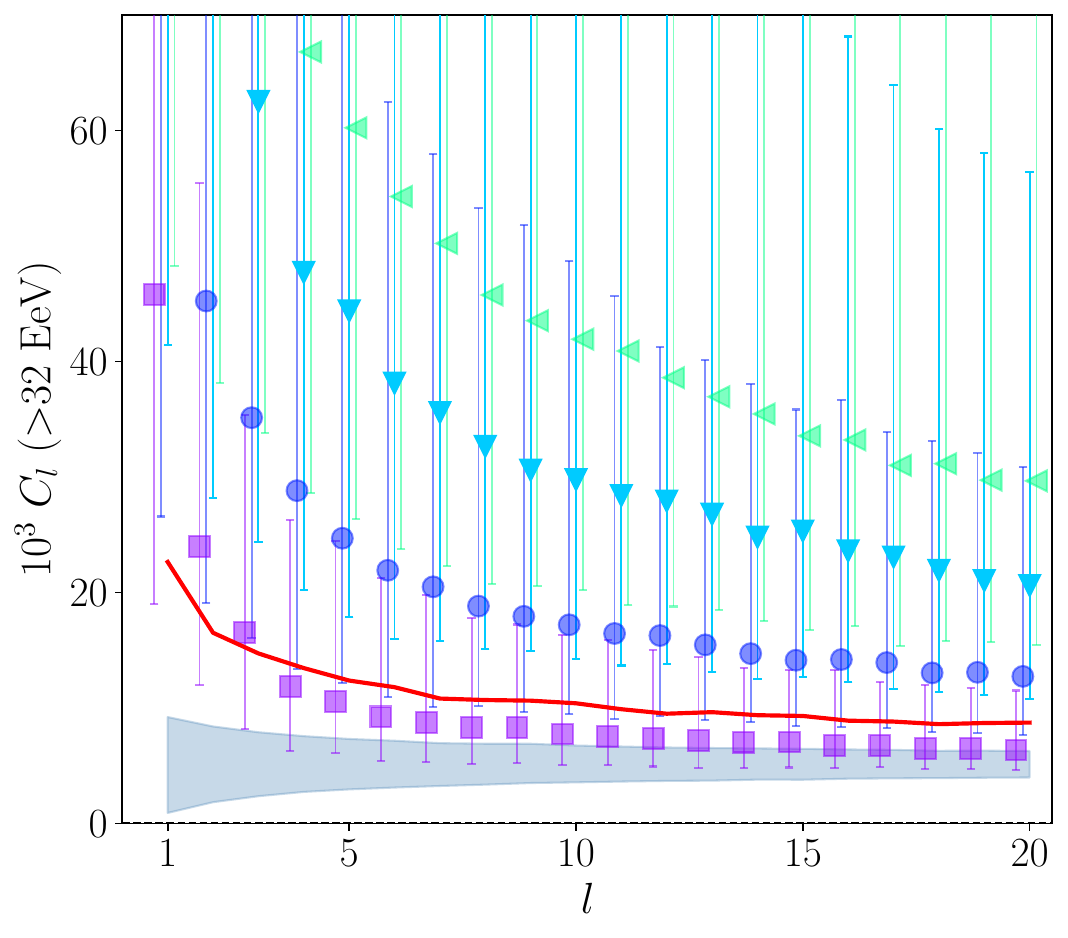}
\caption{Power spectrum for different number densities $n$; same as Fig.~\ref{fig:power_spectrum}, but for the JF12-reg model without turbulent GMF component.}
\label{fig:power_spectrum_jf12reg}
\end{figure*}

In Fig.~\ref{fig:power_spectrum_jf12reg}, the power spectrum for the JF12-reg model without turbulent GMF component is depicted for different sampled number densities $n$. It is visible how, compared to the case with turbulent GMF part (Fig.~\ref{fig:power_spectrum}), the quadrupole and octupole moments are much more pronounced. The pronounced features are also visible in the predicted arrival directions for the continuous JF12-reg model (see Fig.~\ref{fig:dipole_direc}, fourth row).
For no value of $n$ is a simulation found that stays within the 99\% C.L. isotropic expectation for any density.
Thus, we can exclude the case of no turbulence (neither from GMF nor EGMF) at the 99\% C.L..

\section{Alternative to LSS? A homogeneous source distribution} \label{sec:homogeneous}
In~\citet{Auger_dipole_2018}, based on the results of~\citet{Harari_2015}, it was suggested that homogeneously distributed sources (not following the LSS) with a density around $n\simeq10^{-4}\, {\rm Mpc}^{-3}$, could also explain the measured dipole energy evolution. We investigate this by changing the source distribution to a fully homogeneous setup and repeating all steps described in Sec.~\ref{sec:constraints}.
%We use the JF12-rand+reg magnetic field that is also used for the benchmark case.
The best-fit composition and spectrum at the sources in the homogeneous case are similar to the baseline ones, see Table~\ref{tab:results}.

\begin{figure}[ht]
\centering
\includegraphics[width=0.55\textwidth]{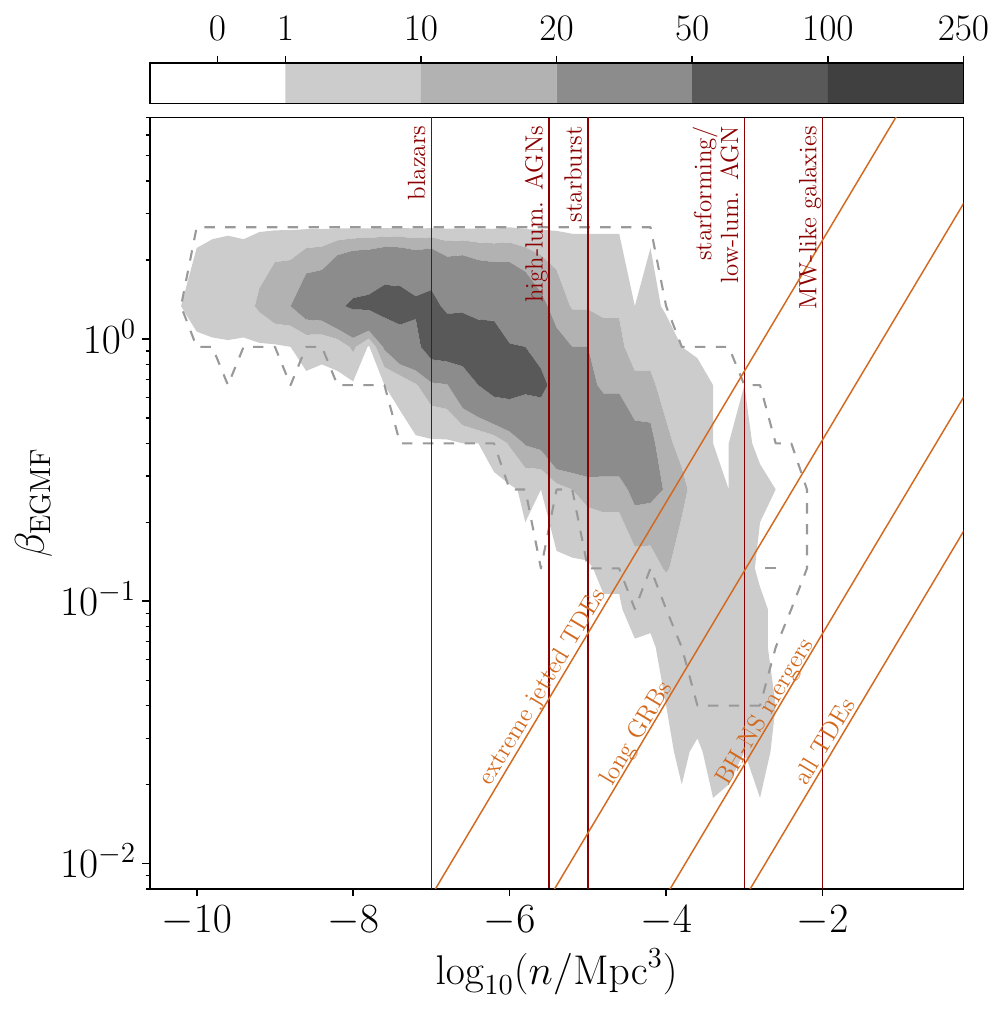}
\caption{Combined constraints on $\beta_\mathrm{EGMF}$ and the number density $n$; same as Fig.~\ref{fig:egmf_dens}, but for a homogeneous source distribution not following the LSS. Again, the dashed region stands for the non-excluded part in the case without Galactic random field while the color scale shows the number of samples for the baseline model with Galactic random field.  This shows that it is possible to explain the magnitude of the dipole and absence of higher multipoles as a statistical fluctuation of homogeneous sources, for a sufficiently rare source population and relatively strong EGMF, albeit in a fairly small fraction of trials (see Fig.~\ref{fig:egmf_dens} for comparison to the LSS model). However, the direction of the dipole is totally random and not predictable or interpretable in this scenario, unlike for the LSS model.}
\label{fig:egmf_dens_iso}
\end{figure}

Analogously to Fig.~\ref{fig:egmf_dens}, we show in Fig.~\ref{fig:egmf_dens_iso} the number of samples that simultaneously fulfill the two criteria explained in Sec.~\ref{sec:dens} (dipole amplitude above 5\% for $E>8$\,EeV and all higher multipole moments inside 99\% isotropic C.L.) for a homogeneous underlying source distribution. As the excluded regions for the models with and without Galactic random field are similar, all following conclusions apply independently of the Galactic turbulence.

\begin{figure*}[ht]
\includegraphics[width=0.45\textwidth]{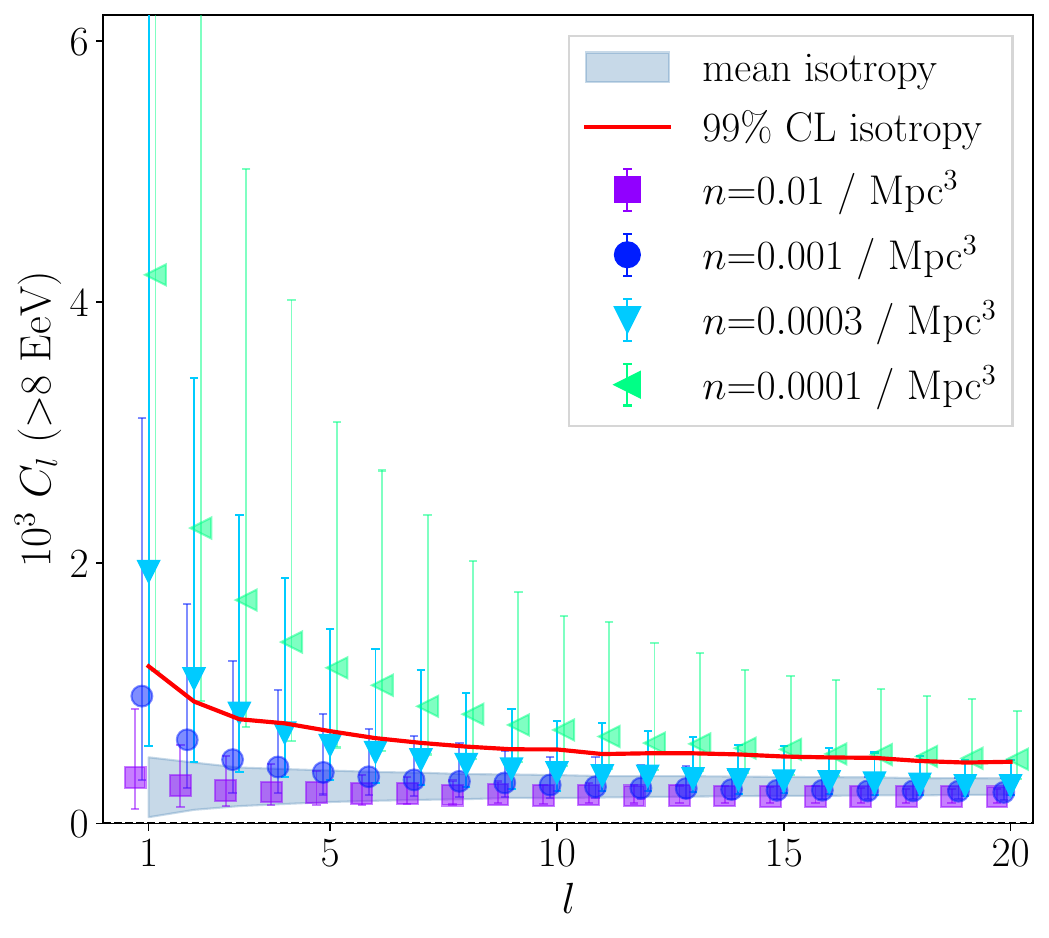}
\includegraphics[width=0.46\textwidth]{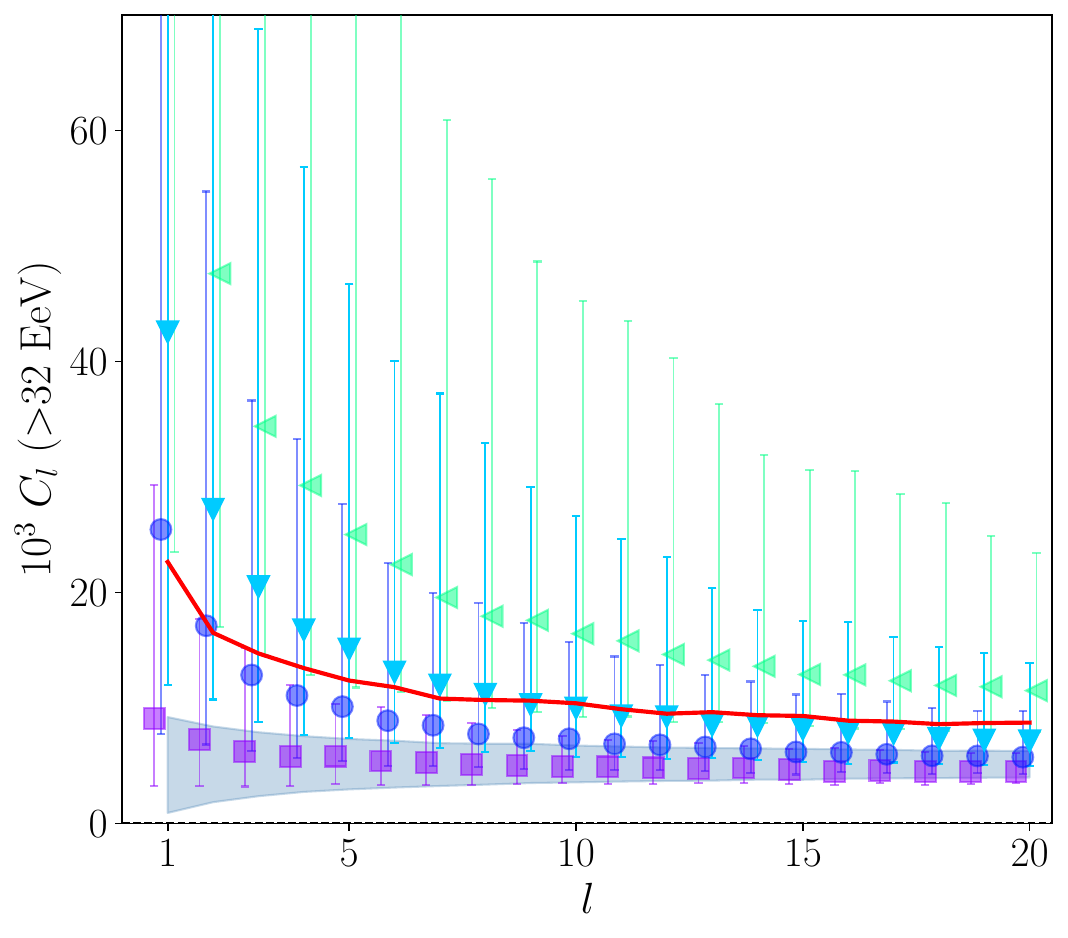}
\caption{Power spectrum for different number densities $n$; same as Fig.~\ref{fig:power_spectrum}, but for a homogeneous source distribution not following the LSS. The EGMF is zero, and the GMF is JF12 reg+rand.}
\label{fig:power_spectrum_iso}
\end{figure*}

In the case of negligible EGMF -- $\beta_\mathrm{EGMF}\lesssim 0.05$ -- no samples are found that fulfill both criteria. This is because relatively small densities of $n\lesssim 10^{-3}\, {\rm Mpc}^{-3}$ are necessary to produce a sizeable dipole from a homogeneous source distribution. But, for these small densities, the higher multipole moments are not compatible with isotropy. This can also be seen in the power spectra in Fig.~\ref{fig:power_spectrum_iso}.

If the EGMF is not negligible -- $\beta_\mathrm{EGMF}\gtrsim0.1$ -- it is possible to fulfill both criteria at the same time (Note that, as in the LSS case, strong EGMFs with $\beta_\mathrm{EGMF}\gtrsim3$ are excluded). It is however significantly less likely than in the LSS case (compare Fig.~\ref{fig:egmf_dens_iso} to Fig.~\ref{fig:egmf_dens}) as it is harder to produce a sizeable dipole from a homogeneous source distribution than from the LSS. The highest number of samples that fulfill both criteria simultaneously is 64 out of 1000 for the homogeneous source model, compared to $200/1000$ for the LSS model.

\begin{figure*}[ht]
\includegraphics[width=0.49\textwidth]{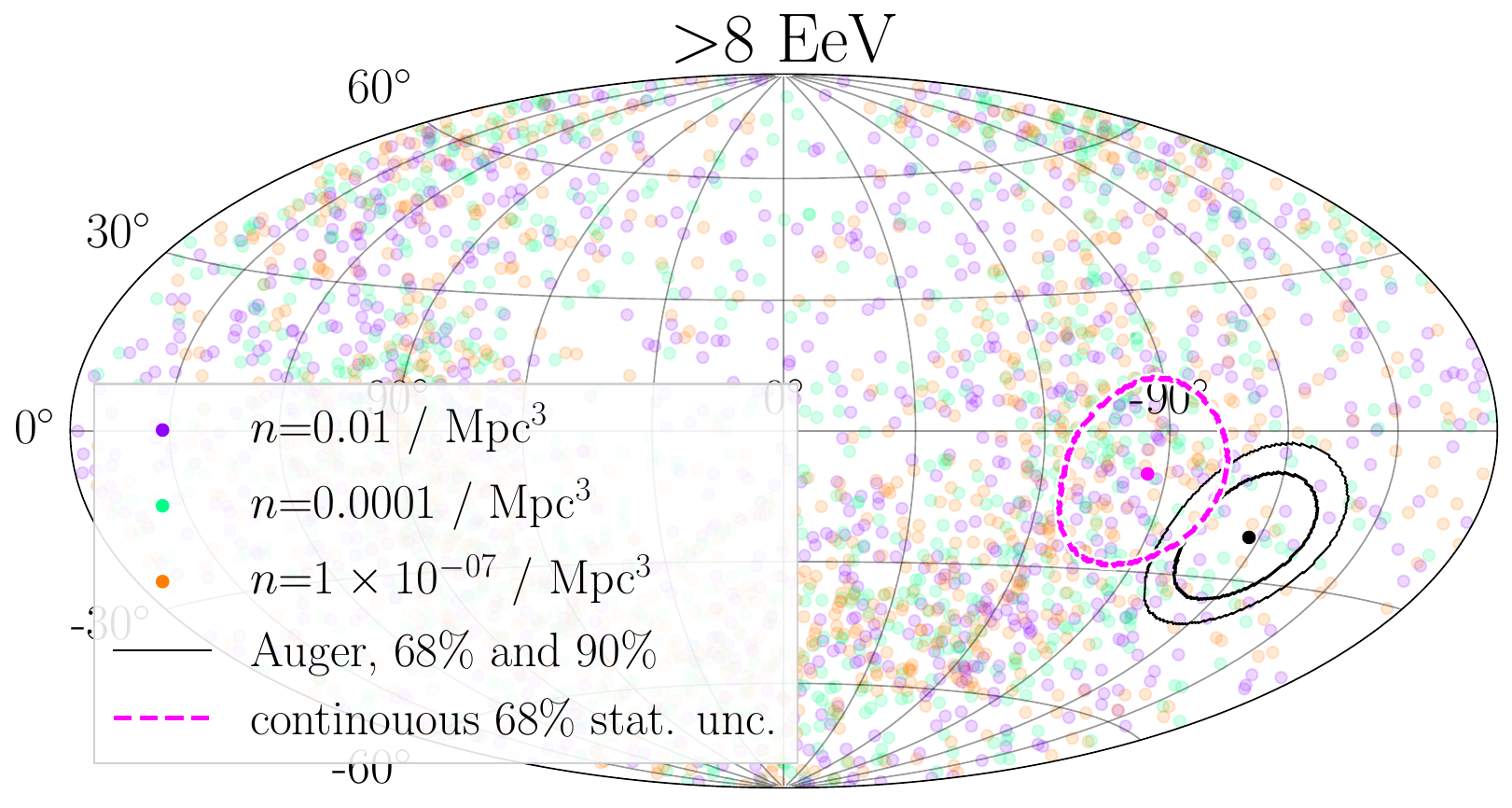}
\includegraphics[width=0.49\textwidth]{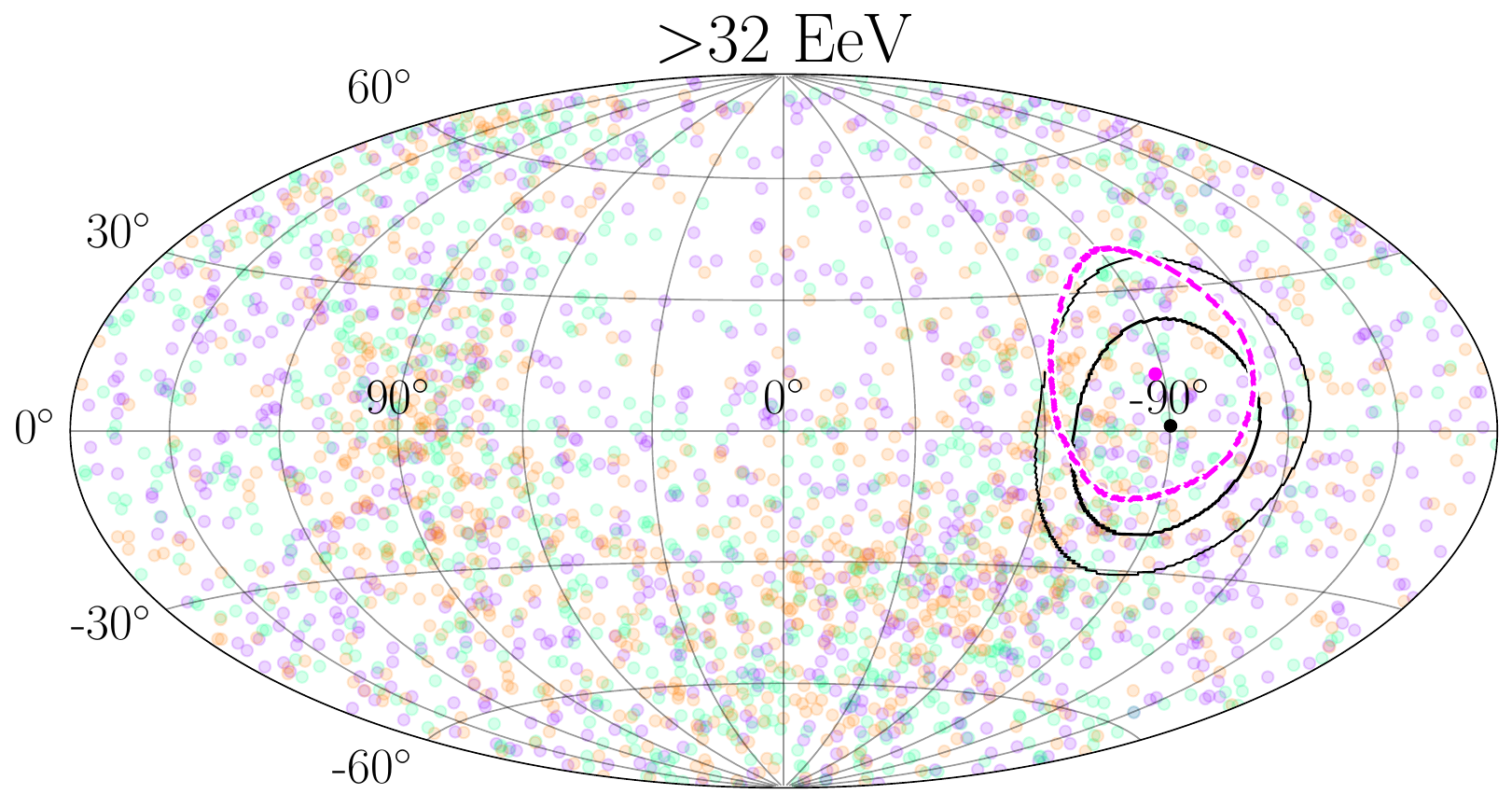}
\caption{Dipole directions of 100 samples for different number densities $n$; same as Fig.~\ref{fig:dens_A_skymaps} (\textit{upper}), but for a homogeneous source distribution not following the LSS. The EGMF is zero, and the GMF is JF12 reg+rand.}
\label{fig:dipole_direc_iso}
\end{figure*}

In addition, the dipole direction is roughly uniformly distributed in the homogeneous source case, which makes it much less likely to align with the measured one. This is visible in Fig.~\ref{fig:dipole_direc_iso}. Note that the dipole directions are not completely uniformly distributed due to the anisotropic extragalactic source distribution from sampling, in combination with preferred deflections of the GMF.

As for the homogeneous model, compared to the LSS one, significantly smaller source densities are necessary, it becomes unlikely that dense source candidates like Milky-Way-like / starforming galaxies or low-luminosity AGNs are the sources of UHECRs in that case. Also, the frequency of all TDEs is too large to be compatible. Even for the extreme jetted case, the likelihood of being compatible with data for a homogeneous underlying distribution is small, same as for long GRBs. Thus, in the case of homogeneous source distribution, sparse or infrequent source candidates like blazars are favored. This is different to the LSS model discussed in sec.~\ref{sec:AstImp}, for which dense or frequent objects are best compatible with the dipole and higher multipoles.

\section{EGMF and source number density constraints}
In table~\ref{tab:dens_egmf}, the number of simulations (out of 1000) that fulfill the two criteria stated in the second column of the table, are given for different values of $\beta_\mathrm{EGMF}$. For more details see sec.~\ref{sec:dens} and sec.~\ref{sec:egmf_dens}. 
\begin{table}[h!]
    \centering
    \begin{tabular}{l| l | l l l l l  l  l  l  l}
         & density $n / {\rm Mpc}^{-3}$ & $10^{-8}$ & $10^{-7}$ & $10^{-6}$ & $10^{-5}$ & $10^{-4}$  & $10^{-3}$  & $10^{-2}$  & $10^{-1}$  \\
         \hline \hline
         $\beta_\mathrm{EGMF}=0$ & $d_\mathrm{8\,EeV}$ >5$\%$ & 995 & 998 & 989 & 906 & 802 & 587 & 416 & 288\\
         & $C_{l>1}$ in $99\%$ isotropic expectation & 0 & 0 & 0 & 0 & 0 & 12 & 273 & 569 \\
         & both combined & \textbf{0} & \textbf{0} & \textbf{0} & \textbf{0} & \textbf{0}  & 5 & 100 & 152\\
         \hline
         $\beta_\mathrm{EGMF}=0.03$ & $d_\mathrm{8\,EeV}$ >5$\%$ & 993 & 998 & 981 & 882 & 789 & 560 & 411 & 293\\
         & $C_{l>1}$ in $99\%$ isotropic expectation & 0 & 0 & 0 & 0 & 0 & 36 & 344 & 635 \\
         & both combined & \textbf{0} & \textbf{0} & \textbf{0} & \textbf{0} & \textbf{0} & 20 & 137 & 176\\
         \hline
         $\beta_\mathrm{EGMF}=0.1$ & $d_\mathrm{8\,EeV}$ >5$\%$ & 989 & 993 & 971 & 836 & 701  & 455 & 293 & 185\\
         & $C_{l>1}$ in $99\%$ isotropic expectation & 0 & 0 & 0 & 1 & 15 & 180 & 547 & 744 \\
         & both combined & \textbf{0} & \textbf{0} & \textbf{0} & 1 & 5  & 66 & 132 & 114\\
         \hline
         $\beta_\mathrm{EGMF}=0.3$ & $d_\mathrm{8\,EeV}$ >5$\%$ & 958 & 959 & 859 & 586 & 394  & 172 & 50 & 9\\
         & $C_{l>1}$ in $99\%$ isotropic expectation & 0 & 0 & 6 & 107 & 287  & 634 & 833 & 944 \\
         & both combined & \textbf{0} & \textbf{0} & 4 & 18 & 39  & 60 & 17 & 8\\
         \hline
         $\beta_\mathrm{EGMF}=1$ & $d_\mathrm{8\,EeV}$ >5$\%$ & 156 & 132 & 100 & 42 & 5  & 5 & 0 & 0\\
         & $C_{l>1}$ in $99\%$ isotropic expectation & 101 & 250 & 606 & 867 & 959 & 985 & 998 & 1000 \\
         & both combined & 59 & 50 & 36 & 15 & 1 & \textbf{0} & \textbf{0} & \textbf{0}\\
         \hline
         $\beta_\mathrm{EGMF}=3$ & $d_\mathrm{8\,EeV}$ >5$\%$ & 0 & 0 & 0 & 0 & 0 & 0 & 0 & 0\\
         & $C_{l>1}$ in $99\%$ isotropic expectation & 186 & 343 & 720 & 959 & 985 & 994 & 995 & 998 \\
         & both combined & \textbf{0} & \textbf{0} & \textbf{0} & \textbf{0} & \textbf{0} & \textbf{0} & \textbf{0} & \textbf{0}\\
    \end{tabular}
    \caption{Same as table~\ref{tab:dens}, for different values of $\beta_\mathrm{EGMF} \equiv B / \mathrm{nG}\,\sqrt{L_c / \mathrm{Mpc}}$.}
    \label{tab:dens_egmf}
\end{table}

\section{Additional model predictions} \label{sec:predictions}
In the following, we will discuss predictions following from the LSS model, and what these would mean for UHECR measurements. The arrival-direction-dependent composition anisotropy presented in Sec.~\ref{sec:composition_ani} has already been investigated on data. The energy-flux correlation and the rigidity dependency of the dipole discussed in sec.~\ref{sec:energy_ani} respective sec~\ref{sec:rigidity_dep} could be used for future checks of the model's correctness.

\subsection{Directional composition anisotropy} \label{sec:composition_ani}
% - show that the LSS + GMF alone does not produce a composition anisotropy (as always stated by Eric): need nearby sources (?) at specific positions (?) (future paper?)
In the Auger data~\citep{Mayotte_ICRC_2021, Mayotte_ICRC_2023} there is an indication for a composition anisotropy, where the lighter events arrive predominantly in the Galactic North and South, while the heavier events arrive closer to the Galactic plane. To investigate, if this anisotropy could be an effect of the Galactic magnetic field, e.g. from the larger field strengths closer to the Galactic plane, we investigate the size of possible composition anisotropies over the sphere for the LSS model. 

\begin{figure}[ht!]
\includegraphics[width=0.48\textwidth]{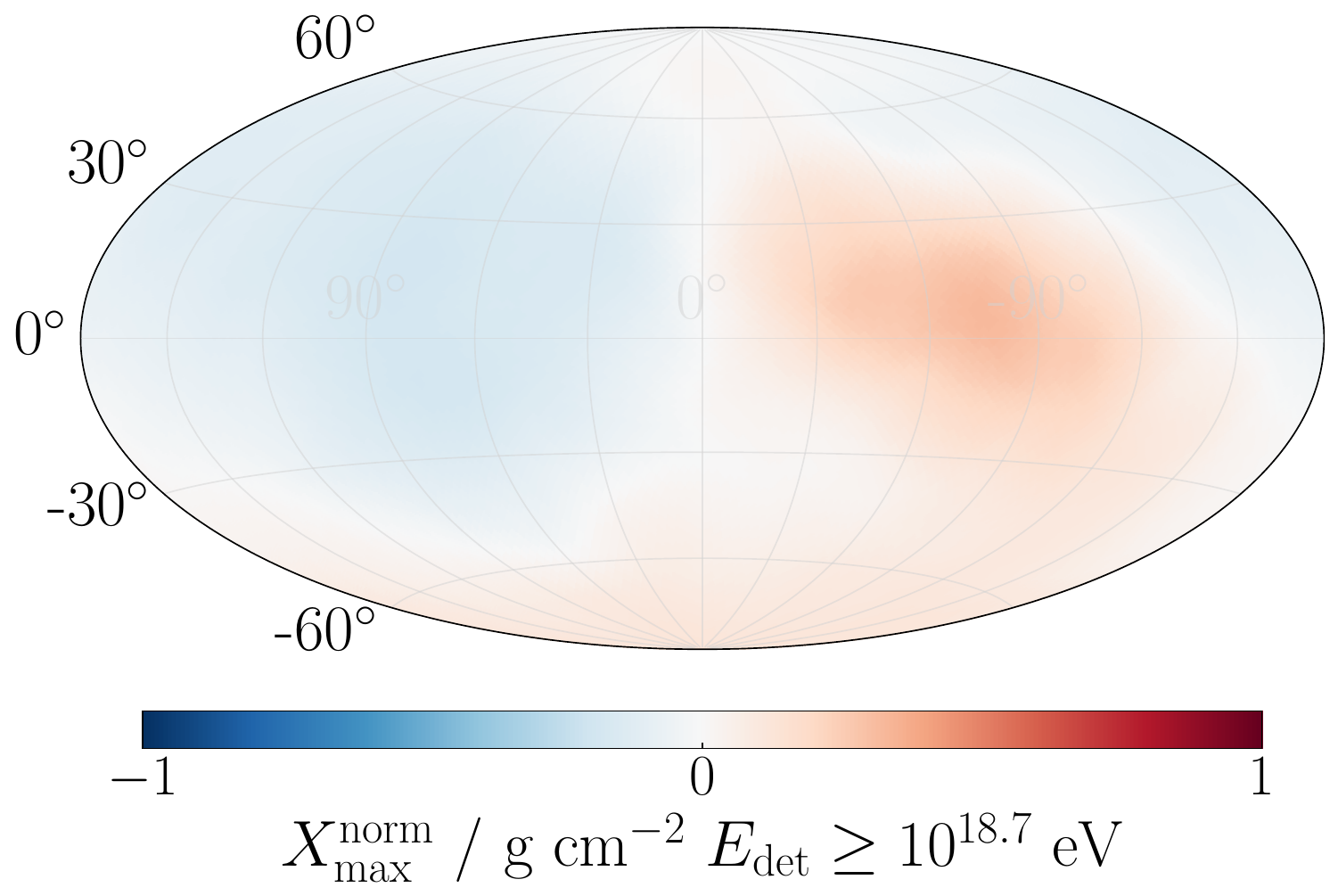}
\includegraphics[width=0.48\textwidth]{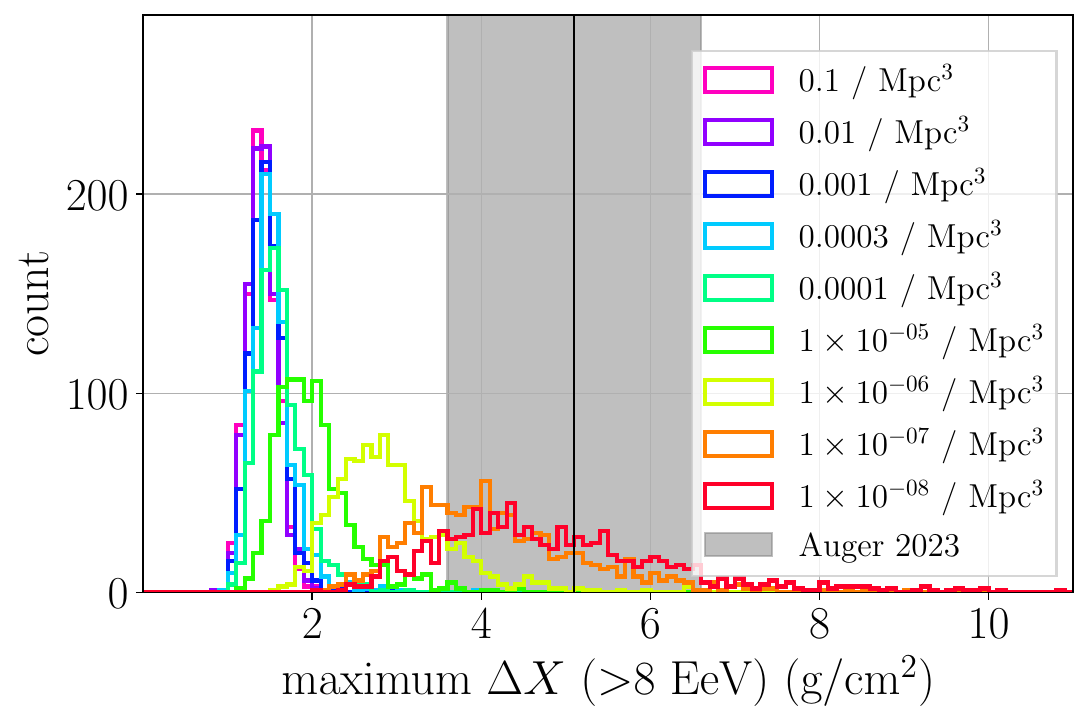}
\caption{\textit{Left:} Normalized (corrected for the energy evolution) maximum shower depth $\xmax^\mathrm{norm}$ for the continuous baseline model, with a $30^\circ$ tophat smoothing; in analogy to~\citet{Mayotte_ICRC_2021}. \textit{Right:} Histogram of the maximum difference $\Delta X$ between smallest and largest $\xmax^\mathrm{norm}$ in $30^\circ$ tophat smoothed maps such as the one on the left, as a function of the source number density $n$, assuming 4000 events.}
\label{fig:heaviness_map}
\end{figure}

The normalized shower depth~\citep{Mayotte_ICRC_2021} (corrected for the energy evolution effect) as a function of the direction, is displayed in Fig.~\ref{fig:heaviness_map} (\textit{left}). For the baseline continuous model, the composition does not vary greatly over the sky, with a maximum difference of $\Delta X \lesssim2$ g/cm$^2$. The lighter part (red) is correlated with the flux overdensity seen in Fig.~\ref{fig:dipole_direc}, which is expected since the lighter (low $Z$) component diffuses less in the GMF. The correlation of the heavier part of the events with the Galactic plane region, as observed in Auger data, is not seen for the LSS model. From this, it can be concluded that it is not a generic feature induced by the GMF, but instead, if not just a coincidence, must be due to specific source directions. This could be investigated in a future publication.

The effect of the source number density on the composition anisotropy is visualized in Fig.~\ref{fig:heaviness_map} (\textit{right}) for the case of no EGMF. The maximum difference between the smallest and largest $\xmax^\mathrm{norm}$ in $30^\circ$ tophat smoothed maps (like Fig.~\ref{fig:heaviness_map}) of the 1000 randomly drawn realizations is depicted. For the maps, we use a smaller event statistic of $N_\mathrm{events}=4000$, as the analysis is based on data by the fluorescence detector with a limited duty cycle. It is visible how only very small densities $n\lesssim10^{-7}\, {\rm Mpc}^{-3}$ can lead to differences between on- and off-plane regions approaching the value of $9.8$ g/cm$^2$ reported in~\citet{Mayotte_ICRC_2021}. To reach the updated value of $5.1$ g/cm$^2$~\citep{Mayotte_ICRC_2023}, still a density as small as $n\approx10^{-6}\, {\rm Mpc}^{-3}$, is necessary. As described above, such small densities lead to very large flux anisotropies which are not in agreement with the data for negligible EGMF.

We have also investigated the possibility of arrival-direction-dependent composition anisotropies for the homogeneous source model (Appendix~\ref{sec:homogeneous}), and have found that the directions of possible anisotropies in the sky are more random in that case. The size of the maximum $\Delta X$, however, does not depend on the source model.

\subsection{Directional energy anisotropy} ~\label{sec:energy_ani}
\begin{figure}[ht]
\includegraphics[width=0.24\textwidth]{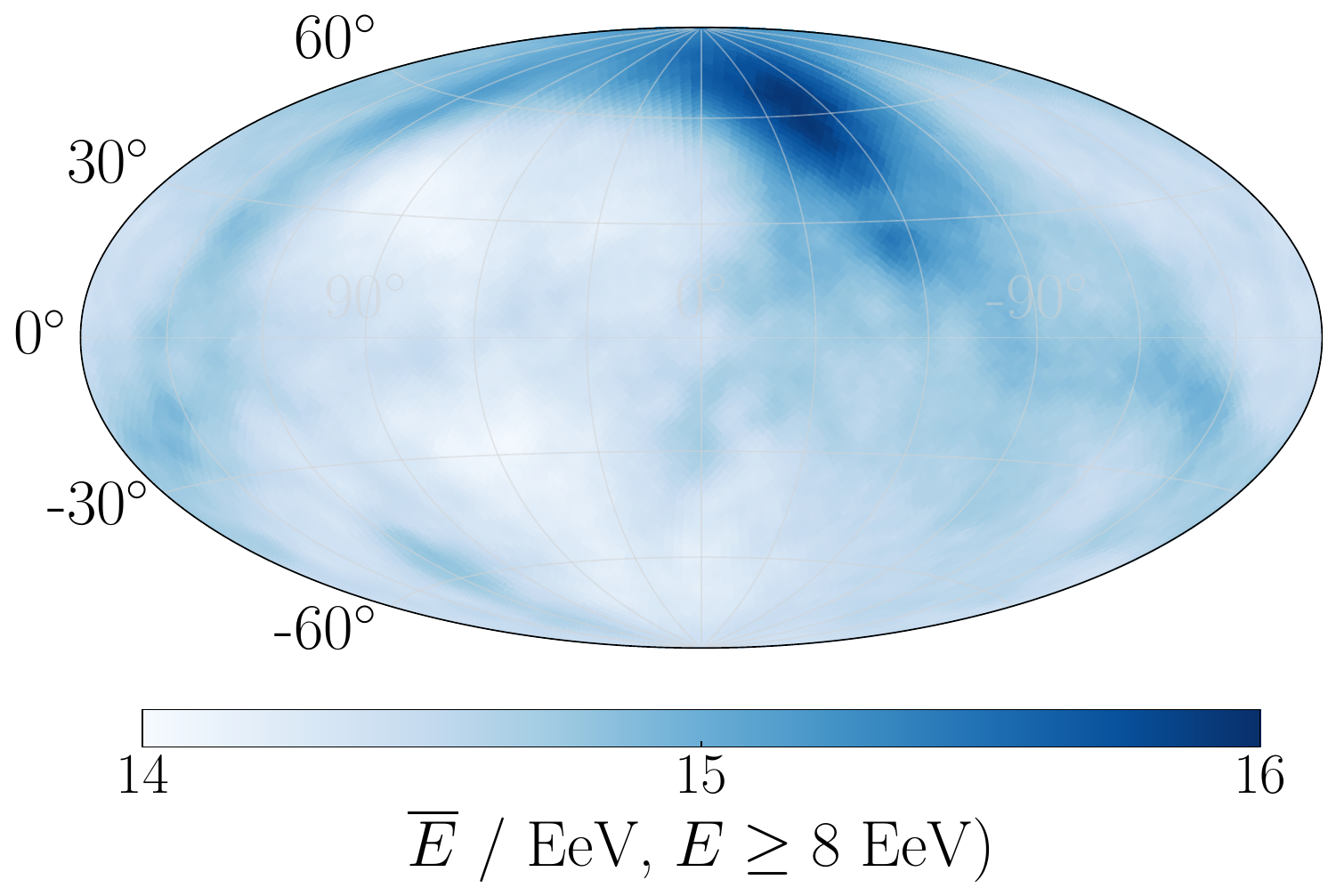}
\includegraphics[width=0.24\textwidth]{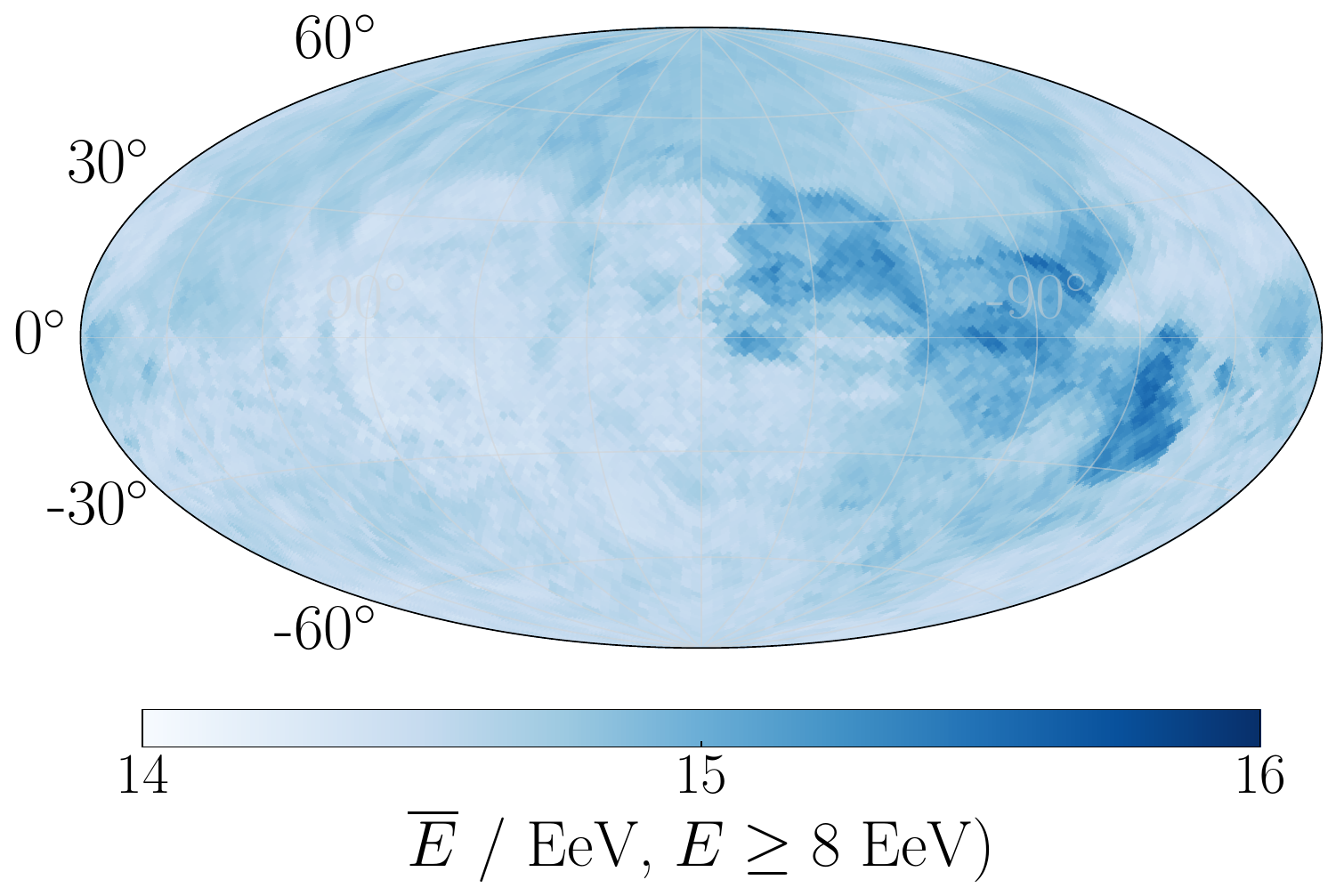}
\includegraphics[width=0.24\textwidth]{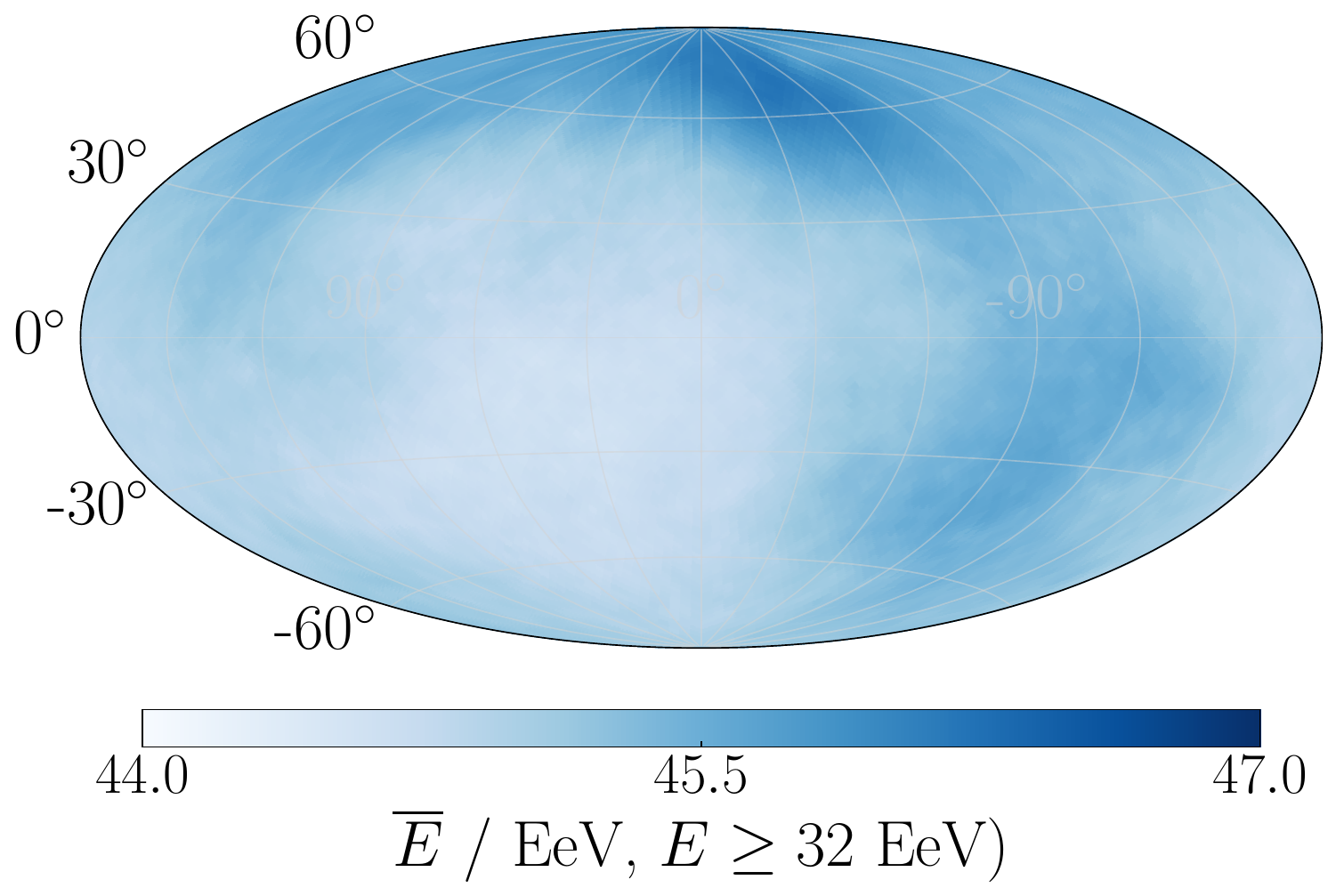}
\includegraphics[width=0.24\textwidth]{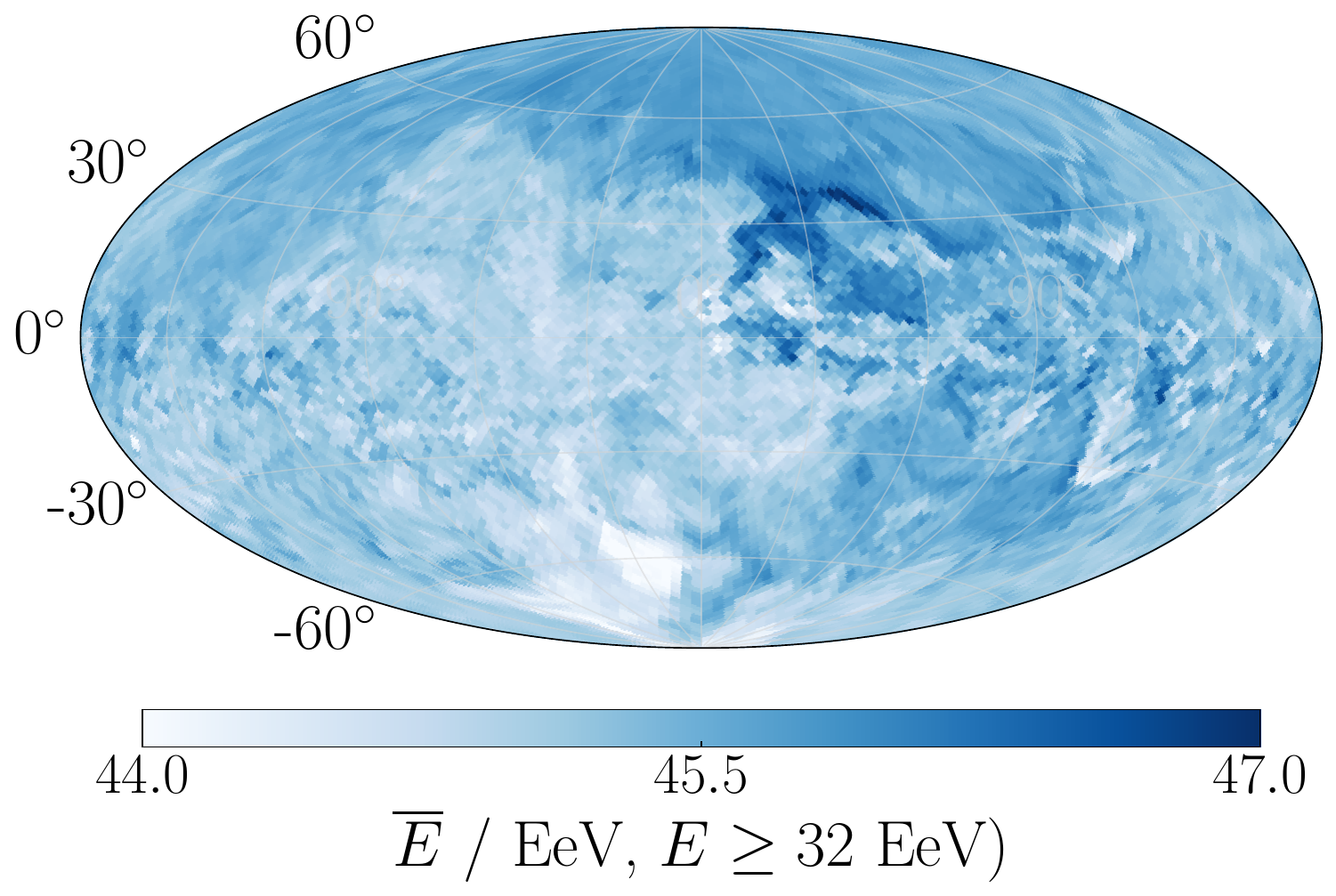}
\includegraphics[width=0.244\textwidth]{best_fit_dipole__I_map_9_-1.pdf}
\includegraphics[width=0.244\textwidth]{best_fit_dipole__map_9_-1noTH.pdf}
\includegraphics[width=0.244\textwidth]{best_fit_dipole__I_map_15_-1.pdf}
\includegraphics[width=0.244\textwidth]{best_fit_dipole__map_15_-1noTH.pdf}
\caption{The \textit{upper row} shows the mean energy per pixel, for the illumination (\textit{first column} for $E>8$ EeV, \textit{third column} for $E>32$ EeV) and the arrival (\textit{second column} for $E>8$ EeV, \textit{fourth column} for $E>32$ EeV). For comparison, the flux maps for the respective energy thresholds are depicted in the \textit{lower row}. A correlation between larger flux and higher energy is visible.} 
\label{fig:e_sphere}
\end{figure}

We have also tested the possibility of a directional energy anisotropy using our model. In Fig.~\ref{fig:e_sphere}, we show the mean energy per pixel (\textit{upper row}) for the illumination and the arrival maps for two energy thresholds for the baseline model. For comparison, the respective flux maps are also shown (\textit{lower row}). A clear correlation is visible: the regions with larger flux also exhibit higher energies which is an effect of the extragalactic propagation. Note that the mean energy is dominated more by nearby structures than the flux, as can be seen in comparison to the illumination maps for different distances shown in Fig.~\ref{fig:illum}. 

We have estimated roughly if the correlation between energy and flux could be tested by Auger by using the simulated event samples described in sec.~\ref{sec:dens}. We have found that the expected correlation coefficient is relatively small $\mathcal{O}(\lesssim5\%)$ which is mostly due to limited statistics. Thus, such a correlation could be a target for future observatories like GCOS~\citep{AlvesBatista_GCOS}.

\subsection{Rigidity dependency of the dipole} ~\label{sec:rigidity_dep}
\begin{figure}[ht]
\includegraphics[width=0.24\textwidth]{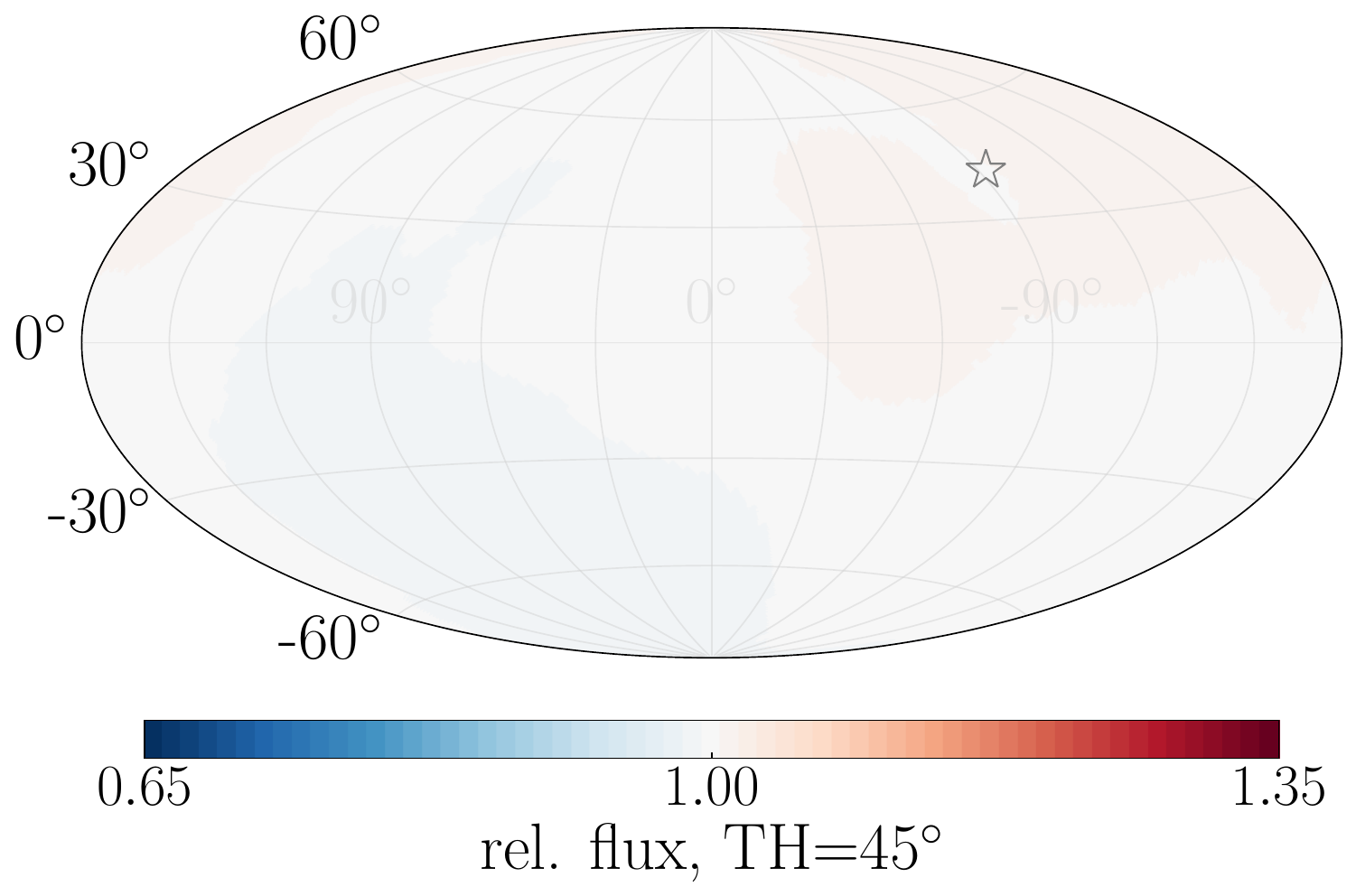}
\includegraphics[width=0.24\textwidth]{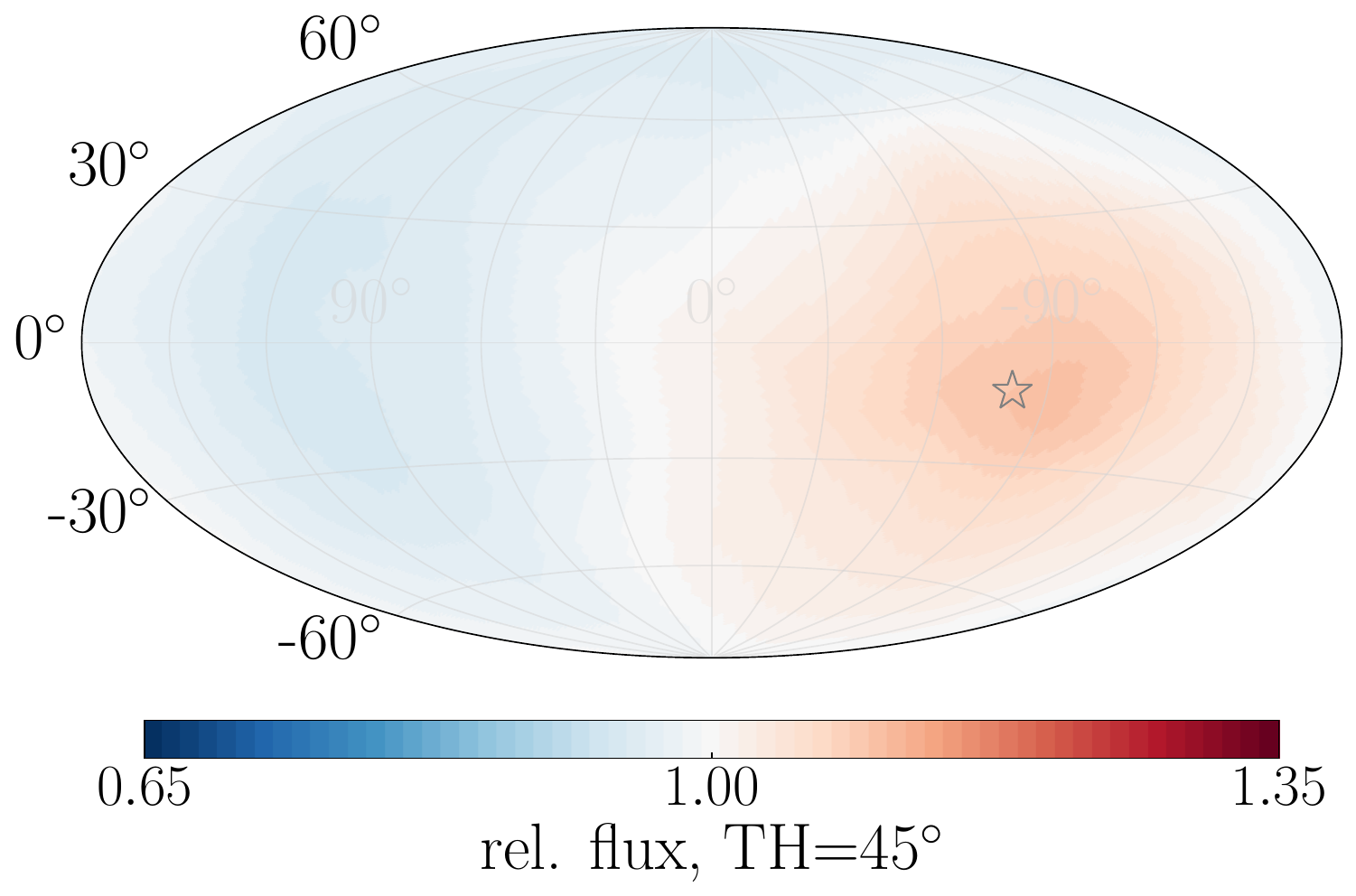}
\includegraphics[width=0.24\textwidth]{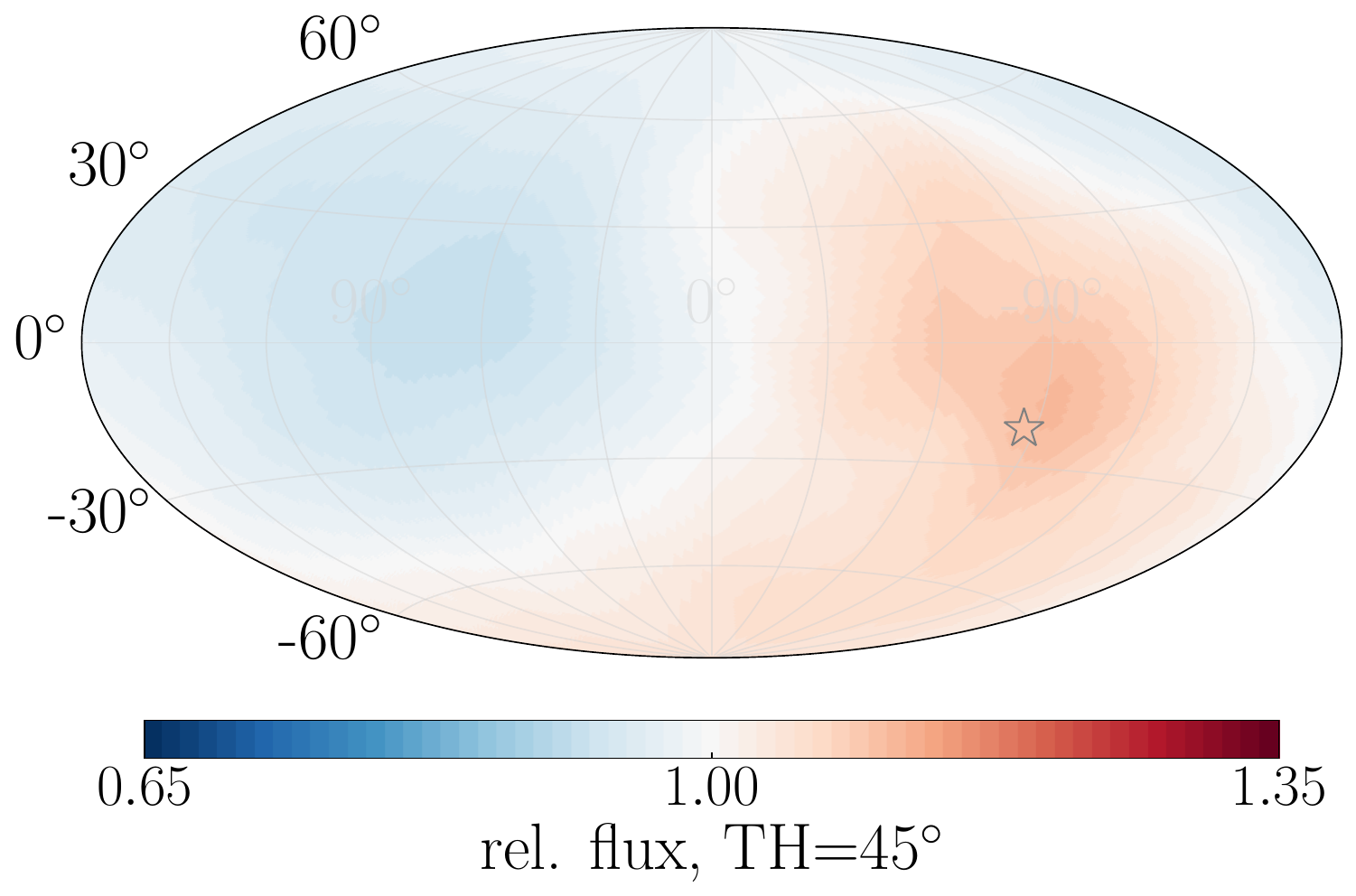}
\includegraphics[width=0.24\textwidth]{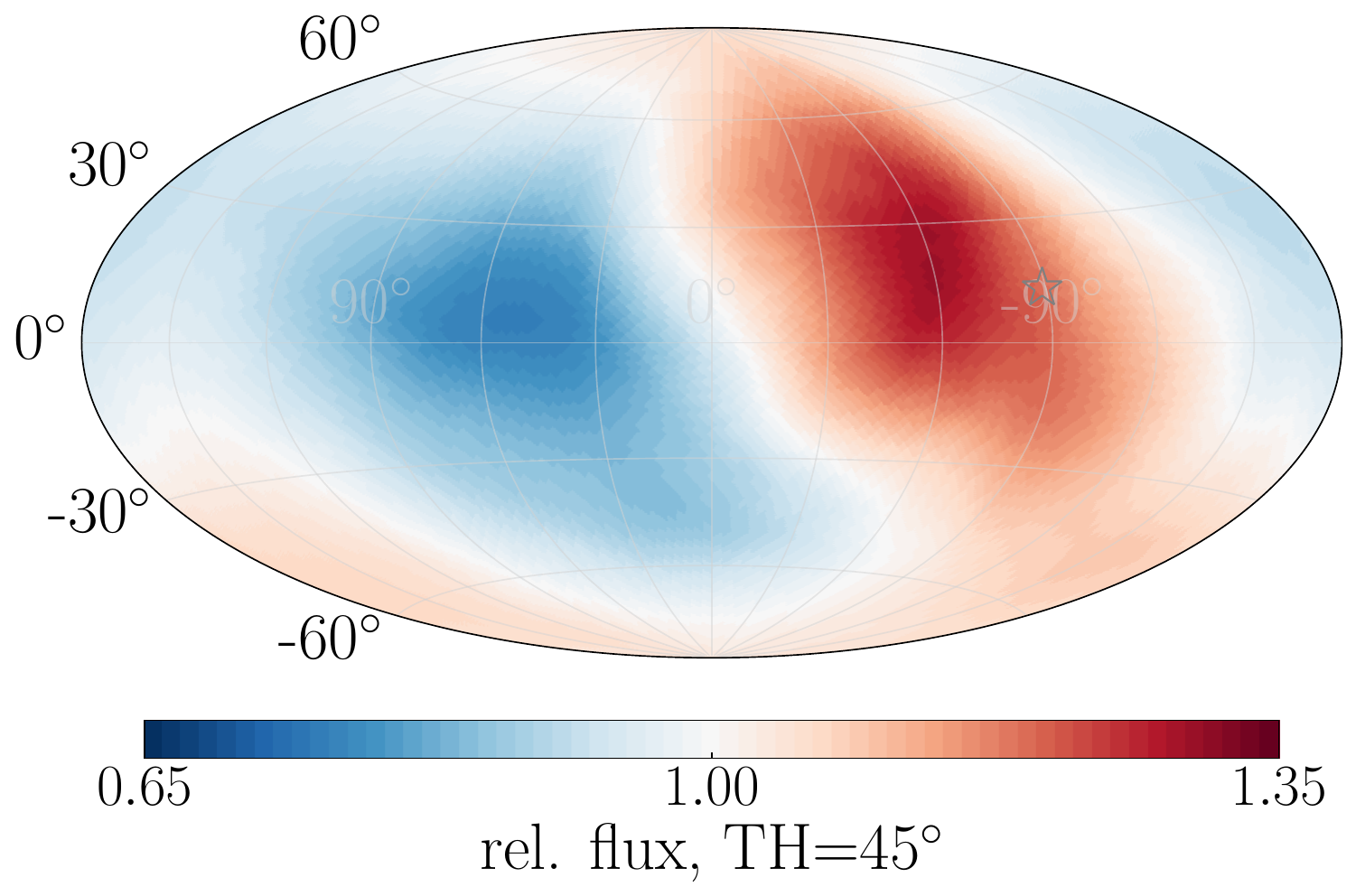}
\caption{Predicted rigidity dependency of the flux $E>8\,\text{EeV}$. Each map contains 25\% of the expected events ordered in rigidity from lowest to highest. The demarcation rigidities are $R=1.7\,\mathrm{EV}$, $R=4.3\,\mathrm{EV}$, and $R=5.6\,\mathrm{EV}$. The star marker indicates the direction of the dipole component.} 
\label{fig:rig_dipole}
\end{figure}

Due to the improving sensitivity of UHECR observatories to the masses and charges of individual events thanks to deep learning methods and detector upgrades~\citep{Castellina_2019, Jonas_ICRC2023}, it can be expected that effects like a possible evolution of the dipole amplitude and direction with the rigidity can be tested in the near future. The predicted dipole evolution with the rigidity based on our LSS model is shown in Fig.~\ref{fig:rig_dipole}. Here, it is clearly visible how the highest rigidity events lead to a much stronger dipole amplitude as they are less affected by diffusion in the GMF. The effect is significantly stronger as a function of rigidity than it is seen for the different energy bins in Fig.~\ref{fig:dipole_direc} and~\ref{fig:dipole_amp}, as expected because magnetic field deflections scale with rigidity. The events with the lowest rigidity even show almost no dipolar anisotropy at all, while the highest rigidity ones arrive significantly more in the Galactic North. 
% The highest rigidities are mostly reached by helium nuclei with $R\lesssim10\,\text{EV}$ as almost no protons are expected at $E>8\,\text{EeV}$ (see Fig.~\ref{fig:best_fit_EXmax}).

Testing for a rigidity evolution of the dipole, instead of just the energy evolution would allow us to disentangle the effects of propagation and GMF deflection, and would thus work as a proof of UHECR deflections in the GMF. Additionally, identifying regions where the highest rigidity events are expected could be advantageous for backtracking to identify individual source candidates. 

\bibliography{bibliography}{}
\bibliographystyle{aasjournal}



\end{document}